\documentclass[fleqn,usenatbib]{mnras}

\usepackage{newtxtext,newtxmath}

\usepackage[T1]{fontenc}

\DeclareRobustCommand{\VAN}[3]{#2}
\let\VANthebibliography\thebibliography
\def\thebibliography{\DeclareRobustCommand{\VAN}[3]{##3}\VANthebibliography}

\usepackage{graphicx}	
\usepackage{amsmath}	
\usepackage{natbib} 

\usepackage{xcolor}
\usepackage{hyperref}
\newcommand{\orc}{\includegraphics[height=\fontcharht\font`A]{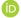}}
\newcommand{\orcid}[1]{\href{https://orcid.org/#1}{\orc}}

\usepackage[normalem]{ulem}

\usepackage{indentfirst}

\usepackage{newtxtext,newtxmath}

\graphicspath{{./}{figures/}}

\def\gsim{\;\rlap{\lower 2.5pt
 \hbox{$\sim$}}\raise 1.5pt\hbox{$>$}\;}
\def\lsim{\;\rlap{\lower 2.5pt
   \hbox{$\sim$}}\raise 1.5pt\hbox{$<$}\;}

\title[Emission-Line Surface Brightness in a Multiphase Galactic Wind]{Modeling Emission-Line Surface Brightness in a Multiphase Galactic Wind: An O VI Case Study}

\author[Z. Chen \& Z. Peng et al.]{
Zirui Chen\orcid{0000-0001-8755-3836}$^{1}$\thanks{E-mail: ziruichen@ucsb.edu}\thanks{Z. Chen and Z. Peng contributed equally to this work and are co-first authors.}, 
Zixuan Peng\orcid{0000-0003-3467-6810}$^{1}$\thanks{E-mail: zixuanpeng@ucsb.edu}\footnotemark[2],
Kate H. R. Rubin\orcid{0000-0001-6248-1864}$^{2}$,
Timothy M. Heckman\orcid{0000-0001-6670-6370}$^{3,4}$,
Matthew J. Hayes\orcid{0000-0001-8587-218X}$^{5}$,
\newauthor
Yakov Faerman\orcid{0000-0003-3520-6503}$^{6,7}$,
Crystal L. Martin\orcid{0000-0001-9189-7818}$^{1}$,
S. Peng Oh\orcid{0000-0002-1013-4657}$^{1}$,
and Drummond B. Fielding\orcid{0000-0003-3806-8548}$^{8}$
\\
$^{1}$Department of Physics, University of California at Santa Barbara, Santa Barbara, CA 93106, USA
\\
$^{2}$Department of Astronomy, San Diego State University, San Diego, CA 92182, USA 
\\
$^{3}$Center for Astrophysical Sciences, Department of Physics \& Astronomy, Johns Hopkins University, Baltimore, MD 21218, USA
\\
$^{4}$School of Earth and Space Exploration, Arizona State University, Tempe, AZ 85287, USA
\\
$^{5}$Stockholm University, Department of Astronomy and Oskar Klein Centre for Cosmoparticle Physics, AlbaNova University Centre, SE-10691, Stockholm, Sweden
\\
$^{6}$University of Washington, Department of Astronomy, 3910 15th Avenue NE, Seattle, WA 98195, USA
\\
$^{7}$School of Physics and Astronomy, Tel Aviv University, Tel Aviv 69978, Israel
\\
$^{8}$Department of Physics, New York University, 726 Broadway, New York, NY 10003, USA
}

\date{Accepted 2026 March 27. Received 2026 March 24; in original form 2025 October 2}

\pubyear{2026}

\begin{document}
\label{firstpage}
\pagerange{\pageref{firstpage}--\pageref{lastpage}}
\maketitle

\begin{abstract}
We present a fast and robust analytic framework for predicting surface brightness (SB) of emission lines in galactic winds as a function of radius up to $\sim 100$ kpc out in the circum-galactic medium. 
We model multiphase structure in galactic winds by capturing emission from both the volume-filling hot phase (T $\sim 10^{6-7}$ K) and turbulent radiative mixing layers that host intermediate temperature gas at the boundaries of cold clouds (T $\sim 10^4$ K). 
Our multiphase framework makes significantly different predictions of emission signatures compared to traditional single-phase models and explains the paucity of O\,{\footnotesize VI} SB measurements in the literature. 
After accounting for ram pressure equilibrium between the cold clouds and hot wind in supersonic outflows, non-equilibrium ionization effects, and energy budgets other than mechanical energy from core-collapse supernovae, our O\,{\footnotesize VI} SB predictions qualitatively match observational results. 
Our framework provides constraints on the optimal galactic wind properties that facilitate O\,{\footnotesize VI} emission observations, including star formation rate surface density, hot phase mass loading factor, and thermalization efficiency factor. These constraints are consistent with existing observations and can help inform future target selections.
\end{abstract}




\defcitealias{FB_2022}{FB22}
\defcitealias{Thompson_2016}{T16}
\defcitealias{Hayes_2016}{H16}
\defcitealias{PS_2020}{PS20}

\section{Introduction}
Galaxy evolution is shaped by how gas is exchanged between the galactic disk and the surrounding circumgalactic medium (CGM) \citep{Tumlinson:2013, tumlinson17, FGOH23}. One main mechanism of this gas exchange is galactic-scale outflows known as galactic winds, which are ubiquitously observed in star-forming galaxies at low and high redshifts \citep{Martin:1999, Pettini:2001, Shapley:2003, Rubin_2014, Martin_2015, Peng_2025}. 
Galactic winds are multiphase, consisting of a volume-filling phase of hot ionized gas ($\sim$ $10^{6-7}$ K) traced by X-ray emission \citep[e.g.,][]{Lopez_2020} and a much colder and denser phase ($\sim$ $10^4$ K) traced by optical emission and UV absorption lines \citep[e.g.,][]{Heckman_2015, Xu_2022, Peng_2025}, where most observations have been conducted.

A rich literature of ``down-the-barrel'' absorption-line spectroscopy studies of galactic winds \citep[e.g.,][]{Heckman_2015, Xu_2022, Xu_2023, Perrotta_2023} provided valuable insights into determining mass, momentum, and energy outflow rates of galaxies and how these outflow rates depend on global galactic properties like star formation rate (SFR). 
On the other hand, while mapping emission lines is more challenging due to the diffuse nature of the CGM, doing so often provides valuable insights into the spatial structure of outflows. 
Recent work has successfully mapped these outflows in various emission-line tracers, including Mg\,{\footnotesize II} \citep[e.g.,][]{Burchett2021}, [O\,{\footnotesize II}] \citep[e.g.,][]{Rupke2019, Shaban_2022}, and [O\,{\footnotesize III}] or Balmer lines \citep[e.g.,][]{Reichardt_Chu2022, Nielsen_2024, Herenz_2025}.
Of particular interest are the very few observations of O\,{\footnotesize VI} emission from extragalactic sources \citep[][]{Otte03, Grimes07, Hayes_2016, Chung21, Kim_2024, Ha_2025}, as O\,{\footnotesize VI} traces warm-hot gas at $\sim 10^{5.5}$ K, which plays an important role in radiative cooling in the CGM and is crucial for determining the CGM phase structure.
Among this small handful of studies, only \citet[][hereafter H16]{Hayes_2016} and \cite{Ha_2025} are sensitive to the O\,{\footnotesize VI} surface brightness (SB) well beyond the stellar components of the galaxies they target ($\gtrsim 10$ kpc).

In parallel with observations, simulations and theoretical works of galactic outflows have seen significant progress in recent years. Galaxy-scale simulations revealed the complex, multiphase nature of outflows \citep[e.g.,][]{Schneider:2020}. The formation mechanism of such multiphase structure, specifically how cold, dense clouds can survive hydrodynamic instabilities induced by the relative motion of phases, has received widespread theoretical interest and has been explored in several recent works \citep{scannapieco15, schneider17, zhang17}.
Cold gas structures in the galactic disk and how they are launched in outflows have been studied by numerous interstellar medium (ISM) simulations \citep{Joung_2006, Walch_2015, Kim_2017, Marinacci_2019, Kim_2021, Tan_2023}.
On smaller scales, physics governing the fate of cold gas 
in hot outflows was studied in 
wind tunnel simulations, where a single cold cloud is subject to an impinging hot wind \citep{armillotta17, gronke18, gronke20-cloud, li20, kanjilal21, abruzzo22, chen24, kaul25}. 
These works highlight the importance of turbulent mixing and radiative cooling in facilitating cloud survival and growth. 

These studies greatly advanced our understanding of the formation and evolution of multiphase galactic outflows. However, connecting theory with observations and making predictions of observables like absorption and emission line signatures requires zooming in even further to turbulent radiative mixing layers (TRMLs) that form at the interface between cold clouds and the hot galactic outflow
\citep{begelman1990}.
Simulations of TRMLs \citep{kwak10, Fielding_2020, Tan_2021, tan21-lines} allow for detailed studies of the mass, momentum, and energy exchange between the phases. 

Despite recent advancements in observing galactic outflows via absorption and emission, as well as understanding their sub-parsec scale TRML structure through 
simulations, uncertainties still exist in both observations and simulations, particularly regarding the bulk properties of galactic outflows, including mass, momentum, and energy outflow rates for both hot and cold phases. 

Observational studies \citep[e.g.,][]{Martin_2006, Rubin_2014, Heckman_2015, Chisholm_2016, Xu_2022} primarily use blueshifted absorption lines to estimate the cold phase mass ($\dot{M}_{\rm{out,cl}} \propto M_{\rm{out,cl}} \ v_{\rm{out,cl}} \ / \ r_{\rm{out}}$) and energy ($\dot{E}_{\rm{out,cl}} \propto \dot{M}_{\rm{out,cl}} \ v_{\rm{out,cl}}^2$) outflow rates. 
However, these estimates suffer from huge systematic uncertainties because many studies that utilize integrated spectra often rely on the assumption of a mass-conserving outflow with a constant outflow velocity, an outflow radius, $r_{\rm{out}}$ assumed to be several times of the half-light radius $r_{\rm{50}}$, and a $r^{-2}$ density profile. These assumptions are not necessarily well-justified and can lead to overestimating 
the actual outflow rates by a factor of 10 compared to more valid treatments of these outflow properties for UV absorption lines \citep{Chisholm_2016}.
Spatially resolved outflow studies using IFU data can trace cold-phase outflows by detecting broad wings in emission lines and thereby relax some of the assumptions made in integrated absorption-line studies.
Nevertheless, emission-line studies lacking reliable constraints on filling factor measurements of cold clouds \citep{Xu_2023b, Martin_2024} can overestimate the wind density by up to 2 dex and the corresponding mass outflow rate by 4 dex \citep{Howatson_2025}.
Moreover, absorption ($N \propto \int n(r) \ dl$) and emission ($\rm{EM} \propto \int n(r)^2 \ dl$) lines have different density dependencies; therefore, absorption lines are more likely to trace diffuse gas at larger radii with higher outflow velocities, compared to emission lines that trace denser gas near the galactic disk at relatively low outflow velocities \citep{Wood_2015, Xu_2024}. 
\citet{Wood_2015} find that the derived $\dot{M}_{\rm{out}}$ from UV absorption lines is approximately one dex higher than that determined from H$\alpha$. In order to understand and eliminate these observational uncertainties, utilizing insights about the structure and properties of galactic winds from simulations is crucial.

On the theory/simulation side, we lack a self-consistent theory for determining the hot-phase mass loading factor, $\eta_{\rm{M,hot}}$, and the thermalization efficiency factor, $\eta_{\rm{E}}$, which control the mass and energy input rates of hot winds \citep{TH_2024}.
Cosmological simulations use simple subgrid feedback models to reproduce the observed scaling relations of galaxies, such as the mass-metallicity relation; however, they are unable to produce physical outflow rates \citep{SD_2015, NO_2017, Torrey_2019}.
Recent advances in hydrodynamic simulations of isolated/single dwarf galaxies, with the mass resolution improved to $\lesssim 10 \ M_{\odot}$ scale, can alleviate systematic biases due to subgrid feedback models \citep[][]{Smith_2021, Natalia_2023,SG_2025} but are still constrained to specific model setups and are too computationally expensive to compare with observations.

Bridging the gap between observations and theory of galactic winds and resolving the existing uncertainties in both fields are crucial. The first steps in this direction have already been undertaken by analytic models of mixing layers in galactic winds \citep{Tan_2021, Chen_2023}, which can predict observables like absorption line column densities and emission line SB. However, such theoretical predictions are formulated in terms of mixing layer parameters like relative shear velocity and density contrast between the phases, which are not directly observable and thus pose challenges when comparing with observations. \citet[][hereafter FB22]{FB_2022} presented an analytic model for multiphase galactic winds that fills this gap by taking bulk outflow properties like SFR, $\eta_{\rm E}$, and $\eta_{\rm M,hot}$ as input. 
Their model predicts radial profiles of important outflow properties like temperature and density for both hot and cold phases, but additional work is needed to convert these outputs into absorption and emission signatures seen in observations.

This paper is motivated by recent advances in our understanding of galactic outflows surveyed above. 
To better connect observational and theoretical outflow studies and constrain outflow rates in galaxies, we present an analytic multiphase galactic wind framework, \texttt{WInterPhase}\footnote{\url{https://github.com/jasonpeng17/WInterPhase}}, which allows generating emission line SB and line ratio predictions as a function of bulk galactic outflow properties. 
Our framework combines the models of \citetalias{FB_2022} and \cite{Chen_2023}. Unlike traditional single-phase galactic outflow models \citep[e.g.,][]{CC85, Thompson_2016, Danehkar_2021, Danehkar_2022, Sarkar_2022} that only consider the volume-filling hot phase as it adiabatically expands and/or radiatively cools, we model a galactic outflow as cold, dense clouds embedded in a volume-filling, outflowing hot phase (as in \citetalias{FB_2022}) and account for the mass, momentum, and energy exchange between the hot and cold phases using the mixing layer model in \cite{Chen_2023}. This allows us to capture emission from both the volume-filling hot phase and TRMLs.

This work is organized as follows. 
In \autoref{sec:model}, we present the methodology for generating emission-line SB predictions from our multiphase galactic wind framework. 
In \autoref{sec:results}, we demonstrate the predictions of our multiphase framework for the O\,{\footnotesize VI} $\lambda\lambda 1031,1037$ doublet and compare them with those of a traditional single-phase model \citepalias{Thompson_2016},
while examining the effects of varying hot- and cold-phase parameters on the SB radial profiles. 
In \autoref{sec:discussions}, we compare our predicted O\,{\footnotesize VI} SB profiles with observations from the literature, discuss potential physical mechanisms that may account for the discrepancies, and outline the implications arising from this comparison. 
We conclude by summarizing our findings in \autoref{sec:conclusion}.
Throughout this paper, we adopt a Flat $\Lambda$CDM cosmology with $\Omega_m = 0.3$, $\Omega_{\Lambda} = 0.7$, and $H_0 = 70 \ \rm{km \ s^{-1} \ Mpc^{-1}}$.

\section{Multiphase Galactic Wind Framework Setup}\label{sec:model}

We present a framework for generating predictions of emission signatures from galactic winds. Our framework produces emission line SB and ratios as a function of galaxy parameters (e.g., SFR and wind mass loading factor) and can be summarized as a two-step process. First, we employ both single‐phase \citep[][hereafter T16]{Thompson_2016} and multiphase \citepalias[][]{FB_2022} galactic wind models to derive radial profiles of density and temperature for galactic outflows. These profiles, combined with the state‐of‐the‐art cooling function from \citet[][hereafter PS20]{PS_2020}, allow us to calculate line emission from the hot wind. Second, we compute the fraction of emission flux originating from different emission lines in TRMLs using the analytic model developed by \cite{Chen_2023}. TRMLs reside at the interface between cold clouds and the hot wind, and given the distribution of cold clouds in a galactic outflow, we can translate the emission flux fraction per TRML into a profile of TRML emission as a function of radius. Finally, we combine the emission contributions from the hot wind and TRMLs to obtain the emission profile of a galactic outflow.

In this section, we briefly describe the galactic wind models \citepalias{Thompson_2016, FB_2022} and TRML models \citep{Chen_2023} that are essential ingredients of our framework.

\begin{figure*}
\centering
\includegraphics[width=\textwidth]{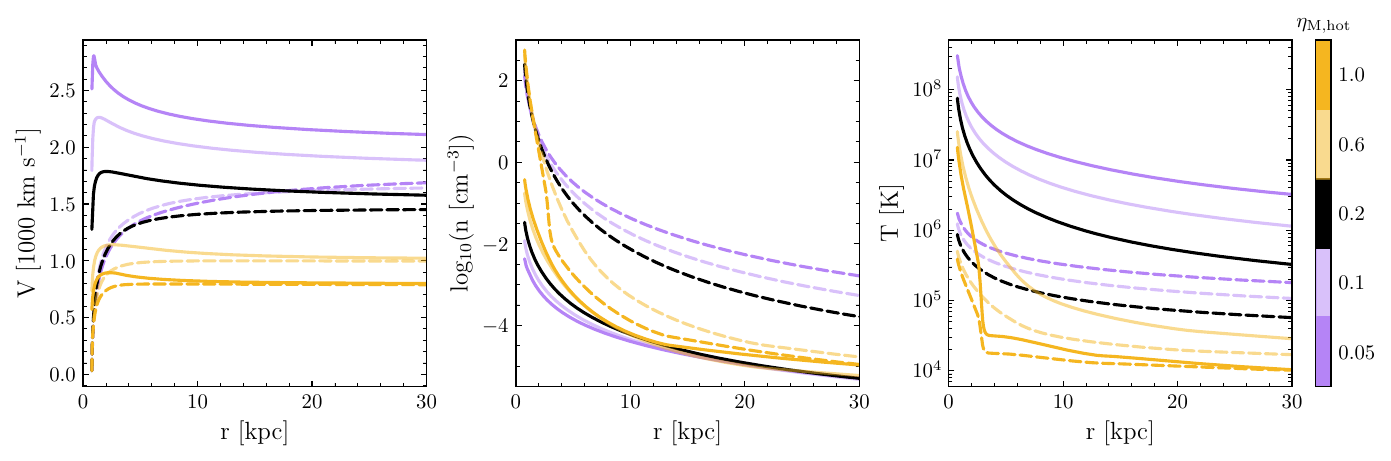}
\caption{Radial profiles of velocity, density, and temperature for 
a multiphase galactic wind generated by the \citetalias{FB_2022} model for different $\eta_{\rm M,hot}$ values. 
In each panel, the solid lines represent the hot wind. In the velocity and density panels, the dashed lines denote the cold clouds. In the temperature panel, the dashed lines instead represent the mixing temperature, $T_{\rm mix} = \sqrt{T_{\rm hot} \, T_{\rm cl}}$, where $T_{\rm cl}$ is fixed at $10^4\,\rm{K}$.
For these profiles, we choose $\mathrm{SFR} = 40\,\rm M_{\odot}\,yr^{-1}$, $\eta_{\rm E} = 1.0$, $\eta_{\rm M,cold} = 0.1$, and $M_{\rm cloud} = 10^5\,M_{\odot}$ (see \autoref{sec:results} for the justifications for these adopted values). The velocities of the two phases converge with increasing radius as cold clouds get entrained in the hot wind. This convergence is slower for lower values of $\eta_{\rm{M,hot}}$ because entrainment is harder in a more tenuous wind. Additionally, the number density and temperature of both phases decrease as the wind expands. The decrease in density and temperature is the steepest for large $\eta_{\rm M,hot}$ and at $r \lesssim 5$kpc. This feature directly affects the surface brightness profiles in the middle panel of \autoref{fig:SB_profile_vs_hot_phase_params}.
}
\label{fig:fb22_solution_example}
\end{figure*}

\subsection{Single-Phase and Multi-Phase Galactic Wind Models}

Traditionally, galactic wind has been modeled as a single-phase fluid consisting entirely of hot gas \citepalias[e.g.,][]{Thompson_2016}. However, recent observations have revealed the multiphase nature of these winds (see \citealp{tumlinson17} and \citealp{TH_2024} for reviews), and prompted the development of a novel multiphase model \citepalias{FB_2022} that simultaneously accounts for both the volume-filling hot phase and the cold gas clouds embedded in it. We first consider both of these models in this work in order to compare the distinct predictions of emission signatures they produce, and then focus on the \citetalias{FB_2022} multiphase model.

Both \citetalias{Thompson_2016} and \citetalias{FB_2022} assume a \cite{CC85} solution of the hot wind inside the sonic point and integrate the equations of mass, momentum, and energy conservation from the sonic point to obtain radial profiles of velocity, density, pressure, and temperature for the hot wind. Additionally, \citetalias{FB_2022} introduce a population of cold clouds moving with some velocity relative to the hot wind and account for the exchange of mass, energy, and momentum between the phases. The interaction between the phases is set by a competition between turbulent mixing, which shears and destroys the cloud, and radiative cooling, which allows the cloud to grow in mass. 
This competition is captured by an important dimensionless ratio $\left. \tau_{\rm mix} \right/ \tau_{\rm cool,mix}$, where $\tau_{\rm mix}$ is the turbulent mixing time or eddy turnover time, and $\tau_{\rm cool,mix}$ is a characteristic cooling time at the mixed temperature. Clouds grow and maintain long-term survival only when $\left. \tau_{\rm mix} \right/ \tau_{\rm cool,mix}>1$ and get destroyed otherwise. 
Both of these outcomes are captured by the \citetalias{FB_2022} model, which means the hot wind can impart mass, energy, and momentum to the cold cloud as the clouds accelerate and survive, and conversely the cold clouds can have a back-reaction on the hot wind as well. 
To model these sub-parsec scale cloud--wind interactions, \citetalias{FB_2022} introduce several parameters calibrated against numerical simulations.
In particular, the cold cloud mass growth rate is given by $\dot{M}_{\rm cl,grow}=\rho_{\rm hot} A_{\rm cool} v_{\rm in}$, where the effective area for cooling is parameterized as $A_{\rm cool}=f_{\rm cool} \, 4 \pi r_{\rm cl}^2 \, \chi^{1/2}$ ($\chi \equiv n_{\rm cold}/n_{\rm hot}$) with fiducial $f_{\rm cool} \sim 2$, capturing the elongation of the cloud in the wind direction during entrainment 
and is calibrated against wind tunnel simulations 
\citep{gronke20-cloud, abruzzo22}. 
$v_{\rm in}$, the inflow velocity from the hot phase onto the cloud, is parameterized by $v_{\rm in}=v_{\rm turb} \left(\left. \tau_{\rm mix}\right/ \tau_{\rm cool,mix}\right)^\alpha$, where $v_{\rm turb}=f_{\rm turb}\,v_{\rm rel}$ ($v_{\rm rel} \equiv v_{\rm hot} - v_{\rm cold}$) is the turbulent velocity with fiducial $f_{\rm turb}\sim 0.1$, 
and $\alpha=1/4$ or $1/2$ depending on whether cloud grows 
or get destroyed. 
Both $f_{\rm turb}$ and $\alpha$ are calibrated against turbulent mixing layer simulations \citep{Fielding_2020, Tan_2021}. 

Although these choices of $f_{\rm cool}$, $f_{\rm turb}$, and $\alpha$ serves as good first-order approximations of the underlying physical processes, several recent works have revealed uncertainties related to these parameters. 
\cite{Afruni_2026} find
that Milky Way wind observations are better reproduced by $f_{\rm cool}\sim 0.3$ (roughly consistent with the result in a supersonic flow; \citealp{scannapieco15}), substantially below the fiducial \citetalias{FB_2022} value. 
\cite{NG_2024} use a time-dependent $A_{\rm cool}$ to capture the cloud tail growth and adopt simulation-motivated scalings of $v_{\rm turb}$ with $v_{\rm rel}$, $r_{\rm cl}$, and $\tau_{\rm cool,cold}$ \citep[$\tau_{\rm cool,cold} = \tau_{\rm cool} (T=T_{\rm cold}, Z=Z_{\rm cold}, P)$;][]{Tan_2021} instead of simply applying the constant $f_{\rm turb}$ in \citetalias{FB_2022}. 
They also introduce a finite floor for $v_{\rm in}$, motivated by the fact that cloud growth need not cease entirely as the cloud approaches entrainment. A similar conclusion is reached by \cite{Dutta_2025}, who show that allowing $v_{\rm in}$ to saturate at a nonzero value yields substantially better agreement with cloud-crushing simulations in an expanding wind \citep{CC85} than the original \citetalias{FB_2022} prescription in which $v_{\rm in}\rightarrow 0$ as $v_{\rm rel}\rightarrow 0$. We plan on incorporating these recent insights into our framework in future works.

We refer readers to the original works of \citetalias{Thompson_2016} and \citetalias{FB_2022} for more details of the model setups. Here, we define two key hot-phase parameters that are crucial to galactic wind structure: the thermalization efficiency factor $\eta_{\rm{E}}$ and the hot-phase mass loading factor $\eta_{\rm{M,hot}}$. Assuming one core-collapse supernova (CCSN) occurs per 100 $M_{\odot}$ of star formation and that each CCSN deposits $E_{\rm{SN}} = 10^{51} \ \rm{erg}$ of mechanical energy, the hot-phase energy input rate, $\dot{E}_{\rm{hot}}$, can be parameterized as a function of $\eta_{\rm{E}}$\footnote{If stellar winds are included, $\dot{E}_{\rm hot}$ increases by a factor of $\sim 1.05 - 1.3$ (for stellar metallicities ranging from 0.2 to 2.0 $Z_\odot$; \citealp{Leitherer_1999}), but this does not qualitatively affect our subsequent arguments. For consistency with \citetalias{Thompson_2016} and \citetalias{FB_2022}, we therefore adopt the value given in \autoref{eq:3.1}.}:
\begin{align}
    \label{eq:3.1} \dot{E}_{\rm{{hot}}} = 3 \times 10^{41} \ \rm{erg \ s^{-1}} \ \eta_{\rm{E}} \ \frac{\rm{SFR}}{M_{\odot} \ \rm{yr^{-1}}}.
\end{align}
Similarly, the hot-phase mass input rate, $\dot{M}_{\rm{hot}}$, is parameterized as a function of $\eta_{\rm{M,hot}}$:
\begin{align}
    \label{eq:3.2} \dot{M}_{\rm{{hot}}} = \eta_{\rm{M,hot}} \ \rm{SFR}.
\end{align}

The galactic wind models we use here are useful because they convert bulk galaxy properties like SFR, $\eta_{\rm{E}}$, and $\eta_{\rm{M,hot}}$ into radial profiles of velocity, density, and temperature of the galactic wind. In \autoref{fig:fb22_solution_example}, we provide examples of such profiles for both the hot and cold phases in the galactic wind generated using the \citetalias{FB_2022} multiphase model at selected values of $\eta_{\rm M,hot}$. 
As the wind expands, the hot and cold phase velocities converge as cold clouds get entrained in the hot wind. 
This convergence is slower for lower values of $\eta_{\rm{M,hot}}$ because entrainment is harder in a more tenuous wind. 
At the same time, the number density and temperature of both phases decrease with radius. This drop is the steepest for large $\eta_{\rm M,hot}$ and at $r \lesssim 5$kpc. As we will later demonstrate, this feature of the density and temperature profiles directly affects the emission signatures of the galactic wind.

Although galactic wind radial profiles like the ones shown in \autoref{fig:fb22_solution_example} are useful for developing intuition of the wind structure and dynamics, these profiles are often not directly observable. To make the connection with observations, we still need to convert these profiles into SB of emission lines as a function of radius.

In a \citetalias{Thompson_2016} single-phase galactic wind, the only source of line emission is the hot wind, whose emissivity $\epsilon_{\lambda, \rm{wind}} (r)$ can be obtained from the gas cooling function given the galactic wind density and temperature profiles. However, things are more complicated in a \citetalias{FB_2022} multiphase galactic wind, where line emission comes from both the hot wind and TRMLs between the hot wind and embedded cold clouds\footnote{We note here that the cloud interior can also be a source of emission for lines that trace $\sim 10^4$ K or even molecular temperature gas. However, in this work, we focus on O\,{\footnotesize VI}, whose emissivity peaks at $\sim 10^{5.5}$ K. The cloud interior has negligible contribution to O\,{\footnotesize VI} emission.}. To reliably calculate the emission originating from TRMLs ($\epsilon_{\lambda, \rm{trml}}$), we adopt the analytic TRML model developed by \cite{Chen_2023}.

\subsection{Turbulent Radiative Mixing Layer Model}

To understand how TRMLs at the cloud-wind interface contribute to emission signatures, we utilize the analytic, 1.5 dimensional TRML model in \cite{Chen_2023}, which numerically integrates the fluid equations to obtain the phase structure of TRMLs, including temperature, density, and pressure profiles. This model includes a simple parameterization of turbulent conductivity and viscosity that is proportional to the shear velocity gradient. 
In practice, the temperature, density, and pressure profiles are numerically integrated in the direction perpendicular to the shear flow, but a shear velocity profile from TRML simulations is also used to compute values of the turbulent conductivity and viscosity (hence a 1.5 dimensional model). 
This analytic model is shown to reproduce the mass flux, total cooling luminosity, and phase structure of 3D TRML simulations with minimal computational cost, which makes it useful for our framework. In particular, we use the temperature and density profiles of TRMLs obtained from the analytic model to determine the line emissivities from TRMLs, $\epsilon_{\lambda, \rm{trml}}$. These emissivities are functions of the pressure $P$, the relative Mach number $\mathcal{M}_{\rm rel}$, and the hot phase temperature $T_{\rm hot}$ in the mixing layer (see \autoref{appendix:ovi_flux_frac_trml} for details of how these TRML parameters affect emission signatures), hence $\epsilon_{\lambda, \rm{trml}}(r)$ vary across TRMLs at different radii in a galactic outflow. Using the cloud population information obtained from the \citetalias{FB_2022} galactic wind model, we can determine the number of TRMLs at each radius and the corresponding $\epsilon_{\lambda, \rm{trml}}$ given $P$, $\mathcal{M}_{\rm rel}$, and $T_{\rm hot}$ in the galactic wind.


Finally, we follow Appendix A in \citet{Danehkar_2021} to determine the line SB at each projected radius $R$ on the 2D projected plane (SB$_{\lambda} (R)$) by integrating the volume line emissivities, $\epsilon_{\lambda, \rm{wind}} (r)$ and $\epsilon_{\lambda, \rm{trml}} (r)$, as an Abel integral. 
These predictions of emission‐line SB radial profiles facilitate direct comparison between analytical galactic wind models and observations.



\section{Surface Brightness Predictions}\label{sec:results}

In this section, we compare the SB predictions from our multiphase framework to those from a single-phase model, incorporating radiative cooling and an isothermal gravitational field \citepalias[][\autoref{subsec:single_vs_multi}]{Thompson_2016}. 
We then examine how variations in key parameters of the hot (\autoref{subsec:hot_phase_pars}) and cold (\autoref{subsec:cold_phase_pars}) phases influence the normalization and shape of the emission-line SB profiles. 
These parameter studies are essential to assess whether changes in SB profiles are sufficiently pronounced to constrain model parameters and outflow properties (e.g., $\dot{M}_{\rm{out,cl}}$ and $\dot{E}_{\rm{out,cl}}$) when compared to observations.
To enable a meaningful comparison between our SB predictions and observations (see \autoref{sec:discussions}), we adopt an SFR of $40 \ \rm{M_{\odot} \ yr^{-1}}$ and a half-light radius ($r_{50}$) of $ 0.75 \ \rm{kpc}$ (derived from the UV continuum) corresponding to the sonic point in our setup.
These values are based on the observed properties of J1156+5008, one of the few galaxies with O\,{\footnotesize VI} emission-line SB measurements available in the literature \citepalias[][]{Hayes_2016}. The fiducial values we assume for other parameters in our framework include $\eta_{\rm{E}} = 1.0$, $\eta_{{\rm M,hot}} = 0.2$, an initial cold phase mass loading factor $\eta_{{\rm M,cold,i}} = 0.1$, and an initial individual cloud mass of $M_{\rm cloud,i}=10^5 \ M_{\odot}$. Hereafter, we drop the subscript ``i'' for the cold-phase parameters $\eta_{{\rm M,cold,i}}$ and $M_{\rm cloud,i}$ for simplicity. The remaining parameters are fixed to the following values: initial cloud metallicity $Z_{\rm{cold, i}} = 0.25 \ Z_{\odot}$, initial hot wind metallicity $Z_{\rm{hot, i}} = 2.0 \ Z_{\odot}$, and initial cloud velocity $v_{\rm{cold, i}} = 30 \ \rm{km \ s^{-1}}$. Note that among all the parameters mentioned above, the hot phase parameters apply to both the single-phase and multiphase setups, while the cold-phase parameters are only relevant for our multiphase framework.

Our choices of fiducial values are inspired by recent observational constraints. 
The hot wind parameters ($\eta_{\rm{E}} = 1.0$, $\eta_{{\rm M,hot}} = 0.2$) successfully reproduce the X-ray Fe K$\alpha$ line luminosity observed in M82 \citep{TH_2024}. 
The cold cloud mass of $ \sim 10^5 \ M_{\odot}$ aligns with estimates for M82 \citep{Lopez_2025} and Makani \citep{Rupke_2023}, while the cold-phase mass-loading factor of $\sim 0.1$ is motivated by analyses of very-broad (FWHM $\sim 1000 \, \rm{km \ s^{-1}}$; \citealp{Peng_2025}) velocity components in strong emission lines ([O\,{\footnotesize III}] and H$\alpha$), which trace galactic winds in local star-forming dwarf galaxies (Peng et al., in prep.).
$Z_{\rm cold,i}$ is similar to the inferred ISM gas-phase metallicity of J1156+5008 \citepalias{Hayes_2016}, while $v_{\rm cold,i}$ is estimated from the characteristic ISM turbulent velocity, $\sigma_{\rm gas} \sim \sqrt{P/\rho}$.
For the latter, we adopt the best-fit pressure constraint of $P/k_{B} \simeq 7\times10^{4}\ \rm K\,cm^{-3}$ reported in \citetalias{Hayes_2016} and assume a number density of order unity, which is consistent with the upper limit derived from the [S\,{\footnotesize II}] $\lambda\lambda 6717,6731$ doublet in \citetalias{Hayes_2016}.

In what follows, we present the model predictions for the O\,{\footnotesize VI} emission-line SB. 
O\,{\footnotesize VI} traces warm-hot gas in the CGM at $\sim 10^{5.5}$ K, dominates radiative cooling and is crucial for differentiating between single-phase and multiphase galactic outflow models. 
We directly compare the predictions of our framework with the observed O\,{\footnotesize VI} SB profile in \citetalias{Hayes_2016}.
While we focus on O\,{\footnotesize VI} in this work, we emphasize that the framework we present is highly flexible and can easily make SB predictions of \textit{any} other emission-line of interest.

\subsection{Comparing Our Multiphase Galactic Wind Framework with a Traditional Single-phase Model}\label{subsec:single_vs_multi}

Intuitively, single- and multi-phase galactic wind models should yield distinct predictions on emission signatures because multiphase models account for TRMLs, which should have significant contributions to emission because they host intermediate temperature gas between $\sim 10^4$ K and $\sim 10^6$ K, where the emissivity of most metal lines peaks \citep{Piacitelli_2022}. For example, O\,{\footnotesize VI} emissivity peaks at $\sim 10^{5.5}$ K, which means a large portion of O\,{\footnotesize VI} emitting gas is expected to reside in mixing layers, and accounting for this contribution significantly boosts the total O\,{\footnotesize VI} SB. 
The radial emission profile is also sensitive to the multiphase nature of galactic outflows. 
In the single-phase picture, the volume-filling hot phase adiabatically expands and radiatively cools, $\sim 10^{5.5}$ K O\,{\footnotesize VI} emitting gas is only located in a narrow radius range where the hot phase cools through that temperature. 
On the other hand, the multiphase picture has mixing layers containing $\sim 10^{5.5}$ K gas distributed throughout the outflow, leading to distinct emission profiles.

\begin{figure}
\centering
\includegraphics[width=\columnwidth]{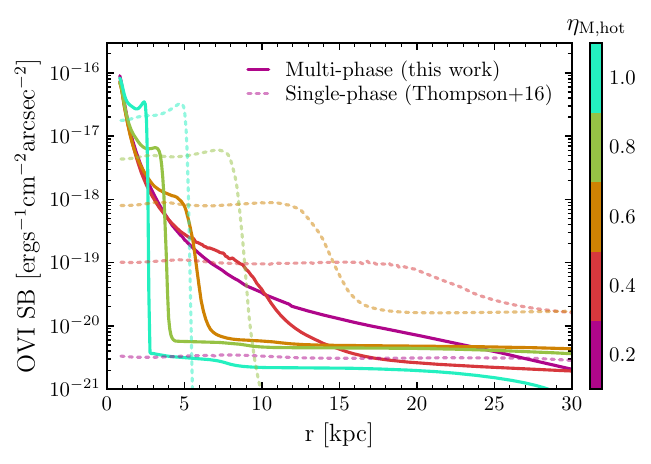}
\caption{O VI SB profile predictions at different $\eta_{\rm{M,hot}}$ for the single-phase (dotted) galactic wind model in \citetalias{Thompson_2016} and the multiphase (solid) framework presented in this work based on \citetalias{FB_2022} and \citet{Chen_2023}. The single-phase model predicts a region of constant O VI SB at small radius. The multiphase framework does not exhibit this feature and instead produces a profile that declines smoothly with radius. 
We explain key features in detail and compare these two setups in \autoref{subsec:single_vs_multi}.
Emission signatures like SB profiles can serve as diagnostics for the phase structure of galactic winds seen in observations and help distinguish between single- and multi-phase models.}
\label{fig:OVI_SB_multi_phase_vs_single_phase}
\end{figure}

\begin{figure*}
\centering
\includegraphics[width=\textwidth]{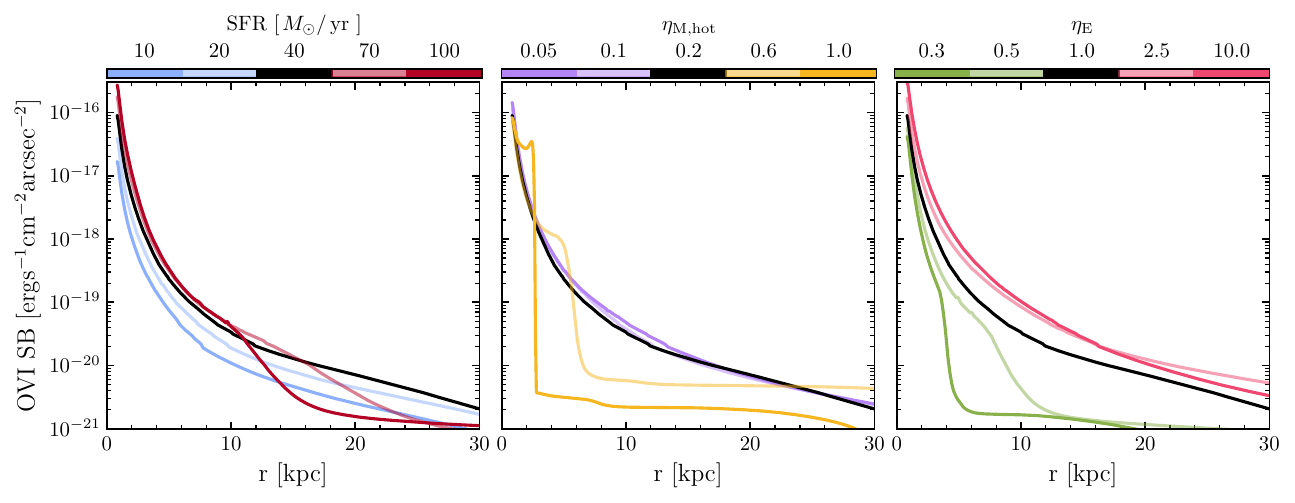}
\caption{O VI SB profile predictions from our multiphase galactic wind framework is sensitive to hot-phase parameters including SFR (\textit{left}),  $\eta_{\rm{M,hot}}$ (\textit{middle}), and $\eta_{\rm{E}}$ (\textit{right}). 
The fiducial parameter choices are SFR = $40 \ \rm{M_{\odot} \ yr^{-1}}$, $\eta_{\rm{M,hot}}=0.2$, $\eta_{\rm{E}}=1.0$, $\eta_{\rm{M,cold}}=0.1$, and $M_{\rm cloud} = 10^5 \ M_{\odot}$ except when a parameter is explicitly varied. We motivated these fiducial parameter choices in \autoref{sec:results}. The black curves across the 3 panels are identical and show the O VI SB profile generated using the fiducial parameters. 
We explain how O VI SB profile depend on these hot-phase parameters in \autoref{subsec:hot_phase_pars}. These dependencies have profound consequences on observations, which will be explored in \autoref{sec:discussions}.
}
\label{fig:SB_profile_vs_hot_phase_params}
\end{figure*}

These ideas motivate a quantitative comparison between the single-phase model and our multiphase framework. 
In \autoref{fig:OVI_SB_multi_phase_vs_single_phase}, we plot the SB of O\,{\footnotesize VI} doublet as a function of radius at a range of $\eta_{\rm M,hot}$ calculated from both the \citetalias{Thompson_2016} single-phase model (dotted lines) and our multiphase framework (solid lines). 
We start by interpreting results from the single-phase model, which is more straightforward to understand. Starting from the sonic point, the initial portion of the SB profile calculated from the single-phase model is flat for all choices of $\eta_{\rm M,hot}$. How can we understand this? O\,{\footnotesize VI} SB mainly depends on two factors, O\,{\footnotesize VI} emissivity and density of the hot phase, both of which are changing with radius. At small radii, the temperature of the hot phase gradually cools to approach the peak O\,{\footnotesize VI} emissivity temperature at $\sim 10^{5.5}$ K, so O\,{\footnotesize VI} emissivity increases with radius. At the same time, density is decreasing as the hot phase expands. These two effects balance each other out to produce the plateau in the SB profile. 
Eventually, the wind cools below $10^{5.5}$ K, and from there on, O\,{\footnotesize VI} SB starts to drop off with respect to radius since both emissivity and density are decreasing. 
Since we fixed the SFR, outflows with larger $\eta_{\rm M,hot}$ imply a larger mass outflow rate and larger wind density, which cools faster with respect to radius and thus has a higher initial value but a shorter plateau in the SB profile. 

The SB profile calculated from our multiphase framework, on the other hand, looks completely different. There is no extended region of flat SB profile, and the SB profiles with different $\eta_{\rm M,hot}$ look similar in both shape and normalization at small radii. In this regime, O\,{\footnotesize VI} SB in the multiphase framework is dominated by contributions from the mixing layers, which are less sensitive to hot phase parameters compared to emission from the hot phase itself. 
Further note that SB at the sonic point is at least 10 times larger after accounting for mixing layer contributions in our multiphase framework, and this effect is more pronounced for smaller $\eta_{\rm M,hot}$. 
Moving to larger radii, we can see a bump in the SB profiles when $\eta_{\rm M, hot} \gtrsim 0.6$. This is where the hot phase cooled to $10^{5.5}$ K, providing a boost to the overall O\,{\footnotesize VI} SB. Although this bump occur due to a similar reason as the inflection point seen in the O\,{\footnotesize VI} SB profile predicted by the single-phase model (both are because the hot phase cooled through $10^{5.5}$ K), the location of this bump is at a smaller radius because the multiphase model cools more rapidly with the help of TRMLs.
Beyond this bump, the hot phase further cools, and O\,{\footnotesize VI} flux fraction drops dramatically in the mixing layers because they can no longer host O\,{\footnotesize VI} emitting gas at $\sim 10^{5.5}$ K. 
This leads to a rapid decline in the SB profile, which occurs at larger radii for smaller $\eta_{\rm M, hot}$.

Since single- and multi-phase models yield distinct predictions for SB profiles, comparing against observations can potentially shed light on which model better describes real-world galactic outflows. Our multiphase framework predictions shown in \autoref{fig:OVI_SB_multi_phase_vs_single_phase} are closer to the observed O\,{\footnotesize VI} SB profiles \citepalias[][see \autoref{sec:discussions} for details]{Hayes_2016}. Constant SB regions extending out to $\sim 10$ kpc, as predicted by the single-phase model, are not observed, and observed O\,{\footnotesize VI} SB profiles follow a smooth, gradual decent beyond $\sim 10$ kpc, similar to multiphase framework predictions with $\eta_{\rm M, hot} \lesssim 0.4$, as shown in \autoref{fig:OVI_SB_multi_phase_vs_single_phase}. 
We will focus on our multiphase framework for the remainder of this work. 

\subsection{Hot-phase Wind Parameters}\label{subsec:hot_phase_pars}

\autoref{eq:3.1} and \autoref{eq:3.2} show that the key parameters that control the mass and energy input rates for the hot phase are $\eta_{\rm{E}}$, $\eta_{\rm{M,hot}}$, and SFR. In \autoref{fig:SB_profile_vs_hot_phase_params}, we use the SB profile of O\,{\footnotesize VI} as an example to show how changes in each of these three parameters affect the resulting emission signature.

The effects of varying $\eta_{\rm E}$ and SFR on the SB profile are similar in that increasing both parameters increases O\,{\footnotesize VI} SB at small radii (as shown in the left and right panels of \autoref{fig:SB_profile_vs_hot_phase_params}) since $\dot{E}_{\rm hot}$ is proportional to $\eta_{\rm E}$ and SFR (\autoref{eq:3.1}). Another key feature in the O\,{\footnotesize VI} SB profile is the radius at which the hot phase cools through $10^{5.5}$K, the temperature where O\,{\footnotesize VI} emissivity peaks. As discussed in \autoref{subsec:single_vs_multi}, this corresponds to a rapid drop in the O\,{\footnotesize VI} SB profile. In \autoref{fig:SB_profile_vs_hot_phase_params}, we can see that this drop happens at smaller radii for larger SFR and smaller $\eta_{\rm E}$. Why is the parameter dependence opposite in this case? This 
can be understood through how the hot wind density, $n_{\rm{hot}}(r)$, depend on these parameters. We can express $n_{\rm{hot}}(r)$ in terms of $\eta_{\rm E}$, $\eta_{\rm M, hot}$, and SFR \citep{CC85} as:
\begin{align}
    \label{eq:3.3} 
    n_{\rm{hot}}(r) \propto \frac{\dot{M}_{\rm hot}^{3/2} \ \dot{E}_{\rm hot}^{-1/2}}{R^{2}} \propto \frac{\eta_{\rm M, hot}^{3/2}}{\eta_{\rm E}^{1/2}} \ \frac{\rm SFR}{R^{2}} \propto \frac{\eta_{\rm M, hot}^{3/2}}{\eta_{\rm E}^{1/2}} \ \dot{\Sigma}_\ast.
\end{align}
Consequently, increasing SFR increases $n_{\rm{hot}}(r)$ and thus boosts the cooling emissivity (which scales as $n_{\rm{hot}}(r)^2$) of O\,{\footnotesize VI}. This means that with a larger SFR, the outflow cools faster and reaches the critical temperature of $10^{5.5}$K at a smaller radius, as shown by the location of the rapid drop of the O\,{\footnotesize VI} SB profile in the left panel of \autoref{fig:SB_profile_vs_hot_phase_params}.
On the other hand, \autoref{eq:3.3} shows that $\eta_{\rm E}$ is negatively correlated with $n_{\rm{hot}}(r)$, which means decreasing $\eta_{\rm E}$ makes the outflow cool faster, as shown in the right panel of \autoref{fig:SB_profile_vs_hot_phase_params}. 
As for $\eta_{\rm M, hot}$, although it does not affect $\dot{E}_{\rm hot}$ in \autoref{eq:3.1}, it is positively correlated with $n_{\rm{hot}}(r)$ in \autoref{eq:3.3}, just like SFR. Thus, increasing $\eta_{\rm M, hot}$ also makes the outflow cool faster and allows the rapid drop in O\,{\footnotesize VI} SB to happen at a smaller radius, 
as shown in \autoref{fig:SB_profile_vs_hot_phase_params}. 
The dependence of $n_{\rm{hot}}(r)$ on $\eta_{\rm M, hot}$ is also visualized in the middle panel of \autoref{fig:fb22_solution_example}, where we can see that a larger $\eta_{\rm M, hot}$ corresponds to a larger value of $n_{\rm{hot}}$ at small r and a more rapid drop of $n_{\rm{hot}}(r)$ at $r \lesssim 5$kpc. Additionally, the right panel of \autoref{fig:fb22_solution_example} shows that the hot phase temperature drops below $10^{5.5}$K at a smaller radii for larger $\eta_{\rm M, hot}$, and the sharp drop in $T_{\rm hot}$ at $\eta_{\rm M, hot}=1.0$ corresponds to a similar sharp drop in O\,{\footnotesize VI} SB in \autoref{fig:SB_profile_vs_hot_phase_params}.


Notably, varying these hot phase parameters cannot significantly change the overall normalization of the O\,{\footnotesize VI} SB profile. This means O\,{\footnotesize VI} SB at large radii ($\gtrsim 10$ kpc) cannot be significantly boosted by these parameters. 
Increasing SFR and $\eta_{\rm{E}}$ can boost 
O\,{\footnotesize VI} SB in a part of the parameter space, but this effect saturates for SFR $ \gtrsim 70 \ M_{\odot}~\mathrm{yr}^{-1}$ and $\eta_{\rm{E}} \gtrsim 2.5$ (\autoref{fig:SB_profile_vs_hot_phase_params}), at which point O\,{\footnotesize VI} SB is only boosted by at most a factor of a few compared to the fiducial case with SFR $= 40 \ M_{\odot}~\mathrm{yr}^{-1}$ and $\eta_{\rm{E}} = 1.0$. As for $\eta_{\rm M, hot}$, O\,{\footnotesize VI} SB at $\sim 10-20$ kpc is maximized for small $\eta_{\rm M, hot}$, but this effect also saturates for $\eta_{\rm M, hot} \lesssim 0.2$. We discuss the implications of these perhaps surprising conclusions in \autoref{sec:discussions} when we compare O\,{\footnotesize VI} SB profiles generated by our framework with observational results.

\subsection{Cold-phase Wind Parameters}\label{subsec:cold_phase_pars}

\begin{figure}
\centering
\includegraphics[width=\columnwidth]{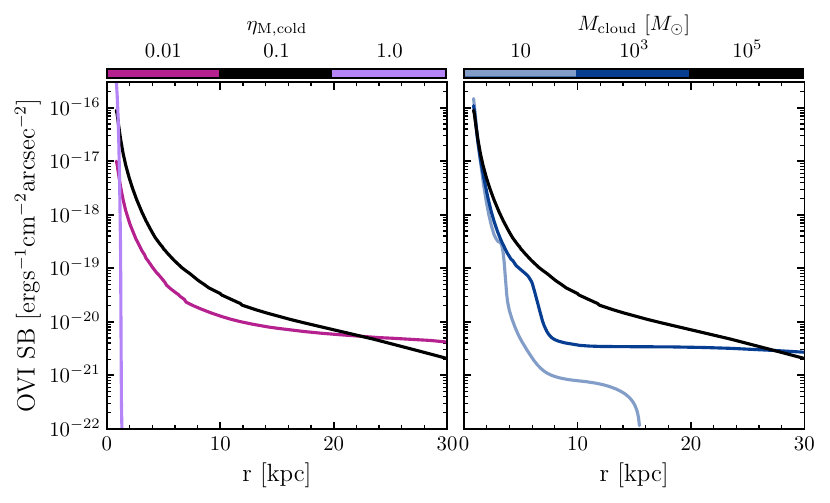}
\caption{Similar to \autoref{fig:SB_profile_vs_hot_phase_params}, but here we vary cold-phase parameters in our multiphase framework including the cold phase mass loading factor $\eta_{\rm{M,cold}}$ (\textit{left}) and single cloud mass $M_{\rm{cloud}}$ (\textit{right}). Increasing $\eta_{\rm{M,cold}}$ steepens the slope of the SB profiles, while increasing $M_{\rm{cloud}}$ has the opposite effect. These trends are explained in detail in \autoref{subsec:cold_phase_pars}.}
\label{fig:SB_profile_vs_cold_phase_params}
\end{figure}

The two main cold-phase parameters that regulate the cooling luminosity within TRMLs in our framework are $\eta_{\rm{M,cold}}$ and $M_{\rm{cloud}}$.  
A higher value of $\eta_{\rm{M, cold}}$ corresponds to increased cloud number fluxes at each radius (i.e., $\dot{N}_{\mathrm{cl}} \propto \eta_{\mathrm{M, cold}}$), resulting in enhanced cooling luminosities from TRMLs.  
As shown in the left panel of \autoref{fig:SB_profile_vs_cold_phase_params}, outflows with larger $\eta_{\rm{M, cold}}$ produce higher O\,{\footnotesize VI} SB at small radii ($r \lesssim 2 \ \rm{kpc}$) and exhibit steeper radial declines due to more rapid cooling.  
Since the total cooling luminosity in our framework is the sum of contributions from both the hot winds and the TRMLs, an increase in $\eta_{\rm{M, cold}}$ enhances the luminosity from TRMLs but reduces that from the hot wind \citepalias[i.e., the loss of hot mass flux arises as hot material cools onto the entrained cold clouds;][]{FB_2022}.  
This compensation results in only a modest net increase in the total cooling luminosity \citep{Peng_2025}. 

Outflows with more massive clouds are harder to accelerate. 
Hence, the relative velocity, $v_{\rm{rel}} = v_{\rm{hot}} - v_{\rm{cold}}$, increases, boosting the kinetic-energy thermalization term \citepalias[$\propto v_{\rm{rel}}$; see Eq. 26 in][]{FB_2022} and the corresponding cooling luminosities within the TRMLs.  
Moreover, the timescale for cold cloud acceleration in a hot wind is given by the drag time\footnote{
Note that the predominant process responsible for cloud acceleration is not drag, but mixing between the hot and cold phases and the momentum transfer from the hot phase to the newly accreted cold gas of the cloud. We use the term ``drag time'' here to stay consistent with the conventions in \cite{gronke18}.
}, $\tau_{\rm drag} \sim \chi \left. r_{\rm cloud}\right/ v_{\rm hot}$ \citep[$\chi = n_{\rm cold} / n_{\rm hot}$;][]{gronke18}, which increases with cloud size and mass. 
Thus, more massive cold clouds take longer to accelerate, radiating more gradually and producing a much shallower SB profile, as shown in the right panel of \autoref{fig:SB_profile_vs_cold_phase_params}.

Similar to what we saw in \autoref{subsec:hot_phase_pars}, changing these cold phase parameters also cannot significantly boost O\,{\footnotesize VI} SB at large radii ($\gtrsim 10$ kpc), as shown in \autoref{fig:SB_profile_vs_cold_phase_params}. Note that we did not increase the value of $M_{\rm{cloud}}$ beyond the fiducial value of $10^5 M_{\odot}$ because this fiducial value is already close to the maximum cloud mass that can be accelerated by the hot wind given fiducial parameters. Thus, we do not expect increasing $M_{\rm{cloud}}$ to significantly affect the O\,{\footnotesize VI} SB profile. These conclusions motivate our discussions in \autoref{sec:discussions}.

\section{Discussions}\label{sec:discussions}

In \autoref{sec:results}, we explored the sensitivity of O\,{\footnotesize VI} SB profiles generated from our multiphase galactic wind to several hot and cold phase parameters. How do these predictions generated by our framework compare with observations? In \autoref{fig:Model_vs_Hayes16_Observation}, we compare the O\,{\footnotesize VI} SB observation in \citetalias{Hayes_2016} (black) with predictions from our framework (solid blue) using the same galaxy parameters and fiducial hot and cold phase parameters. 
Our framework systematically underestimates O\,{\footnotesize VI} SB by $\sim2$ orders of magnitude. This difference is too large to be accounted for by changing model parameters, which boosts the SB by at most a factor of a few as discussed in \autoref{sec:results}. 

In this section, we discuss several physical processes not accounted for in our fiducial framework--including ram pressure equilibrium between the hot wind and cold clouds (\autoref{subsec:ram_pressure}), non-equilibrium ionization (\autoref{subsec:non_equil}), and additional energy sources beyond CCSNe mechanical feedback (\autoref{subsec:extra_energy})--and discuss how they could help resolve the discrepancy between our predictions and observations.
Finally, we summarize the insights from our multiphase framework
and attempt to explore
why so few star-forming galaxies have detected O\,{\footnotesize VI} emission-line SB (\autoref{subsec:connect_to_obs}). 

Our framework utilizes idealized theoretical models of galactic winds and does not account for complications like wind expansion in a pre-existing CGM (\citetalias{FB_2022} and \citetalias{Thompson_2016} assume wind expansion in vacuum), non-spherical wind geometry \citep[e.g.,][]{Heckman_1990, Nguyen_21, Carr_2021, Peng_2023, McPherson_2023}, and turbulent pressure support for the cold cloud\footnote{
Although it is hard to track turbulent pressure in our setup, \cite{Tan_2023} find in their simulations that clouds are supported approximately equally by turbulent and thermal pressure, with their combined pressure balancing the hot wind thermal pressure. This means accounting for cloud turbulent pressure could change the density and temperature profiles of the phases and thus affect the SB profile predictions.
}. 
The results we present here should be treated as a qualitative exploration of several important physical effects under ideal circumstances, which is useful for guiding intuition and connecting bulk galaxy properties with observable emission signatures. However, follow-up work that accounts for the caveats mentioned above is required in order to make quantitative comparisons with observational data.

\subsection{Ram Pressure Equilibrium}\label{subsec:ram_pressure}

\begin{figure}
\centering
\includegraphics[width=\columnwidth]{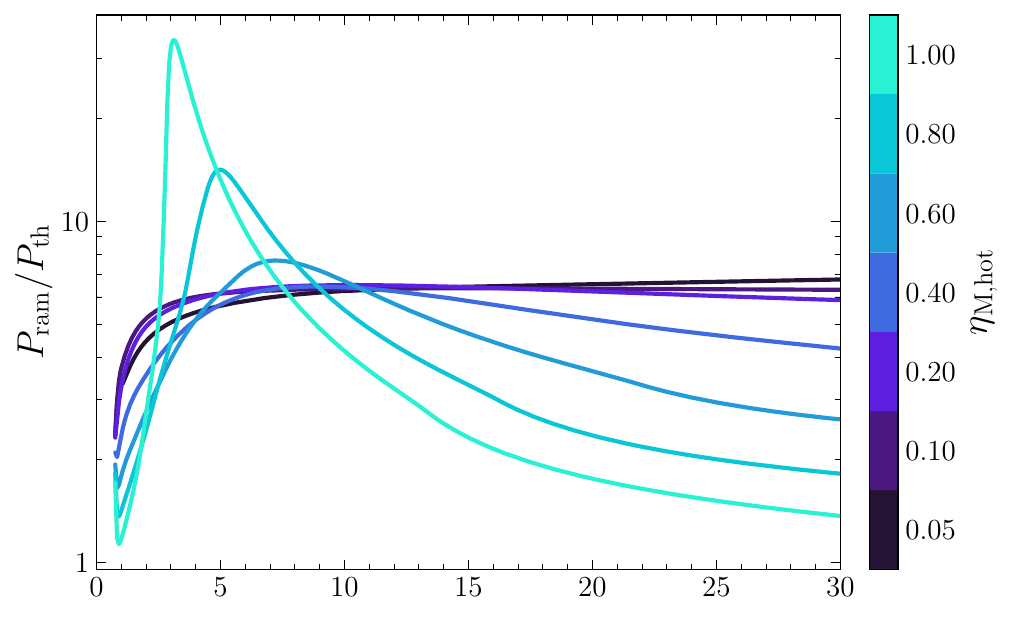}
\includegraphics[width=\columnwidth]{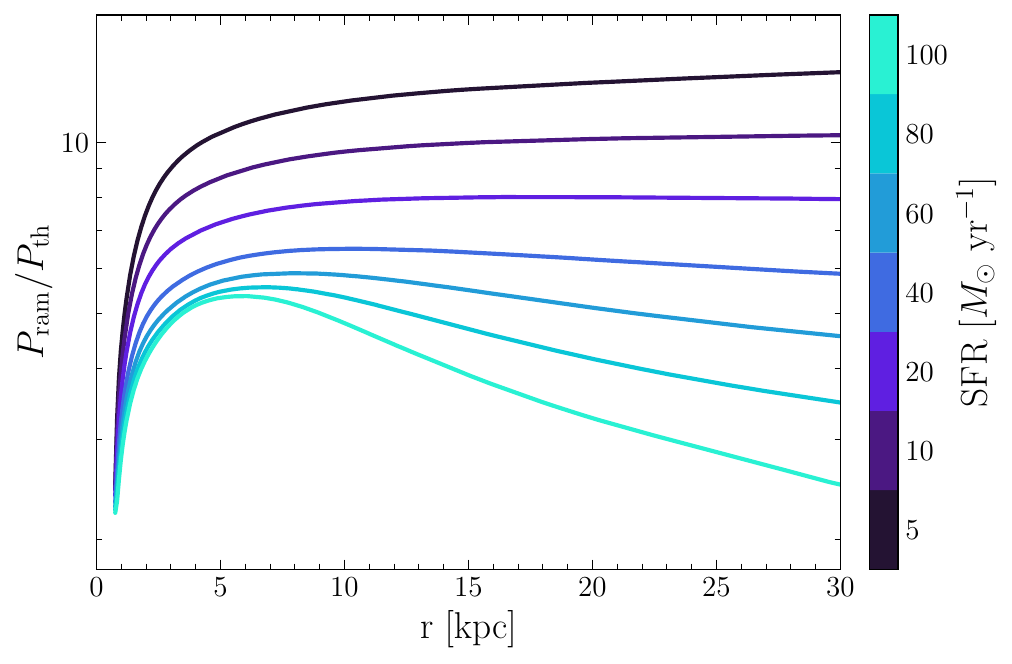}
\caption{The ratio of ram and thermal pressure $\left. P_{\rm ram}\right/ P_{\rm th}$ profile
computed from the \citetalias{FB_2022} model with different $\eta_{\rm M,hot}$ (\textit{top}; $\eta_{\rm{E}} = 1.0$ and SFR = 40 $M_{\odot} \ \rm{yr^{-1}}$) and SFR (\textit{bottom}; $\eta_{\rm{E}} = 1.0$ and $\eta_{\rm{M,hot}} = 0.2$). 
$\left. 
P_{\rm ram}\right/ P_{\rm th} = \gamma \mathcal{M}_{\rm rel}^2$, where $\mathcal{M}_{\rm rel} = v_{\rm{rel}} / c_{\rm{s,hot}}$ is the \textit{relative} Mach number between the cold clouds and the hot wind, and $\gamma$ is the adiabatic index. 
Both panels indicate that ram pressure is always a crucial component of the pressure balance between the cold cloud and the hot wind. Since O VI SB $\propto n^2 \propto P^2$, accounting for ram pressure significantly boosts O VI SB and is crucial for explaining observational results in \citetalias{Hayes_2016}.
}
\label{fig:fb22_pressure_mrel_profiles}
\end{figure}

SB of emission lines is proportional to the square of the number density of the emitting gas. In \citetalias{FB_2022}, the density of cold clouds is controlled by the fact that they are in thermal pressure equilibrium with the volume-filling hot phase. As the wind expands and cools with radius, its thermal pressure drops steeply, leading to a drop in the pressure and density 
(see \autoref{fig:fb22_solution_example} for the density and temperature profiles)
of the cold clouds and the mixing layers. This explains the steep decrease in O\,{\footnotesize VI} SB as predicted by our multiphase framework. However, observations by \citetalias{Hayes_2016} show a much higher SB value at a radius of $\sim 10-20$ kpc. How can we understand this?

\citet{Heckman_1990} found that cold clouds are confined by the thermal pressure ($P_{\rm{th}}$) of the hot wind within the sonic point but by the \textit{ram} pressure ($P_{\rm ram} = \gamma \, \mathcal{M}_{\rm rel}^2 \, P_{\rm{th}}$) at larger radii in starburst galaxies.
Here, $\mathcal{M}_{\rm rel} = v_{\rm{rel}} / c_{\rm{s,hot}}$ is the Mach number of the relative shear velocity between the phases and $\gamma$ is the adiabatic index.
Following this pioneering work, \cite{Xu_2023} showed that the observed density profile of cold clouds in M82 does not follow the steep drop of $P_{\rm{th}}$ with radius. 
Instead, it is better explained if the cold clouds are in equilibrium with $P_{\rm ram}$ of the hot wind. 
Thus, in a highly supersonic outflow, accounting for ram pressure can significantly boost the pressure and density of the cold clouds as well as the mixing layers. Intuitively, this makes sense because a cold cloud outflowing at supersonic velocities creates a bow shock at its head, which compresses gas behind it and subsequently increases the pressure and density at the body of the cloud and the mixing layer. 
In wind tunnel simulations with supersonic winds \citep{scannapieco15}, strong bow shocks lead to compressions and increase the cloud density as a result.\footnote{A caveat here is that the simulations of \cite{scannapieco15} are in the cloud destruction regime. Furthermore, high Mach number wind tunnel simulations are still not numerically converged \citep{FGOH23}.}
Recently, \cite{Lopez_2025} report conspicuous arc-like structures in the outflows of M82 and suggest bow shocks created by supersonic outflows as a possible origin. 


\begin{figure}
\centering
\includegraphics[width=\columnwidth]{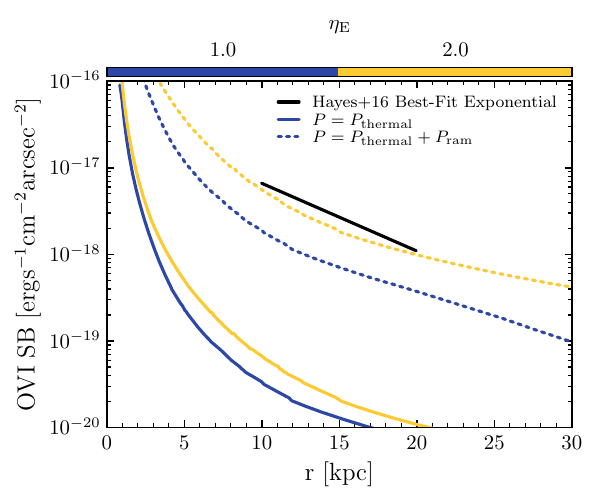}
\caption{Comparison between O VI SB profiles generated by our multiphase wind framework and observational result in \citetalias{Hayes_2016} (black). 
The \citetalias{Hayes_2016} observational result we show here is the best-fit exponential profile they obtained for their spatially resolved O VI emission mapped using \emph{HST}. 
In solid blue, we plot the O VI SB profile generated by our framework using fiducial parameters (SFR = $40 \ \rm{M_{\odot} \ yr^{-1}}$, $\eta_{\rm{M,hot}}=0.2$, $\eta_{\rm{M,cold}}=0.1$, and $M_{\rm cloud} = 10^5 \ M_{\odot}$) and only accounting for thermal pressure. This under-predicts the observation by more than 2 orders of magnitude. As discussed in \autoref{subsec:ram_pressure} and \autoref{fig:fb22_pressure_mrel_profiles}, accounting for ram pressure significantly boosts O VI SB. We model the effect of ram pressure heuristically by multiplying the original thermal-pressure only O VI SB profile by $\left(1 + \left. P_{\rm ram}\right/ P_{\rm th}\right)^2$. This yields the dotted blue curve, which is much closer to observation but still off by a factor of $\sim 3$. One way to bridge this difference is to choose $\eta_{\rm E}=2.0$ (yellow), which yields qualitative agreement with observation. Physically, $\eta_{\rm E}=2.0$ means there are energy sources other than the mechanical energy of supernova explosions that powers the emission. In \autoref{subsec:extra_energy}, we discuss possible energy source including radiation and merger-induced thermal energy and argue that $\eta_{\rm E}=2.0$ is reasonable.}
\label{fig:Model_vs_Hayes16_Observation}
\end{figure}

\autoref{fig:fb22_pressure_mrel_profiles} shows the $P_{\rm ram}/P_{\rm th}$ profiles computed from the \citetalias{FB_2022} model as a function of radius for different $\eta_{\rm M}$ (\textit{top} panel) and SFR (\textit{bottom} panel). 
In general, ram pressure dominates over thermal pressure at all radii across the parameter space explored, and the $P_{\rm ram}/P_{\rm th}$ ratio within the CGM (i.e, $\gtrsim 10 \ \rm{kpc}$) is inversely correlated with both $\eta_{\rm M}$ and SFR. 
Notably, the $P_{\rm ram}/P_{\rm th}$ profiles peak at different radii for different choices of $\eta_{\rm M}$ and SFR: at low values of either parameter (darker colors in \autoref{fig:fb22_pressure_mrel_profiles}), $P_{\rm ram}/P_{\rm th}$ gradually increases with radius and saturates, whereas at high values, it typically peaks at small radii ($\lesssim 5 \ \rm{kpc}$).
Since $P_{\rm ram}/P_{\rm th} \propto \mathcal{M}_{\rm rel}^2 \propto \left(v_{\rm rel}/c_{\rm s,hot}\right)^2 \propto v_{\rm rel}^2/T_{\rm hot}$, the shape of the $P_{\rm ram}/P_{\rm th}$ profile is governed by radial evolution of $v_{\rm rel}^2$ and $T_{\rm hot}$, both of which decrease more rapidly with radius for larger $\eta_{\rm M}$ and SFR, albeit at different rates.
This behavior arises because high-$\eta_{\rm M}$ and high-SFR outflows cool more efficiently (i.e., $T_{\rm hot}$ declines more steeply) due to their higher wind densities, while simultaneously losing substantially more mass and energy through enhanced transfer onto the cold clouds (i.e., the hot wind entrains cold clouds more efficiently, causing $v_{\rm rel}$ to decrease more rapidly).
Moreover, the effect of increasing $\eta_{\rm M}$ is more pronounced than that of increasing SFR because the wind density is more sensitive to $\eta_{\rm M,hot}$ than to SFR (see \autoref{eq:3.3}). A complex interplay between these effects leads to the results shown in \autoref{fig:fb22_pressure_mrel_profiles}.

How significant is this effect in the \citetalias{Hayes_2016} observations? 
\autoref{fig:fb22_pressure_mrel_profiles} shows that for the fiducial parameters of the \citetalias{Hayes_2016} target J1156+5008, $\left. P_{\rm ram}\right/ P_{\rm th} \simeq 6$ beyond a radius of 10 kpc, which corresponds to $\mathcal{M}_{\rm rel} \gtrsim 2$. 
Since O\,{\footnotesize VI} SB $\propto n^2 \propto P^2 \propto \mathcal{M}_{\rm rel}^4$, accounting for the much higher ram pressure boosts O\,{\footnotesize VI} SB by roughly a factor of $\simeq 36$.
To gauge the effects of ram pressure on the emission predictions in our framework,
we 
adopt a heuristic approach by
multiplying our O\,{\footnotesize VI} SB profile--computed assuming only thermal pressure equilibrium--by $\left(1 + \left. P_{\rm ram}\right/ P_{\rm th}\right)^2$ to incorporate the effects of ram pressure. We implement this procedure in \autoref{fig:Model_vs_Hayes16_Observation}, where the solid blue curve shows the O\,{\footnotesize VI} SB profile using our fiducial parameters and only accounting for thermal pressure, and the dotted blue curve adopts the same parameters but includes the effects of ram pressure. 
Indeed, qualitatively estimating the effect of ram pressure significantly boosts O\,{\footnotesize VI} SB and brings the prediction much closer to the observed profile in \citetalias{Hayes_2016} denoted by the black line. However, the magnitude of O\,{\footnotesize VI} SB generated by our framework is still noticeably below the observed value.
This discrepancy is large enough that is suggests the possibility of 
additional energy sources beyond the mechanical energy supplied by CCSNe. We find that increasing $\eta_{\rm E}$ by a factor of 2 allows us to qualitatively match the magnitude of the observed profile, as shown in \autoref{fig:Model_vs_Hayes16_Observation}. 
$\eta_{\rm E}=2$
can possibly be supplied by radiative feedback from massive stars and/or orbital energy from mergers. We discuss these extra energy budgets in detail in \autoref{subsec:extra_energy}.

Although the \citetalias{FB_2022} galactic wind model generates thermal pressure and relative mach number ($\mathcal{M}_{\rm rel} = v_{\rm{rel}} / c_{\rm{s,hot}}$) profiles that allow us to plot $\left. P_{\rm ram}\right/ P_{\rm th}$ as a function of radius (\autoref{fig:fb22_pressure_mrel_profiles}), we emphasize that a rigorous treatment of ram pressure requires self-consistent modeling of its effect on the radial profiles of global wind/cloud properties (e.g., temperature, density and velocity) and is beyond the scope of this work. However, our analysis here suggests it is a promising direction for future studies.

\begin{figure*}
\centering
\includegraphics[width=\textwidth]{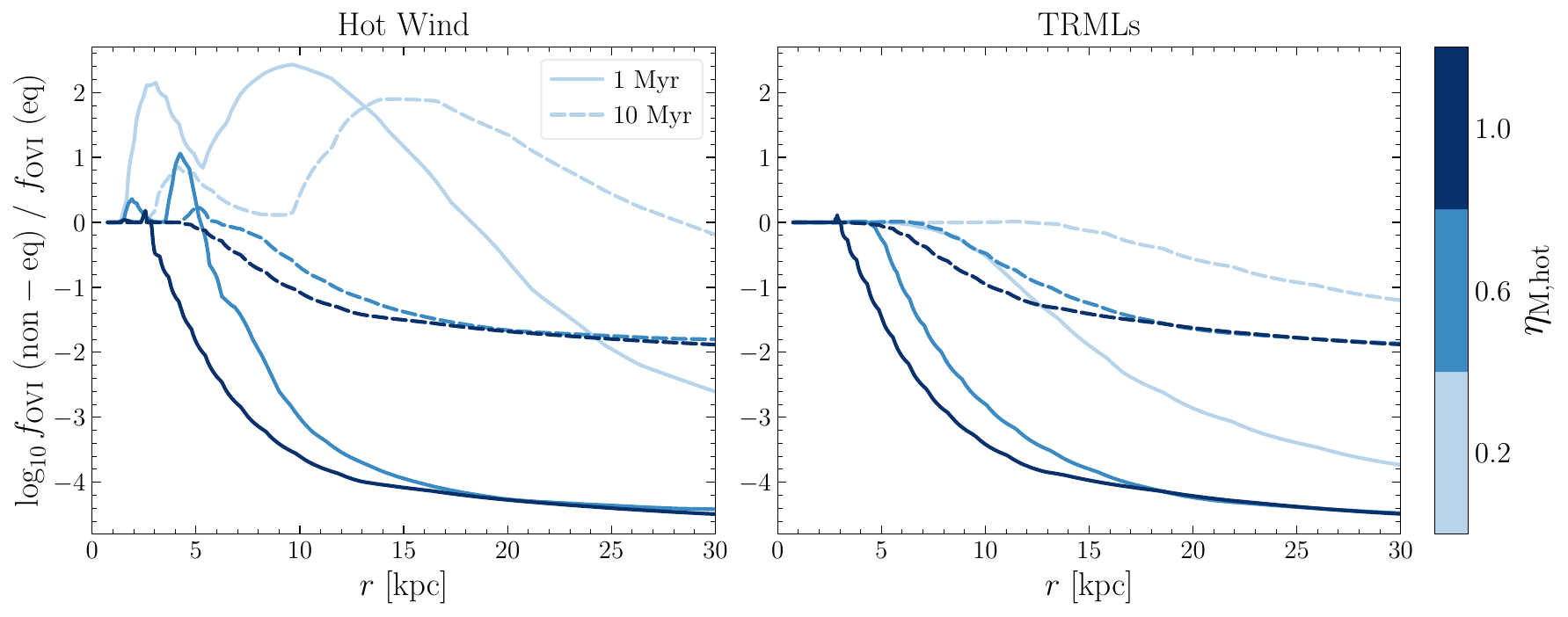}
\caption{Comparisons of the radial profiles of O VI ion fractions ($f_{\rm OVI}$) from non-equilibrium (1 Myr: solid; 10 Myr: dashed) models derived with \texttt{CHIMES} for a range of $\eta_{\rm M,hot}$ values (all other parameters are fiducial). \texttt{CHIMES} computes non-equilibrium ion abundances as a function of density, temperature, and metallicity, which can be converted to the radial profiles shown in the figure given the phase structure of the hot wind (\textit{left}) and TRMLs (\textit{right}) calculated from the \citetalias{FB_2022} model. We demonstrate how the O VI ion abundance directly affects the O VI SB profile in \autoref{fig:Model_vs_Hayes16_Observation_noneq_effects}.}
\label{fig:ovi_eq_noneq_profiles}
\end{figure*}

\begin{figure}
\centering
\includegraphics[width=\columnwidth]{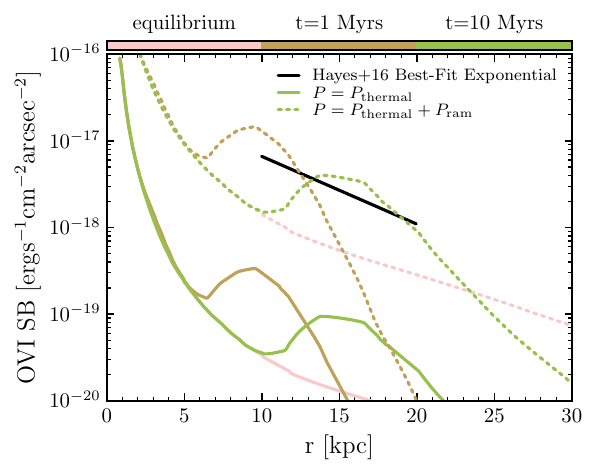}
\caption{Similar to \autoref{fig:Model_vs_Hayes16_Observation} (using fiducial parameters of SFR = $40 \ \rm{M_{\odot} \ yr^{-1}}$, $\eta_{\rm{M,hot}}=0.2$, $\eta_{\rm{M,cold}}=0.1$, and $M_{\rm cloud} = 10^5 \ M_{\odot}$), but now $\eta_{\rm E}$ is fixed at 1 and ion abundances at equilibrium (which is the fiducial choice), $t = 1$ Myrs, and 10 Myrs under non-equilibrium conditions are used. The non-equilibrium abundances are calculated using \texttt{CHIMES} and discussed in detail in \autoref{subsec:non_equil}. As in \autoref{fig:Model_vs_Hayes16_Observation}, ram pressure plays the dominant role in boosting O VI SB. 
Without invoking additional energy sources ($\eta_{\rm E}=1.0$), simple accounting for non-equilibrium abundances can also boost O VI SB by a factor up to $\sim 10$. 
}
\label{fig:Model_vs_Hayes16_Observation_noneq_effects}
\end{figure}

\subsection{Non-Equilibrium Ionization}\label{subsec:non_equil}
Previous studies \citep{Gray_2019_noneq, Danehkar_2022, Sarkar_2022} have found that in a single-phase expanding wind, the wind density gradually decreases with radius ($n_{\rm hot} \propto r^{-2}$), leading to a regime where the recombination time ($\tau_{\rm rec} \sim 1 / n \, \alpha_{\rm rec}$, where $\alpha_{\rm rec}$ is the recombination coefficient) exceeds the radiative cooling time ($\tau_{\rm cool} \propto P / \dot{\epsilon}$, where $\dot{\epsilon} \propto n^2 \, \Lambda$ is the cooling rate) and the advection time ($\tau_{\rm adv} \sim r / v_{\rm hot}$) of the hot wind, thereby breaking the assumption of ionization equilibrium. 
For example, \cite{Danehkar_2022} found that $\tau_{\rm rec}$ exceeds $\tau_{\rm cool}$ for O\,{\footnotesize VI} once the electron density drops to $n_{e} \sim 1 \ \rm{cm^{-3}}$ and the wind temperature cools below $10^6 \ \rm{K}$, pushing the wind into a non-equilibrium ionization state. 
Since these timescales depend on the wind's physical properties, which are functions of radius, it is essential to compute the specific ion fractions at each radius. 
This is typically done in a post-processing manner: the hydrodynamic equations for the wind are solved first (assuming the ions do not influence the wind dynamics), and the resulting properties are then used to determine the non-equilibrium ionization state by coupling to a time-dependent chemical framework \citep{Danehkar_2022, Sarkar_2022}.

However, these previous studies have not explored the effects of non-equilibrium ionization in a multiphase galactic wind, which exhibits distinct radial evolutions for the physical properties of the hot wind (e.g., $T_{\rm hot}$, $n_{\rm hot}$, $Z_{\rm hot}$) and the mixing layers (e.g., $T_{\rm mix} = (T_{\rm hot} T_{\rm cold})^{1/2}$ where $T_{\rm cold}$ is fixed at $10^4 \ \rm{K}$ in the \citetalias{FB_2022} model, $n_{\rm mix} = \chi^{1/2} n_{\rm hot}$, and $Z_{\rm mix} = (Z_{\rm hot} Z_{\rm cold})^{1/2}$). 
Leveraging the chemical network software package \texttt{CHIMES} \citep{Richings_2014a,Richings_2014b}, we can solve for both equilibrium and non-equilibrium ion abundances using a setup similar to that implemented in the \citetalias{PS_2020} cooling table (the \texttt{Colibre} setup in \texttt{CHIMES}). 
This approach allows us to determine whether non-equilibrium ionization is the dominant effect responsible for the discrepancy between our framework's fiducial prediction for O\,{\footnotesize VI} SB and the \citetalias{Hayes_2016} observation (\autoref{fig:Model_vs_Hayes16_Observation}).

\autoref{fig:ovi_eq_noneq_profiles} shows the radial evolution of the O\,{\footnotesize VI} ion fraction ($f_{\rm OVI} = n_{\rm OVI} / n_{\rm O,tot}$) for the hot wind (left panel) and the mixing layers (right panel) by interpolating \texttt{CHIMES} models onto the radial profiles of physical properties derived from the \citetalias{FB_2022} model. 
The results are shown for the \texttt{CHIMES} non-equilibrium models at different evolutionary times (computed with a time step of 1 Myr; 1 Myr: solid line; 10 Myr: dashed line).
These time steps are chosen to roughly reflect the stellar population age constraints from \citetalias{Hayes_2016}.
At small radii ($ r \lesssim 3$ kpc) and for $0.2 \leq \eta_{\rm M,hot} \leq 1.0$, where densities span $10^{-3} \ \rm{cm^{-3}} \lesssim n_{\rm hot} \lesssim 10^{-0.5} \ \rm{cm^{-3}}$,  $10^{-2.5} \ \rm{cm^{-3}} \lesssim n_{\rm cl} \lesssim 10^{3.0} \ \rm{cm^{-3}}$, and the mixing temperature is $10^{4} \ \rm{K} \lesssim T_{\rm mix} \lesssim 10^{6} \ \rm{K}$, the difference in $f_{\rm OVI}$ between the equilibrium and non-equilibrium models is negligible.

At larger radii ($ r \gtrsim 3$ kpc), the properties of the hot wind phase are determined by the scaling relations $v_{\rm hot} \propto (\eta_{\rm E} / \eta_{\rm M,hot})^{1/2}$, $n_{\rm hot} \propto \eta_{\rm M,hot}^{3/2} / \eta_{\rm E}^{1/2}$ (\autoref{eq:3.3}), and $T_{\rm hot} \propto \eta_{\rm E} / \eta_{\rm M,hot}$ \citep{CC85}. Consequently, a smaller $\eta_{\rm M,hot}$ results in a faster, hotter, and lower-density wind. Since the recombination coefficient $\alpha_{\rm rec}$ is inversely correlated to temperature, the combined effects of a lower density ($n_{\rm hot}$) and a smaller $\alpha_{\rm rec}$ at higher temperature ($T_{\rm hot}$) lead to a significantly longer $\tau_{\rm rec}$.

When $\tau_{\rm rec}$ is relatively long compared to $\tau_{\rm cool}$ and $\tau_{\rm adv}$, one might expect the gas to be overionized in the non-equilibrium solution (i.e., $f_{\rm O\,{\footnotesize VI}}\,(\rm non-eq) / f_{\rm O\,{\footnotesize VI}}\,(\rm eq) < 1$). However, the ionization state also depends on the collisional ionization timescale at the specific CHIMES snapshot shown in \autoref{fig:ovi_eq_noneq_profiles}.
At $t=1$ Myrs, collisional ionization has not had sufficient time to drive oxygen all the way up to its equilibrium ionization state, which would be dominated by higher ions (e.g., O\,{\footnotesize VIII}, O\,{\footnotesize IX}), and we are at a frozen snapshot where O\,{\footnotesize VI} exceeds its equilibrium abundance (see the left panel of \autoref{fig:ovi_eq_noneq_profiles}). This effect is less pronounced at later time ($t=10$ Myrs) and larger values of $\eta_{\rm M,hot}$ (which implies a colder, denser wind where collisional ionization can proceed more efficiently). 

Conversely, we find that the trend in the mixing layers is the opposite of that in the hot wind: the non-equilibrium model with a larger $\eta_{\rm M,hot}$ diverges from the equilibrium ionization state more rapidly than the model with a smaller $\eta_{\rm M,hot}$.
This behavior can be explained as follows. Although a lower $\eta_{\rm M,hot}$ results in a higher $T_{\rm hot}$ and consequently a higher $T_{\rm mix}$, it also leads to a higher $n_{\rm mix}$. 
The higher $n_{\rm mix}$ occurs because a hot wind with low mass-loading is less efficient at transferring mass from the hot phase to the cold clouds than a high-$\eta_{\rm M,hot}$ wind.
This efficiency is primarily controlled by the dimensionless parameter $\xi = r_{\rm cloud} / (v_{\rm turb} \tau_{\rm cool,mix})$, which compares the mixing timescale to the cooling timescale in the mixing layers. 
The low-$\eta_{\rm M,hot}$ wind produces a smaller $\xi$ due to its decreased wind density and increased wind temperature, which drives up $\tau_{\rm cool,mix}$ and suppresses efficient mixing.
Consequently, in this regime, $n_{\rm hot}$ does not decrease and converge towards $n_{\rm cl}$ as quickly as it does in the high-$\eta_{\rm M,hot}$ case.
The net effect is a compromise between a slightly smaller recombination coefficient $\alpha_{\rm rec}$ (due to higher $T_{\rm mix}$) and a significantly higher $n_{\rm mix}$. 
When $\eta_{\rm M,hot}$ is low, this results in a shorter $\tau_{\rm rec}$, causing its non-equilibrium solution to diverge more slowly from equilibrium than the high-$\eta_{\rm M,hot}$ case.

Moreover, since the electron thermal energies at $T_{\rm mix} \sim 10^{4}-10^{5.5}\,\mathrm{K}$ are below the ionization thresholds required to produce O\,{\footnotesize VI}, oxygen cannot be collisionally driven into O\,{\footnotesize VI} rapidly enough to attain its equilibrium abundance at the finite \texttt{CHIMES} snapshot time (e.g., $t = 1\,\rm{Myr}$). 
Consequently, the non-equilibrium solution shows an under-abundance of O\,{\footnotesize VI} relative to equilibrium in mixing layers, opposite from the hot-wind scenario.

Because the O\,{\footnotesize VI} cooling emissivity depends on the product of $n_e$ and $n_{\rm OVI}$ and the difference in $n_e$ between the \texttt{CHIMES} equilibrium and non-equilibrium solutions is relatively small ($\lesssim 10 \%$ when $r \lesssim 100 \ \rm{kpc}$) for both the hot wind and the mixing layers, we can approximate the non-equilibrium O\,{\footnotesize VI} SB by scaling our equilibrium radial profile by a factor of $n_{\rm OVI,non-eq} / n_{\rm OVI,eq}$. We show the resulting O\,{\footnotesize VI} SB profiles in \autoref{fig:Model_vs_Hayes16_Observation_noneq_effects}. As a point of comparison, we include SB profiles that account for just non-equilibrium effects, just ram pressure, and both of these effects simultaneously. As we saw in \autoref{subsec:ram_pressure} and \autoref{fig:Model_vs_Hayes16_Observation}, accounting for ram pressure itself provides a significant boost of $\sim 2$ orders of magnitude to O\,{\footnotesize VI} SB but is still insufficient for matching the \citetalias{Hayes_2016} observations. On the other hand, the non-equilibrium models can boost O\,{\footnotesize VI} flux by up to a factor of $\sim10$ at $r \approx 10 \ \rm{kpc}$ (for the 1 Myr model) and $r \approx 15 \ \rm{kpc}$ (for the 10 Myr model). Note that in the non-equilibrium models, we have accounted for the effects of non-equilibrium abundances in \textit{both} the hot wind and the TRMLs. 
In the discussions above and \autoref{fig:ovi_eq_noneq_profiles}, we saw that the radial dependence of the O\,{\footnotesize VI} ion fraction ($f_{\rm OVI} = n_{\rm OVI} / n_{\rm O,tot}$) under non-equilibrium conditions is different for the hot wind and the TRMLs. 
This is reflected in our O\,{\footnotesize VI} SB profile predictions in \autoref{fig:Model_vs_Hayes16_Observation_noneq_effects}. 
At $r \lesssim 5$ kpc, O\,{\footnotesize VI} emission is dominated by the contribution from TRMLs, and since $f_{\rm OVI, non-eq} / f_{\rm OVI, eq} \sim 1$ in TRMLs at small radii (see right panel of \autoref{fig:ovi_eq_noneq_profiles}), the O\,{\footnotesize VI} SB profiles with and without non-equilibrium effects are identical. Moving out to $r \gtrsim 5$ kpc, hot wind starts to overwhelm TRMLs in terms of O\,{\footnotesize VI} emission as the wind temperature drops toward $\sim 10^{5.5}$ K where O\,{\footnotesize VI} emissivity peaks. Since $f_{\rm OVI, non-eq} / f_{\rm OVI, non-eq} \gg 1$ out to $15-25$ kpc (depending on where $t=1$ Myrs or 10 Myrs, see left panel of \autoref{fig:ovi_eq_noneq_profiles}) for the hot wind with our fiducial choice of $\eta_{\rm M,hot} = 0.2$, the O\,{\footnotesize VI} SB profile is consequently boosted by up to a factor of $\sim 10$, as evident in \autoref{fig:Model_vs_Hayes16_Observation_noneq_effects}. 
With both ram pressure and non-equilibrium effects accounted for, the prediction of our framework 
is similar to the \citetalias{Hayes_2016} measurement.
However, we note that the observed O\,{\footnotesize VI} SB follows an exponential profile (albeit with significant scatter at radii $\gtrsim 15\,\rm kpc$) and does not exhibit the ``bump'' at $\sim 10-15$kpc introduced by non-equilibrium effects.

\subsection{Extra Energy Budgets}\label{subsec:extra_energy}

In \autoref{subsec:ram_pressure}, we found that ram pressure plays a dominant role in boosting O\,{\footnotesize VI} SB. However, an additional boost in SB is needed in order to match observational results in \citetalias{Hayes_2016}. 
Besides the effect of non-equilibrium ionization (\autoref{fig:Model_vs_Hayes16_Observation_noneq_effects}), \autoref{fig:Model_vs_Hayes16_Observation} shows that this can be in the form of a larger than unity $\eta_{\rm E}$, where $\eta_{\rm E} \sim 2$ increases O\,{\footnotesize VI} SB by an order of magnitude and is sufficient to match observations. 
This suggests that additional energy sources beyond the mechanical energy from CCSNe might be required to accelerate these cold clouds.

Valuable insights can be drawn from recent studies of the Makani galaxy \citep{Rupke2019, Rupke_2023, Ha_2025, Veilleux_2025}, which is one of the few systems with the O\,{\footnotesize VI} emission-line SB measurement like J1156+5008 in \citetalias{Hayes_2016}. 
Compared to J1156+5008, Makani is a more massive star-forming galaxy ($\log M_{\ast} / M_{\odot} = 11.1 \pm 0.2$) resulting from a two-galaxy merger that triggered extreme star formation (SFR = 225–300 $M_{\odot}~\mathrm{yr}^{-1}$) and hosts a galactic superwind extending to $r \sim 50 \ \rm{kpc}$. 
Makani's outflow consists of a two-stage wind: a slow, outer wind that originated 400 Myr ago (Episode I; $\sim$ 20–50 kpc) and a fast, inner wind that is 7 Myr old (Episode II; $\sim$ 0–20 kpc). \cite{Rupke2019} find that the recent Episode II wind is driven not only by ram pressure from hot ejecta but also by radiation pressure from the central ionizing source (i.e., massive stars). 
This conclusion is based on the observed momentum injection rate, $\dot{p}_{\rm obs} \sim 5 - 6 \times 10^{36} \ \rm{dyn}$, which exceeds that of the hot wind, $\dot{p}_{\rm hot}\simeq (2 \, \dot{E}_{\rm hot} \, \dot{M}_{\rm hot})^{1/2} \simeq (3 - 5) \times 10^{35} \, (\eta_{\rm E})^{1/2} \, \rm{dyn}$ (for $\eta_{\rm M,hot} \sim 0.1$–$0.3$, consistent with M82; \citealt{SH_2009, Rupke_2023}), by a factor of $\sim 10$. 

Given the presence of dusty and molecular phases observed in the Episode II wind \citep{Rupke_2023, Veilleux_2025}, the outflow may approach the optically thick limit (i.e., optically thick to reradiated far-infrared photons). In this regime, the radiation pressure can be as large as $\dot{p}_{\rm rp} \simeq \tau_{\rm R} L / c$, where $\tau_{\rm R} \gg 1$ \citep{TH_2024} and $L / c \simeq \dot{p}_{\rm hot}$ for Makani \citep{Rupke_2023}.
Therefore, the combination of $\dot{p}_{\rm hot}$ and $\dot{p}_{\rm rp}$ in the optically thick limit can potentially account for the observed $\dot{p}_{\rm obs}$. 

Besides the radiative feedback from massive stars, additional O\,{\footnotesize VI} luminosity can arise from oxygen-enriched gas tidally stripped during galaxy mergers and dispersed into the CGM \citep{OSullivan_2009, Baron_2024}. This gas can be shock-heated through the thermalization of the progenitors’ orbital energy \citep{Cox_2004, Martin_2006, Peng_2025}, consistent with the interpretation that the most recent starburst activity in J1156+5008 and Makani was triggered by the final coalescence of two merging galaxies.
Following \cite{Peng_2025}, we can estimate the thermal energy from merger-induced strong shocks, assuming a nearly 50$\%$ transfer efficiency based on Eqs. (2) and (3) in \cite{Cox_2004}:
\begin{align}
      k T_{\rm{sh}} \sim \frac{3}{16} m_p v_{r}^2 \sim \frac{3}{16} m_p \frac{e^2}{1 + e} \frac{G M_{\rm{vir}}}{R_{\rm{peri}}},
     \label{eq:4.2}
\end{align} 
where $T_{\rm{sh}}$ is the gas shock temperature, $m_p$ is the proton mass, $v_{r}$ is the radial velocity of the merger progenitor, and $M_{\rm{vir}}$ is the virial mass.
The initial pericentric distance and the orbit eccentricity are assumed to be $R_{\rm{peri}} \sim 1 - 10 \ \rm{kpc}$ and $e \sim 1$ (i.e., a parabolic trajectory) based on Table 2 in \cite{Cox_2004}.

We can then estimate the total thermal energy via merger-induced shocks as $\frac{3}{16} M_{\rm{g}} v_{r}^2 \sim \frac{3}{16} G M_{\rm{g}} M_{\rm{vir}} / R_{\rm{peri}}$ (\autoref{eq:4.2}), where $M_{\rm{g}}$ is the total gas mass of the system. 
To estimate the gas mass, we first convert the observed SFR surface density $\dot{\Sigma}_{\ast} \sim \rm{SFR}/ (2 \pi r_{50}^2)$ \citepalias[SFR $\simeq 40 \ M_{\odot} \ \rm{yr^{-1}}$ and $r_{50} \simeq 0.75 \ \rm{kpc}$;][]{Hayes_2016} to a gas surface density ($\Sigma_{g}$) by inverting the Kennicutt-Schmidt law \citep{Kennicutt_1998}. 
The gas mass is then given by $M_{\rm{g}} = \mu M_{\ast} / (1 - \mu)$, where the gas fraction is $\mu = \Sigma_{g} / (\Sigma_{g} + \Sigma_{\ast})$ and $\Sigma_{\ast}$ is the stellar mass surface density.
Approximating $M_{\rm vir} \sim 10^{11} \ M_{\odot}$ from the reported stellar mass ($M_{\ast} = 1.5 \times 10^9 \ M_{\odot}$) in \citetalias[][]{Hayes_2016} and the stellar-to-halo mass relation derived by \cite{Girelli_2020}, we estimate the merger-induced thermal energy to be in the range $10^{57} - 10^{58} \ \rm{erg}$ for $R_{\rm{peri}} \sim 1 - 10 \ \rm{kpc}$. 
This value is comparable to the total mechanical energy budget of $7.8 \times 10^{57} \, \rm{erg}$ from O star winds and supernova ejecta \citepalias{Hayes_2016}. 

Consequently, radiative feedback from massive stars and/or the thermalization of merger progenitors’ orbital energy can plausibly enhance the O\,{\footnotesize VI} SB by a factor of $\sim 2$, the level required to match the \citetalias{Hayes_2016} observation in \autoref{fig:Model_vs_Hayes16_Observation}.
We note, however, that these physical mechanisms may dominate at different times. For instance, LyC/UV radiation dominates the momentum and energy budget only for the first $\sim 4$ Myr in a simple stellar population, while CCSNe dominate afterward \citep{Leitherer_1999}. Therefore, a future implementation of the time evolution in our model will be crucial for accurately simulating the combined effects of these feedback mechanisms. 
Furthermore, although the thermalization efficiency factor $\eta_{\rm E}$ is a parameter of  \citetalias{FB_2022} and can be self-consistently accounted for when solving the model, the processes discussed in this section that may enhance $\eta_{\rm E}$ could also violate some of the underlying assumptions of \citetalias{FB_2022}, such as a steady-state and spherically symmetric wind. For this reason, we caution the readers to interpret model outputs with $\eta_{\rm E} > 1$ in a qualitative manner.




\begin{figure*}
\centering
\includegraphics[width=\textwidth]{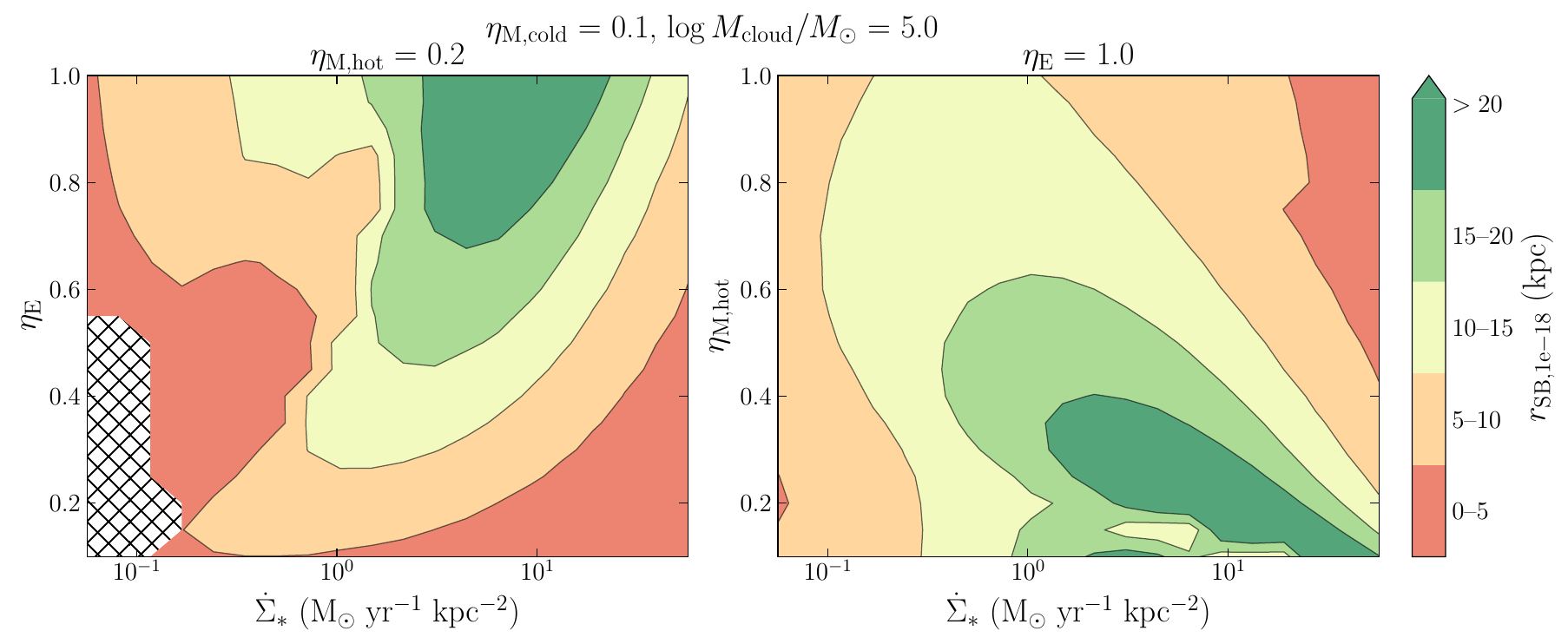}
\caption{O VI detectability study as a function of hot phase parameters in our multiphase galactic wind framework. 
We include the effects of ram pressure equilibrium as discussed in \autoref{subsec:ram_pressure} and non-equilibrium abundances at $t=10$ Myrs as discussed in \autoref{subsec:non_equil}. All parameters are set to their fiducial values unless explicitly varied. We assume a detection limit of $10^{-18} \ \rm{erg \ s^{-1} \ cm^{-2} \ arcsec^{-2}}$ and calculate the radius $r_{\rm{SB,1e-18}}$ where O VI SB falls below the detection limit. 
Green regions indicate large $r_{\rm{SB,1e-18}}$, which is favorable for detection, while red regions indicate small $r_{\rm{SB,1e-18}}$, which is unlikely to be detectable. The hatched region in the left panel represents SB profiles that do not intersect the prescribed detection limit of $10^{-18} \ \rm{erg \ s^{-1} \ cm^{-2} \ arcsec^{-2}}$, i.e., profiles whose amplitudes remain below the detection threshold at all radii for these low-$\dot{\Sigma}_{\ast}$ systems. The ideal regime for O VI SB detection is  $1 \lesssim \dot{\Sigma}_{\ast} / (M_{\odot}\ \rm{yr^{-1}\ kpc^{-2}}) \lesssim 20$, $\eta_{\rm M,hot} \sim 0.2 - 0.4$, and $\eta_{\rm E} \gtrsim 0.8$. These results are consistent with existing O VI SB observations and are extremely informative for guiding future observation programs.}
\label{fig:ovi_detect_limit}
\end{figure*}

\subsection{O VI SB Detectability}\label{subsec:connect_to_obs}

One might ask ``Why are there so few star-forming galaxies with O\,{\footnotesize VI} emission-line SB measurements in the literature besides J1156+5008 and Makani?'' 
For example, the \emph{HST} Cycle-23 proposal 14079 (PI: M. Hayes) used a similar observational setup to \citetalias{Hayes_2016} to map O\,{\footnotesize VI} around four $z \approx 0.25$ starbursts, but no significant emission was detected.
In addition to the fact that the proposed observations might not be as deep as those of J1156+5008 by \citet{Hayes_2016},
a naive explanation is that the proposal did not select sufficiently strong starburst systems; two of the targeted galaxies have $\rm{SFR} \lesssim 5 \ M_{\odot} \ \rm{yr^{-1}}$, and one has $\dot{\Sigma}_{\ast} \simeq 0.05 \ M_{\odot} \ \rm{yr^{-1}} \ \rm{kpc^{-2}}$, making the sample less likely to yield detections.
Is this reasonable? How do other galactic wind parameters affect the detectability of O\,{\footnotesize VI} emission? Our framework allows us to address these questions. 
Without accounting for the additional physics of ram pressure, non-equilibrium ionization, and extra energy sources discussed in \autoref{sec:discussions}, our fiducial framework generally predicts faint O\,{\footnotesize VI} emission at $r \gtrsim 10\, \rm{kpc}$ that is below the detection threshold of current state-of-the-art instruments, largely regardless of the model parameters. 
This is consistent with the lack of O\,{\footnotesize VI} SB measurements reported in the literature. 
However, the fact that accounting for these additional physical effects produces better qualitative agreement with observations in \cite{Hayes_2016} motivates us to examine O\,{\footnotesize VI} detectability with ram pressure and non-equilibrium ionization included.

We examine how key galactic wind parameters including SFR, $\eta_{\rm M,hot}$, and $\eta_{\rm E}$ affect O\,{\footnotesize VI} SB detectability in \autoref{fig:ovi_detect_limit}. For a system to be deemed ``detectable'', O\,{\footnotesize VI} SB needs to be maintained above the detection limit (set by continuum subtraction and exposure time; \citealp{Hayes_2016}) out to sufficiently large radii in the CGM, where contamination from central galactic sources is minimal and clean emission line observations can be taken. 
Using our multiphase framework and including the effects of ram pressure equilibrium and non-equilibrium abundances at $t = 10$ Myrs (as discussed in \autoref{subsec:ram_pressure} and \autoref{subsec:non_equil}, respectively), we generate O\,{\footnotesize VI} SB profiles and determine the radius at which the SB reaches a detection limit of $\sim 10^{-18} \ \rm{erg \ s^{-1} \ cm^{-2} \ arcsec^{-2}}$ \citep[comparable to the limits in \citetalias{Hayes_2016} and][]{Ha_2025}; we denote this radius as $r_{\rm{SB,1e-18}}$. $r_{\rm{SB,1e-18}}$ quantifies the spatial extent of O\,{\footnotesize VI} emission, and a higher value of $r_{\rm{SB,1e-18}}$ means O\,{\footnotesize VI} remains bright and is more likely to 
be detectable at large radii in the CGM.
In the left panel of \autoref{fig:ovi_detect_limit}, we compute $r_{\rm{SB,1e-18}}$ for $\dot{\Sigma}_{\ast}$ varied from $\sim 0.05$ to $55 \ M_{\odot}\ \rm{yr^{-1}\ kpc^{-2}}$ (corresponding to SFR of $0.1$–$100 \ M_{\odot}\ \rm{yr^{-1}}$, with $r_{50}$ fixed at $0.75$ kpc\footnote{We compute the SFR surface density in our framework using $\dot{\Sigma}_{\ast} = \left. {\rm SFR}\right/ \pi r_{50}^2$}) and $\eta_{\rm E}$ between 0.1 and 1.0. 
As we discuss below, the range of $\dot{\Sigma}_{\ast}$ we choose here encapsulates the ideal regime for O\,{\footnotesize VI} emission detection.
Greenish colors indicate that the O\,{\footnotesize VI} SB profile remains bright and likely detectable out to $\gtrsim 20 \ \rm{kpc}$, while reddish colors indicate that the profile falls below the detection limit at $\lesssim 5 \ \rm{kpc}$ and is unlikely to be detectable. 
The ideal regime of O\,{\footnotesize VI} SB detection (colored in green in \autoref{fig:ovi_detect_limit}) appears to be $1 \lesssim \dot{\Sigma}_{\ast} / (M_{\odot}\ \rm{yr^{-1}\ kpc^{-2}}) \lesssim 20$ and $\eta_{\rm E} \gtrsim 0.8$. We can understand this from the parameter studies we conducted in \autoref{sec:results}. 
\autoref{fig:SB_profile_vs_hot_phase_params} shows that increasing $\dot{\Sigma}_{\ast}$ increases the overall normalization of the O\,{\footnotesize VI} SB profile.
We need a large enough $\dot{\Sigma}_{\ast}$ such that O\,{\footnotesize VI} SB is bright enough to be detectable. 
At the same time, \autoref{fig:fb22_pressure_mrel_profiles} demonstrates that $P_{\rm ram} / P_{\rm th}$ is negatively correlated with $\dot{\Sigma}_{\ast}$. 
Moreover, increasing $\dot{\Sigma}_{\ast}$ beyond a certain threshold shifts the peak of the $P_{\rm ram} / P_{\rm th}$ profile to radii $\lesssim 5 \ \rm{kpc}$, in contrast to lower-$\dot{\Sigma}_{\ast}$ models, which exhibit profiles that rise gradually and eventually saturate (see \autoref{subsec:ram_pressure} for details). This means for large $\dot{\Sigma}_{\ast}$, $P_{\rm ram} / P_{\rm th}$ is significantly reduced at $r \gtrsim 10$ kpc in the CGM, which leads to a lower pressure, smaller density, and fainter SB.
This behavior sets the upper limit of $\dot{\Sigma}_{\ast}$ in the green region in \autoref{fig:ovi_detect_limit}. Indeed, both galaxies with existing O\,{\footnotesize VI} SB detections, J1156+5008 and Makani (specifically the Episode II wind), exhibit $10 \lesssim \dot{\Sigma}_{\ast} / (M_{\odot}\ \rm{yr^{-1}\ kpc^{-2}}) \lesssim 20$, sitting right in the region of optimal O\,{\footnotesize VI} detectability we found. 

As for $\eta_{\rm E}$, \autoref{fig:SB_profile_vs_hot_phase_params} shows that increasing it boosts the overall normalization of the SB profile without affecting the slope by much, which is advantageous for detection. Thus, a high value of $\eta_{\rm E} \gtrsim 0.8$ is optimal for detectability, as shown in the left panel of \autoref{fig:ovi_detect_limit}.

The right panel of \autoref{fig:ovi_detect_limit} is similar to the left panel, except that we vary $\eta_{\rm M,hot}$ from 0.1 to 1. 
This panel reveals that $\eta_{\rm M,hot} \sim 0.2 - 0.4$ is also ideal for O\,{\footnotesize VI} emission detection given that we pick optimal values for $\dot{\Sigma}_{\ast}$ and $\eta_{\rm E}$ as discussed above.
Higher $\eta_{\rm M,hot}$ values result in a denser wind that cools faster and exhibits a steeper SB profile (see middle panel of \autoref{fig:SB_profile_vs_hot_phase_params} for details), and for $\eta_{\rm M,hot}$ lower than this range, the wind soon becomes too tenuous to carry the bulky cold phase out in an outflow. 

Our analysis primarily focuses on the \textit{magnitude} of  O\,{\footnotesize VI} SB and at what radius it drops below a prescribed detection limit. At the same time, another relevant factor is the \textit{shape} of the O\,{\footnotesize VI} SB profile. 
In particular, a flat O\,{\footnotesize VI} profile can be mistaken for background sky emission if its spatial extent is comparable to the instrument’s field of view and thus removed during sky subtraction (i.e., in the absence of a ``pure'' sky region), while a profile with a steep radial gradient does not suffer from this problem\footnote{An example of such a flat, spatially-extended SB profile can be found in \autoref{fig:SB_profile_vs_hot_phase_params} when $\eta_{\rm M,hot}$ is large. 
}. We note that the ideal choices of galactic outflow parameters we identified above tend to produce steep O\,{\footnotesize VI} SB profiles, as we show in \autoref{fig:SB_profile_vs_hot_phase_params}. At $r\sim 10-20$ kpc, these ideal parameters produce profiles that drop by at least $\sim 1$ order of magnitude. Furthermore, non-equilibrium effects discussed in \autoref{subsec:non_equil} introduce additional curvatures to the SB profile, as shown in \autoref{fig:Model_vs_Hayes16_Observation_noneq_effects}. 
For these reasons, we believe that the shape of O\,{\footnotesize VI} SB profiles is unlikely to pose challenges to observations.

The results of this parameter study are consistent with existing observations and are extremely informative for guiding future endeavors in studying galactic outflows via emission. 
We refer readers to \autoref{appendix:ovi_sb_detect} for how O\,{\footnotesize VI} SB detectability depends on hot-phase (\autoref{fig:ovi_detect_limit_a1}, with different choices of $\eta_{\rm E}$ and $\eta_{\rm M,hot}$ compared to \autoref{fig:ovi_detect_limit}) and cold-phase parameters (\autoref{fig:ovi_detect_limit_a11}). 
The detection limit of $10^{-18} \ \rm{erg \ s^{-1} \ cm^{-2} \ arcsec^{-2}}$ we use in \autoref{fig:ovi_detect_limit} is based on results in \citetalias{Hayes_2016}. 
However, it is worth noting that upcoming instruments, including \emph{Aspera} \citep{Chung24} and the Habitable Worlds Observatory (\emph{HWO}), will likely allow for orders of magnitude enhancement in O\,{\footnotesize VI} detection sensitivity (e.g., reaching depths of $\sim 10^{-21} \ \rm{erg \ s^{-1} \ cm^{-2} \ arcsec^{-2}}$ with \emph{HWO};  \citealp{Burchett25}). 
In \autoref{fig:ovi_detect_limit_a2}, we demonstrate that increasing detection sensitivity by a factor of 5 can already lead to tremendous improvements on O\,{\footnotesize VI} detectability. 
As more spatially resolved emission studies of galactic outflows in the CGM are enabled by the next generation of instruments, the utility of the framework described in this work will exponentially increase.

\begin{figure}
\centering
\includegraphics[width=\columnwidth]{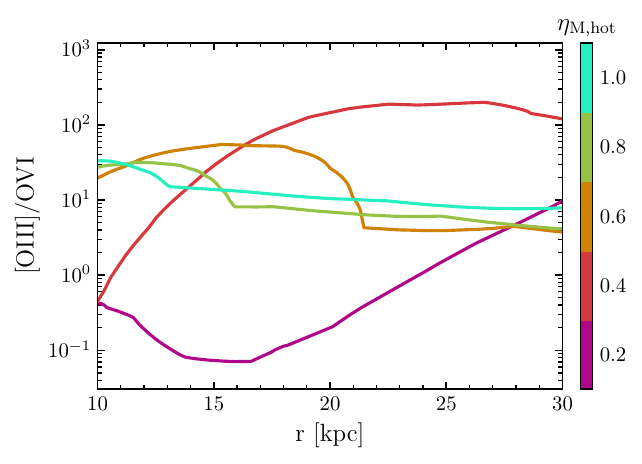}
\caption{[O III]/O VI line ratio predictions at different $\eta_{\rm{M,hot}}$ for the multiphase framework presented in this work, including the effect of \texttt{CHIMES} non-equilibrium abundances at $t=10$ Myrs as discussed in \autoref{subsec:non_equil}. 
}
\label{fig:OIII_over_OVI_ratio_multi_phase_vs_single_phase_noneq}
\end{figure}

Besides using these galactic outflow parameters to gauge O\,{\footnotesize VI} SB detectability, a perhaps more observationally accessible approach is to relate the detectability of O\,{\footnotesize VI} with other emission lines. 
For example, \autoref{fig:OIII_over_OVI_ratio_multi_phase_vs_single_phase_noneq} shows the radial profiles of the [O\,{\footnotesize III}] $\lambda 5007$ to the O\,{\footnotesize VI} doublet ratio ([O\,{\footnotesize III}]/O\,{\footnotesize VI}) as a function of radius for different $\eta_{\rm M,hot}$, including the effect of non-equilibrium abundances at $t = 10$ Myr as discussed in \autoref{subsec:non_equil}. 
The impact of ram-pressure equilibrium on these two lines is identical, since both SB values under thermal-pressure equilibrium are scaled by the same factor, $\left(1 + P_{\rm ram}/P_{\rm th}\right)^{2}$ (\autoref{subsec:ram_pressure}).
Qualitatively, the [O\,{\footnotesize III}]/O\,{\footnotesize VI} profiles are influenced by two key factors: (1) the emissivities of the two lines peak at different temperatures, [O\,{\footnotesize III}] at $\sim 10^{5}$ K and O\,{\footnotesize VI} at $\sim 10^{5.5}$ K, and (2) non-equilibrium chemistry has a more substantial effect on O\,{\footnotesize VI} than on [O\,{\footnotesize III}] \citep{Danehkar_2022, Sarkar_2022}.
Since the [O\,{\footnotesize III}]/O\,{\footnotesize VI} radial profiles differ substantially in both shape and amplitude across values of $\eta_{\rm M,hot}$, observations of [O\,{\footnotesize III}] in the \citetalias{Hayes_2016} target can test whether our framework with fiducial parameter choices (e.g., $\eta_{\rm E} = 1.0$ and $\eta_{\rm M.hot} = 0.2$), which reproduces the O\,{\footnotesize VI} profile (\autoref{fig:Model_vs_Hayes16_Observation} and \autoref{fig:Model_vs_Hayes16_Observation_noneq_effects}), also predicts the correct [O\,{\footnotesize III}]/O\,{\footnotesize VI} profile.
One limitation of our current framework is that it does not incorporate stellar photoionization, which can affect both the total energy budget and the dynamics of the multiphase wind (due to radiative transfer). These processes have a stronger impact on [O\,{\footnotesize III}] than on O\,{\footnotesize VI}, since even hot massive stars produce negligible numbers of photons energetic enough to generate O\,{\footnotesize VI} emission.
In an upcoming work, we will explore the correlation between different emission lines, including [O\,{\footnotesize III}] and O\,{\footnotesize VI}, and use these relationships to make predictions about O\,{\footnotesize VI} detectability.



\section{Conclusion}\label{sec:conclusion}

In this work, we present an analytic framework for predicting emission features of galactic winds. We combine the multiphase galactic wind model by \citetalias{FB_2022} and the TRML model by \cite{Chen_2023} to account for emission from both the volume-filling hot phase and the TRMLs at the interfaces between cold clouds and the hot wind. 
Our framework takes bulk properties of galactic winds (e.g., SFR and mass loading factors of both phases) as inputs and predicts emission features like SB and line ratios as a function of radius up to $\sim 100$ kpc in the CGM. 
After accounting for effects like ram pressure increasing the density cold clouds and TRMLs, non-equilibrium ionization, and energy sources beyond the mechanical energy of supernova explosions (including galaxy mergers and radiative feedback from massive stars), our framework produces O\,{\footnotesize VI} SB profile predictions that are in qualitative agreement with observations by \citetalias{Hayes_2016}.
We also use our framework to investigate how O\,{\footnotesize VI} SB detectability depends on parameters of galactic outflows and attempt to explain the lack of O\,{\footnotesize VI} emission observations to date.

Our key findings are as follows:
\begin{enumerate}
    \item Our multiphase galactic outflow framework makes distinct predictions for emission signatures compared to traditional single-phase models like \citetalias{Thompson_2016}. 
    The normalization and shape of SB profiles are different across multiphase and single-phase models. This provides a diagnostic for the phase structure of galactic winds.
    
    \item Our framework allows for the flexibility of varying several hot and cold phase parameters of the galactic outflow. 
    In particular, we conducted parameter studies on hot phase parameters, including SFR, hot phase mass loading factor ($\eta_{\rm M,hot}$), and thermalization efficiency factor ($\eta_{\rm E}$), as well as the cold phase parameters, including the initial cold phase mass loading factor ($\eta_{\rm M,cold}$) and the initial individual cloud mass ($M_{\rm cloud}$). 
    All of these parameters affect the shape of emission line SB profiles. These results highlight the versatility of our framework and the rich information that can be learned from simple emission line observations.

    \item O\,{\footnotesize VI} SB profile predicted by our framework shows qualitative agreement with observations in \citetalias{Hayes_2016} after we account for the effects of ram pressure equilibrium between the cold clouds and hot wind, non-equilibrium ionization, and extra energy sources like radiation and merger-induced thermal energy. Among these factors, ram pressure (in addition to thermal pressure) has the most significant effect and can boost O\,{\footnotesize VI} SB by $\sim 2$ orders of magnitude.

    \item We perform a model parameter study and find that the optimal galactic wind properties for facilitating O\,{\footnotesize VI} emission observations above a detection limit of $10^{-18} \ \rm{erg \ s^{-1} \ cm^{-2} \ arcsec^{-2}}$ are SFR surface density $1 \lesssim \dot{\Sigma}_{\ast} / (M_{\odot}\ \rm{yr^{-1}\ kpc^{-2}}) \lesssim 20$, $\eta_{\rm M,hot} \sim 0.2 - 0.4$, and $\eta_{\rm E} \gtrsim 0.8$. 
    
\end{enumerate}

Our framework provides a unique avenue for studying galactic winds through emission-line SB radial profiles. 
It has low computational cost and is widely applicable to galaxies with a diverse range of sizes, outflow rates, and star formation activities. Insights from our framework not only help with interpreting existing and upcoming data but can also inform future observational decisions of galactic outflow studies.

Complementary to this work, we will extend our multiphase framework to predict absorption-line features, such as ion column densities as functions of radius and impact parameter, enabling direct comparisons with “down-the-barrel” and quasar absorption-line spectroscopy studies of galactic winds. This absorption-line extension will allow more stringent constraints on galactic outflow rates (e.g., mass, momentum, and energy) and cold-cloud properties (e.g., $\eta_{\rm M,cold}$ and $M_{\rm cloud}$) by jointly fitting emission-line SB profiles and/or line ratios together with absorption-line column densities.

\section*{Acknowledgements}
Z.P. is grateful to Xinfeng Xu for useful discussions.
Z.P. sincerely acknowledges support for this work from NASA FINESST (Future Investigators in NASA Earth and Space Science and Technology) grant 80NSSC23K1450. Zirui Chen and Peng Oh acknowledge support from NSF grant AST240752.
The authors are grateful to the anonymous referee for providing constructive suggestions.

This work made use of the Purdue Anvil supercomputer through allocation TG-PHY240001 from the Advanced Cyberinfrastructure Coordination Ecosystem: Services \& Support (ACCESS) program, which is supported by National Science Foundation grants \#2138259, \#2138286, \#2138307, \#2137603, and \#2138296.

\section*{Data Availability}
The \texttt{PYTHON} scripts associated with our framework available at following GitHub repository: \url{https://github.com/jasonpeng17/WInterPhase}. This allows the readers to reproduce results presented in this work or generate predictions of emission signatures for any emission line and a wide range of galactic and outflow parameters.



\bibliographystyle{mnras}
\bibliography{example} 

@ARTICLE{Piacitelli_2022,
       author = {{Piacitelli}, Daniel R. and {Solhaug}, Erik and {Faerman}, Yakov and {McQuinn}, Matthew},
        title = "{Absorption-based circumgalactic medium line emission estimates}",
      journal = {\mnras},
         year = 2022,
        month = oct,
       volume = {516},
       number = {2},
        pages = {3049-3067},
          doi = {10.1093/mnras/stac2390},
archivePrefix = {arXiv},
       eprint = {2202.11121},
 primaryClass = {astro-ph.GA},
       adsurl = {https://ui.adsabs.harvard.edu/abs/2022MNRAS.516.3049P}
}

@ARTICLE{FB_2022,
       author = {{Fielding}, Drummond B. and {Bryan}, Greg L.},
        title = "{The Structure of Multiphase Galactic Winds}",
      journal = {\apj},
         year = 2022,
        month = jan,
       volume = {924},
       number = {2},
          eid = {82},
        pages = {82},
          doi = {10.3847/1538-4357/ac2f41},
archivePrefix = {arXiv},
       eprint = {2108.05355},
 primaryClass = {astro-ph.GA},
       adsurl = {https://ui.adsabs.harvard.edu/abs/2022ApJ...924...82F}
}

@ARTICLE{Thompson_2016,
       author = {{Thompson}, Todd A. and {Quataert}, Eliot and {Zhang}, Dong and {Weinberg}, David H.},
        title = "{An origin for multiphase gas in galactic winds and haloes}",
      journal = {\mnras},
         year = 2016,
        month = jan,
       volume = {455},
       number = {2},
        pages = {1830-1844},
          doi = {10.1093/mnras/stv2428},
archivePrefix = {arXiv},
       eprint = {1507.04362},
 primaryClass = {astro-ph.GA},
       adsurl = {https://ui.adsabs.harvard.edu/abs/2016MNRAS.455.1830T}
}

@ARTICLE{Chen_2023,
       author = {{Chen}, Zirui and {Fielding}, Drummond B. and {Bryan}, Greg L.},
        title = "{The Anatomy of a Turbulent Radiative Mixing Layer: Insights from an Analytic Model with Turbulent Conduction and Viscosity}",
      journal = {\apj},
         year = 2023,
        month = jun,
       volume = {950},
       number = {2},
          eid = {91},
        pages = {91},
          doi = {10.3847/1538-4357/acc73f},
archivePrefix = {arXiv},
       eprint = {2211.01395},
 primaryClass = {astro-ph.GA},
       adsurl = {https://ui.adsabs.harvard.edu/abs/2023ApJ...950...91C}
}

@ARTICLE{PS_2020,
       author = {{Ploeckinger}, Sylvia and {Schaye}, Joop},
        title = "{Radiative cooling rates, ion fractions, molecule abundances, and line emissivities including self-shielding and both local and metagalactic radiation fields}",
      journal = {\mnras},
         year = 2020,
        month = oct,
       volume = {497},
       number = {4},
        pages = {4857-4883},
          doi = {10.1093/mnras/staa2172},
archivePrefix = {arXiv},
       eprint = {2006.14322},
 primaryClass = {astro-ph.GA},
       adsurl = {https://ui.adsabs.harvard.edu/abs/2020MNRAS.497.4857P}
}

@ARTICLE{Peng_2025,
       author = {{Peng}, Zixuan and {Martin}, Crystal L. and {Chen}, Zirui and {Fielding}, Drummond B. and {Xu}, Xinfeng and {Heckman}, Timothy and {Ramambason}, Lise and {Li}, Yuan and {Carr}, Cody and {Hu}, Weida and {Chen}, Zuyi and {Scarlata}, Claudia and {Henry}, Alaina},
        title = "{Physical Origins of Outflowing Cold Clouds in Local Star-forming Dwarf Galaxies}",
      journal = {\apj},
         year = 2025,
        month = mar,
       volume = {981},
       number = {2},
          eid = {171},
        pages = {171},
          doi = {10.3847/1538-4357/ada606},
archivePrefix = {arXiv},
       eprint = {2412.05371},
 primaryClass = {astro-ph.GA},
       adsurl = {https://ui.adsabs.harvard.edu/abs/2025ApJ...981..171P}
}

@ARTICLE{Danehkar_2021,
       author = {{Danehkar}, Ashkbiz and {Oey}, M.~S. and {Gray}, William J.},
        title = "{Catastrophic Cooling in Superwinds. II. Exploring the Parameter Space}",
      journal = {\apj},
         year = 2021,
        month = nov,
       volume = {921},
       number = {1},
          eid = {91},
        pages = {91},
          doi = {10.3847/1538-4357/ac1a76},
archivePrefix = {arXiv},
       eprint = {2106.10854},
 primaryClass = {astro-ph.GA},
       adsurl = {https://ui.adsabs.harvard.edu/abs/2021ApJ...921...91D}
}

@ARTICLE{CC85,
       author = {{Chevalier}, R.~A. and {Clegg}, A.~W.},
        title = "{Wind from a starburst galaxy nucleus}",
      journal = {\nat},
         year = 1985,
        month = sep,
       volume = {317},
       number = {6032},
        pages = {44-45},
          doi = {10.1038/317044a0},
       adsurl = {https://ui.adsabs.harvard.edu/abs/1985Natur.317...44C}
}

@ARTICLE{Hayes_2016,
       author = {{Hayes}, Matthew and {Melinder}, Jens and {{\"O}stlin}, G{\"o}ran and {Scarlata}, Claudia and {Lehnert}, Matthew D. and {Mannerstr{\"o}m-Jansson}, Gustav},
        title = "{O VI Emission Imaging of a Galaxy with the Hubble Space Telescope: a Warm Gas Halo Surrounding the Intense Starburst SDSS J115630.63+500822.1}",
      journal = {\apj},
         year = 2016,
        month = sep,
       volume = {828},
       number = {1},
          eid = {49},
        pages = {49},
          doi = {10.3847/0004-637X/828/1/49},
archivePrefix = {arXiv},
       eprint = {1606.04536},
 primaryClass = {astro-ph.GA},
       adsurl = {https://ui.adsabs.harvard.edu/abs/2016ApJ...828...49H}
}

@article{Ha_2025,
doi = {10.3847/1538-4357/add0b5},
url = {https://dx.doi.org/10.3847/1538-4357/add0b5},
year = {2025},
month = {jun},
publisher = {The American Astronomical Society},
volume = {986},
number = {1},
pages = {87},
author = {Ha, Triet and Rupke, David S. N. and Caraker, Shane and Harper, Jack and Coil, Alison L. and Li, Miao and Tremonti, Christy A. and Diamond-Stanic, Aleksandar M. and Geach, James E. and Hickox, Ryan C. and Johnson, Sean D. and Leung, Gene C. K. and Moustakas, John and Perrotta, Serena and Rudnick, Gregory H. and Sell, Paul H. and Whalen, Kelly E.},
title = {Deep Ultraviolet, Emission-line Imaging of the Makani Galactic Wind},
journal = {\apj}
}

@article{Fielding_2020,
	adsurl = {https://ui.adsabs.harvard.edu/abs/2020ApJ...894L..24F},
	archiveprefix = {arXiv},
	author = {{Fielding}, Drummond B. and {Ostriker}, Eve C. and {Bryan}, Greg L. and {Jermyn}, Adam S.},
	doi = {10.3847/2041-8213/ab8d2c},
	eid = {L24},
	eprint = {2003.08390},
	journal = {\apjl},
	month = may,
	number = {2},
	pages = {L24},
	primaryclass = {astro-ph.GA},
	title = {{Multiphase Gas and the Fractal Nature of Radiative Turbulent Mixing Layers}},
	volume = {894},
	year = 2020}

@ARTICLE{Tan_2021,
       author = {{Tan}, Brent and {Oh}, S. Peng and {Gronke}, Max},
        title = "{Radiative Mixing Layers: Insights from Turbulent Combustion}",
      journal = {\mnras},
         year = 2021,
        month = jan,
          doi = {10.1093/mnras/stab053},
archivePrefix = {arXiv},
       eprint = {2008.12302},
 primaryClass = {astro-ph.GA},
       adsurl = {https://ui.adsabs.harvard.edu/abs/2021MNRAS.tmp..101T}
}

@ARTICLE{TH_2024,
       author = {{Thompson}, Todd A. and {Heckman}, Timothy M.},
        title = "{Theory and Observation of Winds from Star-Forming Galaxies}",
      journal = {\araa},
         year = 2024,
        month = sep,
       volume = {62},
       number = {1},
        pages = {529-591},
          doi = {10.1146/annurev-astro-041224-011924},
archivePrefix = {arXiv},
       eprint = {2406.08561},
 primaryClass = {astro-ph.GA},
       adsurl = {https://ui.adsabs.harvard.edu/abs/2024ARA&A..62..529T}
}

@ARTICLE{Smith_2021,
       author = {{Smith}, Matthew C. and {Bryan}, Greg L. and {Somerville}, Rachel S. and {Hu}, Chia-Yu and {Teyssier}, Romain and {Burkhart}, Blakesley and {Hernquist}, Lars},
        title = "{Efficient early stellar feedback can suppress galactic outflows by reducing supernova clustering}",
      journal = {\mnras},
         year = 2021,
        month = sep,
       volume = {506},
       number = {3},
        pages = {3882-3915},
          doi = {10.1093/mnras/stab1896},
archivePrefix = {arXiv},
       eprint = {2009.11309},
 primaryClass = {astro-ph.GA},
       adsurl = {https://ui.adsabs.harvard.edu/abs/2021MNRAS.506.3882S}
}

@ARTICLE{Natalia_2023,
       author = {{Lah{\'e}n}, Natalia and {Naab}, Thorsten and {Kauffmann}, Guinevere and {Sz{\'e}csi}, Dorottya and {Hislop}, Jessica May and {Rantala}, Antti and {Kozyreva}, Alexandra and {Walch}, Stefanie and {Hu}, Chia-Yu},
        title = "{Formation of star clusters and enrichment by massive stars in simulations of low-metallicity galaxies with a fully sampled initial stellar mass function}",
      journal = {\mnras},
         year = 2023,
        month = jun,
       volume = {522},
       number = {2},
        pages = {3092-3116},
          doi = {10.1093/mnras/stad1147},
archivePrefix = {arXiv},
       eprint = {2211.15705},
 primaryClass = {astro-ph.GA},
       adsurl = {https://ui.adsabs.harvard.edu/abs/2023MNRAS.522.3092L}
}

@ARTICLE{SG_2025,
       author = {{Steinwandel}, Ulrich P. and {Goldberg}, Jared A.},
        title = "{Some Stars Fade Quietly: Varied Supernova Explosion Outcomes and Their Effects on the Multiphase Interstellar Medium}",
      journal = {\apj},
         year = 2025,
        month = jan,
       volume = {979},
       number = {1},
          eid = {44},
        pages = {44},
          doi = {10.3847/1538-4357/ad98ea},
archivePrefix = {arXiv},
       eprint = {2310.11495},
 primaryClass = {astro-ph.GA},
       adsurl = {https://ui.adsabs.harvard.edu/abs/2025ApJ...979...44S}
}

@ARTICLE{SD_2015,
       author = {{Somerville}, Rachel S. and {Dav{\'e}}, Romeel},
        title = "{Physical Models of Galaxy Formation in a Cosmological Framework}",
      journal = {\araa},
         year = 2015,
        month = aug,
       volume = {53},
        pages = {51-113},
          doi = {10.1146/annurev-astro-082812-140951},
archivePrefix = {arXiv},
       eprint = {1412.2712},
 primaryClass = {astro-ph.GA},
       adsurl = {https://ui.adsabs.harvard.edu/abs/2015ARA&A..53...51S}
}

@ARTICLE{Torrey_2019,
       author = {{Torrey}, Paul and {Vogelsberger}, Mark and {Marinacci}, Federico and {Pakmor}, R{\"u}diger and {Springel}, Volker and {Nelson}, Dylan and {Naiman}, Jill and {Pillepich}, Annalisa and {Genel}, Shy and {Weinberger}, Rainer and {Hernquist}, Lars},
        title = "{The evolution of the mass-metallicity relation and its scatter in IllustrisTNG}",
      journal = {\mnras},
         year = 2019,
        month = apr,
       volume = {484},
       number = {4},
        pages = {5587-5607},
          doi = {10.1093/mnras/stz243},
archivePrefix = {arXiv},
       eprint = {1711.05261},
 primaryClass = {astro-ph.GA},
       adsurl = {https://ui.adsabs.harvard.edu/abs/2019MNRAS.484.5587T}
}

@ARTICLE{Chisholm_2016,
       author = {{Chisholm}, John and {Tremonti Christy}, A. and {Leitherer}, Claus and {Chen}, Yanmei},
        title = "{A robust measurement of the mass outflow rate of the galactic outflow from NGC 6090}",
      journal = {\mnras},
         year = 2016,
        month = nov,
       volume = {463},
       number = {1},
        pages = {541-556},
          doi = {10.1093/mnras/stw1951},
archivePrefix = {arXiv},
       eprint = {1605.05769},
 primaryClass = {astro-ph.GA},
       adsurl = {https://ui.adsabs.harvard.edu/abs/2016MNRAS.463..541C}
}

@ARTICLE{Martin_2024,
       author = {{Martin}, Crystal L. and {Peng}, Zixuan and {Li}, Yuan},
        title = "{Resolving the Mechanical and Radiative Feedback in J1044+0353 with Keck Cosmic Web Imager Spectral Mapping}",
      journal = {\apj},
         year = 2024,
        month = may,
       volume = {966},
       number = {2},
          eid = {190},
        pages = {190},
          doi = {10.3847/1538-4357/ad34ac},
archivePrefix = {arXiv},
       eprint = {2403.11390},
 primaryClass = {astro-ph.GA},
       adsurl = {https://ui.adsabs.harvard.edu/abs/2024ApJ...966..190M}
}

@article{Wood_2015,
    author = {Wood, Corey M. and Tremonti, Christy A. and Calzetti, Daniela and Leitherer, Claus and Chisholm, John and Gallagher, John S., III},
    title = {Supernova-driven outflows in NGC 7552: a comparison of H α and UV tracers},
    journal = {\mnras},
    volume = {452},
    number = {3},
    pages = {2712-2730},
    year = {2015},
    month = {07},
    issn = {0035-8711},
    doi = {10.1093/mnras/stv1471},
    url = {https://doi.org/10.1093/mnras/stv1471},
    eprint = {https://academic.oup.com/mnras/article-pdf/452/3/2712/4924056/stv1471.pdf},
}

@ARTICLE{Xu_2024,
       author = {{Xu}, Xinfeng and {Henry}, Alaina and {Heckman}, Timothy and {Carr}, Cody and {Strom}, Allison L. and {Jones}, Tucker and {Berg}, Danielle A. and {Chisholm}, John and {Erb}, Dawn and {James}, Bethan L. and {Jaskot}, Anne and {Martin}, Crystal L. and {Mingozzi}, Matilde and {Senchyna}, Peter and {Roy}, Namrata and {Scarlata}, Claudia and {Stark}, Daniel P.},
        title = "{Shining a Light on the Connections between Galactic Outflows Seen in Absorption and Emission Lines}",
      journal = {\apj},
         year = 2025,
        month = may,
       volume = {984},
       number = {1},
          eid = {94},
        pages = {94},
          doi = {10.3847/1538-4357/adc302},
archivePrefix = {arXiv},
       eprint = {2409.19776},
 primaryClass = {astro-ph.GA},
       adsurl = {https://ui.adsabs.harvard.edu/abs/2025ApJ...984...94X}
}

@ARTICLE{NO_2017,
       author = {{Naab}, Thorsten and {Ostriker}, Jeremiah P.},
        title = "{Theoretical Challenges in Galaxy Formation}",
      journal = {\araa},
         year = 2017,
        month = aug,
       volume = {55},
       number = {1},
        pages = {59-109},
          doi = {10.1146/annurev-astro-081913-040019},
archivePrefix = {arXiv},
       eprint = {1612.06891},
 primaryClass = {astro-ph.GA},
       adsurl = {https://ui.adsabs.harvard.edu/abs/2017ARA&A..55...59N}
}

@article{Xu_2023,
doi = {10.3847/1538-4357/acfa71},
url = {https://dx.doi.org/10.3847/1538-4357/acfa71},
year = {2023},
month = {oct},
publisher = {The American Astronomical Society},
volume = {956},
number = {2},
pages = {142},
author = {Xu, Xinfeng and Heckman, Timothy and Yoshida, Michitoshi and Henry, Alaina and Ohyama, Youichi},
title = {What Are the Radial Distributions of Density, Outflow Rates, and Cloud Structures in the M82 Wind?},
journal = {\apj}
}

@article{Tan_2023,
    author = {Tan, Brent and Fielding, Drummond B},
    title = {Cloud atlas: navigating the multiphase landscape of tempestuous galactic winds},
    journal = {\mnras},
    volume = {527},
    number = {4},
    pages = {9683-9714},
    year = {2023},
    month = {12},
    issn = {0035-8711},
    doi = {10.1093/mnras/stad3793},
    url = {https://doi.org/10.1093/mnras/stad3793},
    eprint = {https://academic.oup.com/mnras/article-pdf/527/4/9683/54903497/stad3793.pdf},
}

@ARTICLE{Richings_2014a,
       author = {{Richings}, A.~J. and {Schaye}, J. and {Oppenheimer}, B.~D.},
        title = "{Non-equilibrium chemistry and cooling in the diffuse interstellar medium - I. Optically thin regime}",
      journal = {\mnras},
         year = 2014,
        month = jun,
       volume = {440},
       number = {4},
        pages = {3349-3369},
          doi = {10.1093/mnras/stu525},
archivePrefix = {arXiv},
       eprint = {1401.4719},
 primaryClass = {astro-ph.GA},
       adsurl = {https://ui.adsabs.harvard.edu/abs/2014MNRAS.440.3349R}
}

@ARTICLE{Richings_2014b,
       author = {{Richings}, A.~J. and {Schaye}, J. and {Oppenheimer}, B.~D.},
        title = "{Non-equilibrium chemistry and cooling in the diffuse interstellar medium - II. Shielded gas}",
      journal = {\mnras},
         year = 2014,
        month = aug,
       volume = {442},
       number = {3},
        pages = {2780-2796},
          doi = {10.1093/mnras/stu1046},
archivePrefix = {arXiv},
       eprint = {1403.6155},
 primaryClass = {astro-ph.GA},
       adsurl = {https://ui.adsabs.harvard.edu/abs/2014MNRAS.442.2780R}
}

@ARTICLE{Gray_2019_noneq,
       author = {{Gray}, William J. and {Scannapieco}, Evan and {Lehnert}, Matthew D.},
        title = "{Nonequilibrium Ionization States within Galactic Outflows: Explaining Their O VI and N V Column Densities}",
      journal = {\apj},
         year = 2019,
        month = apr,
       volume = {875},
       number = {2},
          eid = {110},
        pages = {110},
          doi = {10.3847/1538-4357/ab1004},
archivePrefix = {arXiv},
       eprint = {1903.06183},
 primaryClass = {astro-ph.GA},
       adsurl = {https://ui.adsabs.harvard.edu/abs/2019ApJ...875..110G}
}

@ARTICLE{Sarkar_2022,
       author = {{Sarkar}, Kartick C. and {Sternberg}, Amiel and {Gnat}, Orly},
        title = "{Self-ionizing Galactic Winds}",
      journal = {\apj},
         year = 2022,
        month = nov,
       volume = {940},
       number = {1},
          eid = {44},
        pages = {44},
          doi = {10.3847/1538-4357/ac9835},
archivePrefix = {arXiv},
       eprint = {2203.15814},
 primaryClass = {astro-ph.GA},
       adsurl = {https://ui.adsabs.harvard.edu/abs/2022ApJ...940...44S}
}

@ARTICLE{Danehkar_2022,
       author = {{Danehkar}, A. and {Oey}, M.~S. and {Gray}, W.~J.},
        title = "{Catastrophic Cooling in Superwinds. III. Nonequilibrium Photoionization}",
      journal = {\apj},
         year = 2022,
        month = oct,
       volume = {937},
       number = {2},
          eid = {68},
        pages = {68},
          doi = {10.3847/1538-4357/ac8cec},
archivePrefix = {arXiv},
       eprint = {2208.12030},
 primaryClass = {astro-ph.GA},
       adsurl = {https://ui.adsabs.harvard.edu/abs/2022ApJ...937...68D}
}

@article{Schneider:2020,
	adsurl = {https://ui.adsabs.harvard.edu/abs/2020ApJ...895...43S},
	archiveprefix = {arXiv},
	author = {{Schneider}, Evan E. and {Ostriker}, Eve C. and {Robertson}, Brant E. and {Thompson}, Todd A.},
	doi = {10.3847/1538-4357/ab8ae8},
	eid = {43},
	eprint = {2002.10468},
	journal = {\apj},
	month = may,
	number = {1},
	pages = {43},
	primaryclass = {astro-ph.GA},
	title = {{The Physical Nature of Starburst-driven Galactic Outflows}},
	volume = {895},
	year = 2020}

@article{scannapieco15,
	adsurl = {http://adsabs.harvard.edu/abs/2015ApJ...805..158S},
	archiveprefix = {arXiv},
	author = {{Scannapieco}, E. and {Br{\"u}ggen}, M.},
	doi = {10.1088/0004-637X/805/2/158},
	eid = {158},
	eprint = {1503.06800},
	journal = {\apj},
	month = jun,
	pages = {158},
	title = {{The Launching of Cold Clouds by Galaxy Outflows. I. Hydrodynamic Interactions with Radiative Cooling}},
	volume = 805,
	year = 2015}

@article{schneider17,
	adsurl = {http://adsabs.harvard.edu/abs/2017ApJ...834..144S},
	archiveprefix = {arXiv},
	author = {{Schneider}, E.~E. and {Robertson}, B.~E.},
	doi = {10.3847/1538-4357/834/2/144},
	eid = {144},
	eprint = {1607.01788},
	journal = {\apj},
	month = jan,
	pages = {144},
	title = {{Hydrodynamical Coupling of Mass and Momentum in Multiphase Galactic Winds}},
	volume = 834,
	year = 2017}

@article{zhang17,
	adsurl = {http://adsabs.harvard.edu/abs/2017MNRAS.468.4801Z},
	archiveprefix = {arXiv},
	author = {{Zhang}, D. and {Thompson}, T.~A. and {Quataert}, E. and {Murray}, N.},
	doi = {10.1093/mnras/stx822},
	eprint = {1507.01951},
	journal = {\mnras},
	month = jul,
	pages = {4801-4814},
	title = {{Entrainment in trouble: cool cloud acceleration and destruction in hot supernova-driven galactic winds}},
	volume = 468,
	year = 2017}

@article{armillotta17,
	adsurl = {https://ui.adsabs.harvard.edu/abs/2017MNRAS.470..114A},
	archiveprefix = {arXiv},
	author = {{Armillotta}, L. and {Fraternali}, F. and {Werk}, J.~K. and {Prochaska}, J.~X. and {Marinacci}, F.},
	doi = {10.1093/mnras/stx1239},
	eprint = {1608.05416},
	journal = {\mnras},
	month = {Sep},
	number = {1},
	pages = {114-125},
	primaryclass = {astro-ph.GA},
	title = {{The survival of gas clouds in the circumgalactic medium of Milky Way-like galaxies}},
	volume = {470},
	year = {2017}}

@article{gronke18,
	adsurl = {http://adsabs.harvard.edu/abs/2018MNRAS.tmpL.135G},
	archiveprefix = {arXiv},
	author = {{Gronke}, M. and {Oh}, S.~P.},
	doi = {10.1093/mnrasl/sly131},
	eprint = {1806.02728},
	journal = {\mnras},
	month = jul,
	title = {{The growth and entrainment of cold gas in a hot wind}},
	year = 2018}

@article{gronke20-cloud,
	adsurl = {https://ui.adsabs.harvard.edu/abs/2020MNRAS.492.1970G},
	archiveprefix = {arXiv},
	author = {{Gronke}, Max and {Oh}, S. Peng},
	doi = {10.1093/mnras/stz3332},
	eprint = {1907.04771},
	journal = {\mnras},
	month = feb,
	number = {2},
	pages = {1970-1990},
	primaryclass = {astro-ph.GA},
	title = {{How cold gas continuously entrains mass and momentum from a hot wind}},
	volume = {492},
	year = 2020}

@article{li20,
	adsurl = {https://ui.adsabs.harvard.edu/abs/2020MNRAS.492.1841L},
	archiveprefix = {arXiv},
	author = {{Li}, Zhihui and {Hopkins}, Philip F. and {Squire}, Jonathan and {Hummels}, Cameron},
	doi = {10.1093/mnras/stz3567},
	eprint = {1909.02632},
	journal = {\mnras},
	month = feb,
	number = {2},
	pages = {1841-1854},
	primaryclass = {astro-ph.GA},
	title = {{On the survival of cool clouds in the circumgalactic medium}},
	volume = {492},
	year = 2020}

@article{kanjilal21,
	adsurl = {https://ui.adsabs.harvard.edu/abs/2021MNRAS.501.1143K},
	archiveprefix = {arXiv},
	author = {{Kanjilal}, Vijit and {Dutta}, Alankar and {Sharma}, Prateek},
	doi = {10.1093/mnras/staa3610},
	eprint = {2009.00525},
	journal = {\mnras},
	month = feb,
	number = {1},
	pages = {1143-1159},
	primaryclass = {astro-ph.GA},
	title = {{Growth and structure of multiphase gas in the cloud-crushing problem with cooling}},
	volume = {501},
	year = 2021}

@article{abruzzo22,
	adsurl = {https://ui.adsabs.harvard.edu/abs/2022ApJ...925..199A},
	archiveprefix = {arXiv},
	author = {{Abruzzo}, Matthew W. and {Bryan}, Greg L. and {Fielding}, Drummond B.},
	doi = {10.3847/1538-4357/ac3c48},
	eid = {199},
	eprint = {2101.10344},
	journal = {\apj},
	month = feb,
	number = {2},
	pages = {199},
	primaryclass = {astro-ph.GA},
	title = {{A Simple Model for Mixing and Cooling in Cloud-Wind Interactions}},
	volume = {925},
	year = 2022}

@article{chen24,
    author = {Chen, Zirui and Oh, S Peng},
    title = {The survival and entrainment of molecules and dust in galactic winds},
    journal = {\mnras},
    volume = {530},
    number = {4},
    pages = {4032-4057},
    year = {2024},
    month = {04},
    issn = {0035-8711},
    doi = {10.1093/mnras/stae1113},
    url = {https://doi.org/10.1093/mnras/stae1113},
    eprint = {https://academic.oup.com/mnras/article-pdf/530/4/4032/57439757/stae1113.pdf},
}

@article{kaul25,
    author = {Kaul, Ish and Tan, Brent and Oh, S Peng and Mandelker, Nir},
    title = {Tales of tension: magnetized infalling cold clouds and streams in the CGM},
    journal = {\mnras},
    volume = {539},
    number = {4},
    pages = {3669-3696},
    year = {2025},
    month = {04},
    issn = {0035-8711},
    doi = {10.1093/mnras/staf706},
    url = {https://doi.org/10.1093/mnras/staf706},
    eprint = {https://academic.oup.com/mnras/article-pdf/539/4/3669/63037497/staf706.pdf},
}

@article{kwak10,
	adsurl = {http://adsabs.harvard.edu/abs/2010ApJ...719..523K},
	archiveprefix = {arXiv},
	author = {{Kwak}, K. and {Shelton}, R.~L.},
	doi = {10.1088/0004-637X/719/1/523},
	eprint = {1006.3811},
	journal = {\apj},
	month = aug,
	pages = {523-539},
	primaryclass = {astro-ph.GA},
	title = {{Numerical Study of Turbulent Mixing Layers with Non-equilibrium Ionization Calculations}},
	volume = 719,
	year = 2010}

@article{tan21-lines,
	adsurl = {https://ui.adsabs.harvard.edu/abs/2021MNRAS.508L..37T},
	archiveprefix = {arXiv},
	author = {{Tan}, Brent and {Oh}, S. Peng},
	doi = {10.1093/mnrasl/slab100},
	eprint = {2105.11496},
	journal = {\mnras},
	month = nov,
	number = {1},
	pages = {L37-L42},
	primaryclass = {astro-ph.GA},
	title = {{A model for line absorption and emission from turbulent mixing layers}},
	volume = {508},
	year = 2021}

@ARTICLE{begelman1990,
       author = {{Begelman}, Mitchell C. and {Fabian}, A.~C.},
        title = "{Turbulent mixing layers in the interstellar and intracluster medium.}",
      journal = {\mnras},
         year = 1990,
        month = may,
       volume = {244},
        pages = {26P-29},
       adsurl = {https://ui.adsabs.harvard.edu/abs/1990MNRAS.244P..26B}
}

@article{tumlinson17,
	adsurl = {http://adsabs.harvard.edu/abs/2017ARA%26A..55..389T},
	archiveprefix = {arXiv},
	author = {{Tumlinson}, J. and {Peeples}, M.~S. and {Werk}, J.~K.},
	doi = {10.1146/annurev-astro-091916-055240},
	eprint = {1709.09180},
	journal = {\araa},
	month = aug,
	pages = {389-432},
	title = {{The Circumgalactic Medium}},
	volume = 55,
	year = 2017}

@ARTICLE{Tumlinson:2013,
       author = {{Tumlinson}, Jason and {Thom}, Christopher and {Werk}, Jessica K. and {Prochaska}, J. Xavier and {Tripp}, Todd M. and {Katz}, Neal and {Dav{\'e}}, Romeel and {Oppenheimer}, Benjamin D. and {Meiring}, Joseph D. and {Ford}, Amanda Brady and {O'Meara}, John M. and {Peeples}, Molly S. and {Sembach}, Kenneth R. and {Weinberg}, David H.},
        title = "{The COS-Halos Survey: Rationale, Design, and a Census of Circumgalactic Neutral Hydrogen}",
      journal = {\apj},
         year = 2013,
        month = nov,
       volume = {777},
       number = {1},
          eid = {59},
        pages = {59},
          doi = {10.1088/0004-637X/777/1/59},
archivePrefix = {arXiv},
       eprint = {1309.6317},
 primaryClass = {astro-ph.CO},
       adsurl = {https://ui.adsabs.harvard.edu/abs/2013ApJ...777...59T}
}

@article{FGOH23,
	adsurl = {https://ui.adsabs.harvard.edu/abs/2023ARA&A..61..131F},
	archiveprefix = {arXiv},
	author = {{Faucher-Gigu{\`e}re}, Claude-Andr{\'e} and {Oh}, S. Peng},
	doi = {10.1146/annurev-astro-052920-125203},
	eprint = {2301.10253},
	journal = {\araa},
	month = aug,
	pages = {131-195},
	primaryclass = {astro-ph.GA},
	title = {{Key Physical Processes in the Circumgalactic Medium}},
	volume = {61},
	year = 2023}

@article{Martin:1999,
doi = {10.1086/306863},
url = {https://dx.doi.org/10.1086/306863},
year = {1999},
month = {mar},
publisher = {},
volume = {513},
number = {1},
pages = {156},
author = {Crystal L. Martin},
title = {Properties of Galactic Outflows: Measurements of the Feedback from Star Formation},
journal = {\apj}
}

@article{Rubin_2014,
doi = {10.1088/0004-637X/794/2/156},
url = {https://dx.doi.org/10.1088/0004-637X/794/2/156},
year = {2014},
month = {oct},
publisher = {The American Astronomical Society},
volume = {794},
number = {2},
pages = {156},
author = {Rubin, Kate H. R. and Prochaska, J. Xavier and Koo, David C. and Phillips, Andrew C. and Martin, Crystal L. and Winstrom, Lucas O.},
title = {EVIDENCE FOR UBIQUITOUS COLLIMATED GALACTIC-SCALE OUTFLOWS ALONG THE STAR-FORMING SEQUENCE AT z ∼ 0.5},
journal = {\apj}
}

@article{Pettini:2001,
doi = {10.1086/321403},
url = {https://dx.doi.org/10.1086/321403},
year = {2001},
month = {jun},
publisher = {},
volume = {554},
number = {2},
pages = {981},
author = {Max Pettini and Alice E. Shapley and Charles C. Steidel and Jean-Gabriel Cuby and Mark Dickinson and Alan F. M. Moorwood and Kurt L. Adelberger and Mauro Giavalisco},
title = {The Rest-Frame Optical Spectra of Lyman Break Galaxies: Star Formation, Extinction, Abundances, and Kinematics*},
journal = {\apj}
}

@article{Shapley:2003,
doi = {10.1086/373922},
url = {https://dx.doi.org/10.1086/373922},
year = {2003},
month = {may},
publisher = {},
volume = {588},
number = {1},
pages = {65},
author = {Alice E. Shapley and Charles C. Steidel and Max Pettini and Kurt L. Adelberger},
title = {Rest-Frame Ultraviolet Spectra of z ∼ 3 Lyman Break Galaxies*},
journal = {\apj}
}

@ARTICLE{Heckman_2015,
       author = {{Heckman}, Timothy M. and {Alexandroff}, Rachel M. and {Borthakur}, Sanchayeeta and {Overzier}, Roderik and {Leitherer}, Claus},
        title = "{The Systematic Properties of the Warm Phase of Starburst-Driven Galactic Winds}",
      journal = {\apj},
         year = 2015,
        month = aug,
       volume = {809},
       number = {2},
          eid = {147},
        pages = {147},
          doi = {10.1088/0004-637X/809/2/147},
archivePrefix = {arXiv},
       eprint = {1507.05622},
 primaryClass = {astro-ph.GA},
       adsurl = {https://ui.adsabs.harvard.edu/abs/2015ApJ...809..147H}
}

@ARTICLE{Martin_2015,
       author = {{Martin}, Crystal L. and {Dijkstra}, Mark and {Henry}, Alaina and {Soto}, Kurt T. and {Danforth}, Charles W. and {Wong}, Joseph},
        title = "{The Ly{\ensuremath{\alpha}} Line Profiles of Ultraluminous Infrared Galaxies: Fast Winds and Lyman Continuum Leakage}",
      journal = {\apj},
         year = 2015,
        month = apr,
       volume = {803},
       number = {1},
          eid = {6},
        pages = {6},
          doi = {10.1088/0004-637X/803/1/6},
archivePrefix = {arXiv},
       eprint = {1501.05946},
 primaryClass = {astro-ph.GA},
       adsurl = {https://ui.adsabs.harvard.edu/abs/2015ApJ...803....6M}
}

@ARTICLE{Xu_2022,
       author = {{Xu}, Xinfeng and {Heckman}, Timothy and {Henry}, Alaina and {Berg}, Danielle A. and {Chisholm}, John and {James}, Bethan L. and {Martin}, Crystal L. and {Stark}, Daniel P. and {Aloisi}, Alessandra and {Amor{\'\i}n}, Ricardo O. and {Arellano-C{\'o}rdova}, Karla Z. and {Bordoloi}, Rongmon and {Charlot}, St{\'e}phane and {Chen}, Zuyi and {Hayes}, Matthew and {Mingozzi}, Matilde and {Sugahara}, Yuma and {Kewley}, Lisa J. and {Ouchi}, Masami and {Scarlata}, Claudia and {Steidel}, Charles C.},
        title = "{CLASSY III. The Properties of Starburst-driven Warm Ionized Outflows}",
      journal = {\apj},
         year = 2022,
        month = jul,
       volume = {933},
       number = {2},
          eid = {222},
        pages = {222},
          doi = {10.3847/1538-4357/ac6d56},
archivePrefix = {arXiv},
       eprint = {2204.09181},
 primaryClass = {astro-ph.GA},
       adsurl = {https://ui.adsabs.harvard.edu/abs/2022ApJ...933..222X}
}

@ARTICLE{Lopez_2020,
       author = {{Lopez}, Laura A. and {Mathur}, Smita and {Nguyen}, Dustin D. and {Thompson}, Todd A. and {Olivier}, Grace M.},
        title = "{Temperature and Metallicity Gradients in the Hot Gas Outflows of M82}",
      journal = {\apj},
         year = 2020,
        month = dec,
       volume = {904},
       number = {2},
          eid = {152},
        pages = {152},
          doi = {10.3847/1538-4357/abc010},
archivePrefix = {arXiv},
       eprint = {2006.08623},
 primaryClass = {astro-ph.HE},
       adsurl = {https://ui.adsabs.harvard.edu/abs/2020ApJ...904..152L}
}

@ARTICLE{Rupke2019,
       author = {{Rupke}, David S.~N. and {Coil}, Alison and {Geach}, James E. and {Tremonti}, Christy and {Diamond-Stanic}, Aleksandar M. and {George}, Erin R. and {Hickox}, Ryan C. and {Kepley}, Amanda A. and {Leung}, Gene and {Moustakas}, John and {Rudnick}, Gregory and {Sell}, Paul H.},
        title = "{A 100-kiloparsec wind feeding the circumgalactic medium of a massive compact galaxy}",
      journal = {\nat},
         year = 2019,
        month = oct,
       volume = {574},
       number = {7780},
        pages = {643-646},
          doi = {10.1038/s41586-019-1686-1},
archivePrefix = {arXiv},
       eprint = {1910.13507},
 primaryClass = {astro-ph.GA},
       adsurl = {https://ui.adsabs.harvard.edu/abs/2019Natur.574..643R}
}

@ARTICLE{Burchett2021,
       author = {{Burchett}, Joseph N. and {Rubin}, Kate H.~R. and {Prochaska}, J. Xavier and {Coil}, Alison L. and {Vaught}, Ryan Rickards and {Hennawi}, Joseph F.},
        title = "{Circumgalactic Mg II Emission from an Isotropic Starburst Galaxy Outflow Mapped by KCWI}",
      journal = {\apj},
         year = 2021,
        month = mar,
       volume = {909},
       number = {2},
          eid = {151},
        pages = {151},
          doi = {10.3847/1538-4357/abd4e0},
archivePrefix = {arXiv},
       eprint = {2005.03017},
 primaryClass = {astro-ph.GA},
       adsurl = {https://ui.adsabs.harvard.edu/abs/2021ApJ...909..151B}
}

@ARTICLE{Reichardt_Chu2022,
       author = {{Reichardt Chu}, Bronwyn and {Fisher}, Deanne B. and {Nielsen}, Nikole M. and {Chisholm}, John and {Girard}, Marianne and {Kacprzak}, Glenn G. and {Bolatto}, Alberto and {Herrera-Camus}, Rodrigo and {Sandstrom}, Karin and {Li}, Miao and {Rickards Vaught}, Ryan and {McPherson}, Daniel K.},
        title = "{The DUVET Survey: Resolved maps of star formation-driven outflows in a compact, starbursting disc galaxy}",
      journal = {\mnras},
         year = 2022,
        month = apr,
       volume = {511},
       number = {4},
        pages = {5782-5796},
          doi = {10.1093/mnras/stac420},
archivePrefix = {arXiv},
       eprint = {2202.04672},
 primaryClass = {astro-ph.GA},
       adsurl = {https://ui.adsabs.harvard.edu/abs/2022MNRAS.511.5782R}
}

@ARTICLE{Rupke_2023,
       author = {{Rupke}, David S.~N. and {Coil}, Alison L. and {Perrotta}, Serena and {Davis}, Julie D. and {Diamond-Stanic}, Aleksandar M. and {Geach}, James E. and {Hickox}, Ryan C. and {Moustakas}, John and {Petter}, Grayson C. and {Rudnick}, Gregory H. and {Sell}, Paul H. and {Tremonti}, Christy A. and {Whalen}, Kelly E.},
        title = "{The Ionization and Dynamics of the Makani Galactic Wind}",
      journal = {\apj},
         year = 2023,
        month = apr,
       volume = {947},
       number = {1},
          eid = {33},
        pages = {33},
          doi = {10.3847/1538-4357/acbfae},
archivePrefix = {arXiv},
       eprint = {2303.00194},
 primaryClass = {astro-ph.GA},
       adsurl = {https://ui.adsabs.harvard.edu/abs/2023ApJ...947...33R}
}

@ARTICLE{Herenz_2025,
       author = {{Herenz}, Edmund Christian and {Kusakabe}, Haruka and {Maulick}, Soumil},
        title = "{The extreme starburst J1044+0353 blows kiloparsec-scale bubbles}",
      journal = {\pasj},
         year = 2025,
        month = jul,
          doi = {10.1093/pasj/psaf073},
archivePrefix = {arXiv},
       eprint = {2502.16969},
 primaryClass = {astro-ph.GA},
       adsurl = {https://ui.adsabs.harvard.edu/abs/2025PASJ..tmp...80H}
}

@ARTICLE{Perrotta_2023,
       author = {{Perrotta}, Serena and {Coil}, Alison L. and {Rupke}, David S.~N. and {Tremonti}, Christy A. and {Davis}, Julie D. and {Diamond-Stanic}, Aleksandar M. and {Geach}, James E. and {Hickox}, Ryan C. and {Moustakas}, John and {Rudnick}, Gregory H. and {Sell}, Paul H. and {Swiggum}, Cameren N. and {Whalen}, Kelly E.},
        title = "{Kinematics, Structure, and Mass Outflow Rates of Extreme Starburst Galactic Outflows}",
      journal = {\apj},
         year = 2023,
        month = may,
       volume = {949},
       number = {1},
          eid = {9},
        pages = {9},
          doi = {10.3847/1538-4357/acc660},
archivePrefix = {arXiv},
       eprint = {2303.07448},
 primaryClass = {astro-ph.GA},
       adsurl = {https://ui.adsabs.harvard.edu/abs/2023ApJ...949....9P}
}

@ARTICLE{Nielsen_2024,
       author = {{Nielsen}, Nikole M. and {Fisher}, Deanne B. and {Kacprzak}, Glenn G. and {Chisholm}, John and {Martin}, D. Christopher and {Reichardt Chu}, Bronwyn and {Sandstrom}, Karin M. and {Rickards Vaught}, Ryan J.},
        title = "{An emission map of the disk-circumgalactic medium transition in starburst IRAS 08339+6517}",
      journal = {Nature Astronomy},
         year = 2024,
        month = dec,
       volume = {8},
        pages = {1602-1609},
          doi = {10.1038/s41550-024-02365-x},
archivePrefix = {arXiv},
       eprint = {2311.00856},
 primaryClass = {astro-ph.GA},
       adsurl = {https://ui.adsabs.harvard.edu/abs/2024NatAs...8.1602N}
}

@ARTICLE{Shaban_2022,
       author = {{Shaban}, Ahmed and {Bordoloi}, Rongmon and {Chisholm}, John and {Sharma}, Soniya and {Sharon}, Keren and {Rigby}, Jane R. and {Gladders}, Michael G. and {Bayliss}, Matthew B. and {Barrientos}, L. Felipe and {Lopez}, Sebastian and {Tejos}, Nicolas and {Ledoux}, C{\'e}dric and {Florian}, Michael K.},
        title = "{A 30 kpc Spatially Extended Clumpy and Asymmetric Galactic Outflow at z   1.7}",
      journal = {\apj},
         year = 2022,
        month = sep,
       volume = {936},
       number = {1},
          eid = {77},
        pages = {77},
          doi = {10.3847/1538-4357/ac7c65},
archivePrefix = {arXiv},
       eprint = {2109.13264},
 primaryClass = {astro-ph.GA},
       adsurl = {https://ui.adsabs.harvard.edu/abs/2022ApJ...936...77S}
}

@ARTICLE{Xu_2023b,
       author = {{Xu}, Xinfeng and {Heckman}, Timothy and {Yoshida}, Michitoshi and {Henry}, Alaina and {Ohyama}, Youichi},
        title = "{What Are the Radial Distributions of Density, Outflow Rates, and Cloud Structures in the M82 Wind?}",
      journal = {\apj},
         year = 2023,
        month = oct,
       volume = {956},
       number = {2},
          eid = {142},
        pages = {142},
          doi = {10.3847/1538-4357/acfa71},
archivePrefix = {arXiv},
       eprint = {2310.00094},
 primaryClass = {astro-ph.GA},
       adsurl = {https://ui.adsabs.harvard.edu/abs/2023ApJ...956..142X}
}

@ARTICLE{Leitherer_1999,
       author = {{Leitherer}, Claus and {Schaerer}, Daniel and {Goldader}, Jeffrey D. and {Delgado}, Rosa M. Gonz{\'a}lez and {Robert}, Carmelle and {Kune}, Denis Foo and {de Mello}, Du{\'\i}lia F. and {Devost}, Daniel and {Heckman}, Timothy M.},
        title = "{Starburst99: Synthesis Models for Galaxies with Active Star Formation}",
      journal = {\apjs},
         year = 1999,
        month = jul,
       volume = {123},
       number = {1},
        pages = {3-40},
          doi = {10.1086/313233},
archivePrefix = {arXiv},
       eprint = {astro-ph/9902334},
 primaryClass = {astro-ph},
       adsurl = {https://ui.adsabs.harvard.edu/abs/1999ApJS..123....3L}
}

@ARTICLE{SH_2009,
       author = {{Strickland}, David K. and {Heckman}, Timothy M.},
        title = "{Supernova Feedback Efficiency and Mass Loading in the Starburst and Galactic Superwind Exemplar M82}",
      journal = {\apj},
         year = 2009,
        month = jun,
       volume = {697},
       number = {2},
        pages = {2030-2056},
          doi = {10.1088/0004-637X/697/2/2030},
archivePrefix = {arXiv},
       eprint = {0903.4175},
 primaryClass = {astro-ph.CO},
       adsurl = {https://ui.adsabs.harvard.edu/abs/2009ApJ...697.2030S}
}

@ARTICLE{Veilleux_2025,
       author = {{Veilleux}, Sylvain and {Shockley}, Steven D. and {Melendez}, Marcio and {Rupke}, David S.~N. and {Coil}, Alison L. and {Diamond-Stanic}, Aleksandar M. and {Geach}, James E. and {Hickox}, Ryan C. and {Moustakas}, John and {Rudnick}, Gregory H. and {Sell}, Paul H. and {Tremonti}, Christy A. and {Cha}, Hojoon},
        title = "{JWST Discovery of Warm Dust in the Circumgalactic Medium of the Makani Galaxy}",
      journal = {arXiv e-prints},
         year = 2025,
        month = jul,
          eid = {arXiv:2507.08098},
        pages = {arXiv:2507.08098},
          doi = {10.48550/arXiv.2507.08098},
archivePrefix = {arXiv},
       eprint = {2507.08098},
 primaryClass = {astro-ph.GA},
       adsurl = {https://ui.adsabs.harvard.edu/abs/2025arXiv250708098V}
}

@article{Lopez_2025,
doi = {10.3847/1538-4357/adec75},
url = {https://dx.doi.org/10.3847/1538-4357/adec75},
year = {2025},
month = {aug},
publisher = {The American Astronomical Society},
volume = {989},
number = {1},
pages = {100},
author = {Lopez, Sebastian and Lopez, Laura A. and Thompson, Todd A. and Leroy, Adam K. and Bolatto, Alberto D.},
title = {Observational Constraints on Cool Gas Clouds in M82’s Starburst-driven Outflow},
journal = {\apj}
}

@ARTICLE{Heckman_1990,
       author = {{Heckman}, Timothy M. and {Armus}, Lee and {Miley}, George K.},
        title = "{On the Nature and Implications of Starburst-driven Galactic Superwinds}",
      journal = {\apjs},
         year = 1990,
        month = dec,
       volume = {74},
        pages = {833},
          doi = {10.1086/191522},
       adsurl = {https://ui.adsabs.harvard.edu/abs/1990ApJS...74..833H}
}

@ARTICLE{Martin_2006,
       author = {{Martin}, Crystal L.},
        title = "{Mapping Large-Scale Gaseous Outflows in Ultraluminous Infrared Galaxies with Keck II ESI Spectra: Spatial Extent of the Outflow}",
      journal = {\apj},
         year = 2006,
        month = aug,
       volume = {647},
       number = {1},
        pages = {222-243},
          doi = {10.1086/504886},
archivePrefix = {arXiv},
       eprint = {astro-ph/0604173},
 primaryClass = {astro-ph},
       adsurl = {https://ui.adsabs.harvard.edu/abs/2006ApJ...647..222M}
}

@ARTICLE{Cox_2004,
       author = {{Cox}, T.~J. and {Primack}, Joel and {Jonsson}, Patrik and {Somerville}, Rachel S.},
        title = "{Generating Hot Gas in Simulations of Disk-Galaxy Major Mergers}",
      journal = {\apjl},
         year = 2004,
        month = jun,
       volume = {607},
       number = {2},
        pages = {L87-L90},
          doi = {10.1086/421905},
archivePrefix = {arXiv},
       eprint = {astro-ph/0402675},
 primaryClass = {astro-ph},
       adsurl = {https://ui.adsabs.harvard.edu/abs/2004ApJ...607L..87C}
}

@ARTICLE{Girelli_2020,
       author = {{Girelli}, G. and {Pozzetti}, L. and {Bolzonella}, M. and {Giocoli}, C. and {Marulli}, F. and {Baldi}, M.},
        title = "{The stellar-to-halo mass relation over the past 12 Gyr. I. Standard {\ensuremath{\Lambda}}CDM model}",
      journal = {\aap},
         year = 2020,
        month = feb,
       volume = {634},
          eid = {A135},
        pages = {A135},
          doi = {10.1051/0004-6361/201936329},
archivePrefix = {arXiv},
       eprint = {2001.02230},
 primaryClass = {astro-ph.CO},
       adsurl = {https://ui.adsabs.harvard.edu/abs/2020A&A...634A.135G}
}

@article{Kennicutt_1998,
	doi = {10.1086/305588},
	url = {https://doi.org/10.1086/305588},
	year = 1998,
	month = {may},
	publisher = {American Astronomical Society},
	volume = {498},
	number = {2},
	pages = {541--552},
	author = {Kennicutt, Jr.},
	title = {The Global Schmidt Law in Star-forming Galaxies},
	journal = {\apj}
}

@ARTICLE{OSullivan_2009,
       author = {{O'Sullivan}, E. and {Giacintucci}, S. and {Vrtilek}, J.~M. and {Raychaudhury}, S. and {David}, L.~P.},
        title = "{A Chandra X-ray View of Stephan's Quintet: Shocks and Star Formation}",
      journal = {\apj},
         year = 2009,
        month = aug,
       volume = {701},
       number = {2},
        pages = {1560-1568},
          doi = {10.1088/0004-637X/701/2/1560},
archivePrefix = {arXiv},
       eprint = {0812.0383},
 primaryClass = {astro-ph},
       adsurl = {https://ui.adsabs.harvard.edu/abs/2009ApJ...701.1560O}
}

@ARTICLE{Baron_2024,
       author = {{Baron}, Dalya and {Netzer}, Hagai and {Lutz}, Dieter and {Davies}, Ric I. and {Prochaska}, J. Xavier},
        title = "{Not So Windy After All: MUSE Disentangles AGN-driven Winds from Merger-induced Flows in Galaxies along the Starburst Sequence}",
      journal = {\apj},
         year = 2024,
        month = jun,
       volume = {968},
       number = {1},
          eid = {23},
        pages = {23},
          doi = {10.3847/1538-4357/ad39e9},
archivePrefix = {arXiv},
       eprint = {2401.09576},
 primaryClass = {astro-ph.GA},
       adsurl = {https://ui.adsabs.harvard.edu/abs/2024ApJ...968...23B}
}

@article{Otte03,
doi = {10.1086/374731},
url = {https://dx.doi.org/10.1086/374731},
year = {2003},
month = {feb},
publisher = {},
volume = {586},
number = {1},
pages = {L53},
author = {Otte, Birgit and Van Dyke Dixon, W. and Sankrit, Ravi},
title = {The Far Ultraviolet Spectroscopic Explorer Detection of Galactic O VI Emission in the Halo above the Perseus Arm},
journal = {\apj}
}

@ARTICLE{Chung21,
       author = {{Chung}, Haeun and {Vargas}, Carlos J. and {Hamden}, Erika},
        title = "{Revisiting FUSE O VI Emission in Galaxy Halos}",
      journal = {\apj},
         year = 2021,
        month = jul,
       volume = {916},
       number = {1},
          eid = {7},
        pages = {7},
          doi = {10.3847/1538-4357/ac04af},
archivePrefix = {arXiv},
       eprint = {2103.05008},
 primaryClass = {astro-ph.GA},
       adsurl = {https://ui.adsabs.harvard.edu/abs/2021ApJ...916....7C}
}

@ARTICLE{Grimes07,
       author = {{Grimes}, J.~P. and {Heckman}, T. and {Strickland}, D. and {Dixon}, W.~V. and {Sembach}, K. and {Overzier}, R. and {Hoopes}, C. and {Aloisi}, A. and {Ptak}, A.},
        title = "{Feedback in the Local Lyman-break Galaxy Analog Haro 11 as Probed by Far-Ultraviolet and X-Ray Observations}",
      journal = {\apj},
         year = 2007,
        month = oct,
       volume = {668},
       number = {2},
        pages = {891-905},
          doi = {10.1086/521353},
archivePrefix = {arXiv},
       eprint = {0707.0693},
 primaryClass = {astro-ph},
       adsurl = {https://ui.adsabs.harvard.edu/abs/2007ApJ...668..891G}
}

@INPROCEEDINGS{Chung24,
       author = {{Chung}, Haeun and {Vargas}, Carlos J. and {Hamden}, Erika and {McMahon}, Tom and {Tanquary}, Hannah and {Khan}, Aafaque R. and {Coronado}, Fernando and {Verts}, Bill and {Melso}, Nicole and {Choi}, Heejoo and {Agarwal}, Simran and {Corliss}, Jason and {Hoadley}, Keri and {Keppler}, Miriam and {Hamara}, Dave and {Garcia}, Elijah and {Truong}, Daniel and {Yescas}, Naomi and {Augustin}, Ramona and {Batkis}, Mateo and {Behroozi}, Peter and {Bradley}, Harrison and {Brendel}, Trenton and {Burchett}, Joseph N. and {Martinez Castillo}, Jasmine and {Chambers}, Jacob and {Corlies}, Lauren and {Davis}, Greyson and {Dettmar}, Ralf-J{\"u}rgen and {Douglas}, Ewan and {Ghidoli}, Giulia and {Goodwin}, Alfred and {Harris}, Walter and {Hennessy}, John J. and {Hergenrother}, Carl and {Howk}, J. Christopher and {Kerkeser}, Nazende Ipek and {Kidd}, John N. and {Kim}, Daewook and {Li}, Jessica S. and {Martin}, Adrian and {Noenickx}, Jamison and {Noriega}, Gabe and {Park}, Sooseong and {Pecha}, Ryan and {Quijada}, Manuel and {Rodriguez de Marcos}, L. and {Sauve}, Cork and {Schiminovich}, David and {Selznick}, Sanford and {Siegmund}, Oswald and {Su}, Rebecca and {Uppnor}, Sumedha and {Wolcott}, Ellie and {Zaritsky}, Dennis},
        title = "{Aspera payload design overview: UV SmallSat mission to detect and map warm-hot halo gas around the nearby galaxies}",
    booktitle = {Space Telescopes and Instrumentation 2024: Ultraviolet to Gamma Ray},
         year = 2024,
       editor = {{den Herder}, Jan-Willem A. and {Nikzad}, Shouleh and {Nakazawa}, Kazuhiro},
       series = {Society of Photo-Optical Instrumentation Engineers (SPIE) Conference Series},
       volume = {13093},
        month = aug,
          eid = {1309302},
        pages = {1309302},
          doi = {10.1117/12.3017274},
       adsurl = {https://ui.adsabs.harvard.edu/abs/2024SPIE13093E..02C}
}

@ARTICLE{Burchett25,
       author = {{Burchett}, Joseph N. and {Lokhorst}, Deborah M. and {Faerman}, Yakov and {France}, Kevin and {Rubin}, Kate H.~R. and {Rupke}, David S.~N. and {Borthakur}, Sanchayeeta},
        title = "{Mapping Galactic Winds and Small-scale Structure in the Circumgalactic Medium with Habitable Worlds Observatory}",
      journal = {arXiv e-prints},
         year = 2025,
        month = jul,
          eid = {arXiv:2507.03750},
        pages = {arXiv:2507.03750},
          doi = {10.48550/arXiv.2507.03750},
archivePrefix = {arXiv},
       eprint = {2507.03750},
 primaryClass = {astro-ph.GA},
       adsurl = {https://ui.adsabs.harvard.edu/abs/2025arXiv250703750B}
}

@ARTICLE{Kim_2024,
       author = {{Kim}, Jin-Ah and {Chung}, Haeun and {Vargas}, Carlos J. and {Hamden}, Erika},
        title = "{UV Cooling via O VI Emission in the Superwind of M82 Observed with the Far Ultraviolet Spectroscopic Explorer (FUSE)}",
      journal = {\aj},
         year = 2024,
        month = jul,
       volume = {168},
       number = {1},
          eid = {11},
        pages = {11},
          doi = {10.3847/1538-3881/ad4887},
archivePrefix = {arXiv},
       eprint = {2406.01742},
 primaryClass = {astro-ph.GA},
       adsurl = {https://ui.adsabs.harvard.edu/abs/2024AJ....168...11K}
}

@ARTICLE{Kim_2017,
       author = {{Kim}, Chang-Goo and {Ostriker}, Eve C.},
        title = "{Three-phase Interstellar Medium in Galaxies Resolving Evolution with Star Formation and Supernova Feedback (TIGRESS): Algorithms, Fiducial Model, and Convergence}",
      journal = {\apj},
         year = 2017,
        month = sep,
       volume = {846},
       number = {2},
          eid = {133},
        pages = {133},
          doi = {10.3847/1538-4357/aa8599},
archivePrefix = {arXiv},
       eprint = {1612.03918},
 primaryClass = {astro-ph.GA},
       adsurl = {https://ui.adsabs.harvard.edu/abs/2017ApJ...846..133K}
}

@article{Kim_2021,
doi = {10.3847/1538-4357/abe934},
url = {https://doi.org/10.3847/1538-4357/abe934},
year = {2021},
month = {apr},
publisher = {The American Astronomical Society},
volume = {911},
number = {2},
pages = {128},
author = {Kim, Jeong-Gyu and Ostriker, Eve C. and Filippova, Nina},
title = {Star Formation Efficiency and Dispersal of Giant Molecular Clouds with UV Radiation Feedback: Dependence on Gravitational Boundedness and Magnetic Fields},
journal = {The Astrophysical Journal}
}

@article{Joung_2006,
doi = {10.1086/508795},
url = {https://doi.org/10.1086/508795},
year = {2006},
month = {dec},
publisher = {},
volume = {653},
number = {2},
pages = {1266},
author = {Joung, M. K. Ryan and Mac Low, Mordecai-Mark},
title = {Turbulent Structure of a Stratified Supernova-driven Interstellar Medium},
journal = {The Astrophysical Journal}
}

@article{Walch_2015,
    author = {Walch, S. and Girichidis, P. and Naab, T. and Gatto, A. and Glover, S. C. O. and Wünsch, R. and Klessen, R. S. and Clark, P. C. and Peters, T. and Derigs, D. and Baczynski, C.},
    title = {The SILCC (SImulating the LifeCycle of molecular Clouds) project – I. Chemical evolution of the supernova-driven ISM},
    journal = {Monthly Notices of the Royal Astronomical Society},
    volume = {454},
    number = {1},
    pages = {238-268},
    year = {2015},
    month = {09},
    issn = {0035-8711},
    doi = {10.1093/mnras/stv1975},
    url = {https://doi.org/10.1093/mnras/stv1975},
    eprint = {https://academic.oup.com/mnras/article-pdf/454/1/238/3924820/stv1975.pdf},
}

@article{Marinacci_2019,
    author = {Marinacci, Federico and Sales, Laura V and Vogelsberger, Mark and Torrey, Paul and Springel, Volker},
    title = {Simulating the interstellar medium and stellar feedback on a moving mesh: implementation and isolated galaxies},
    journal = {Monthly Notices of the Royal Astronomical Society},
    volume = {489},
    number = {3},
    pages = {4233-4260},
    year = {2019},
    month = {09},
    issn = {0035-8711},
    doi = {10.1093/mnras/stz2391},
    url = {https://doi.org/10.1093/mnras/stz2391},
    eprint = {https://academic.oup.com/mnras/article-pdf/489/3/4233/30037059/stz2391.pdf},
}

@article{Nguyen_21,
    author = {Nguyen, Dustin D and Thompson, Todd A},
    title = {Mass-loading and non-spherical divergence in hot galactic winds: implications for X-ray observations},
    journal = {Monthly Notices of the Royal Astronomical Society},
    volume = {508},
    number = {4},
    pages = {5310-5325},
    year = {2021},
    month = {10},
    issn = {0035-8711},
    doi = {10.1093/mnras/stab2910},
    url = {https://doi.org/10.1093/mnras/stab2910},
    eprint = {https://academic.oup.com/mnras/article-pdf/508/4/5310/40960582/stab2910.pdf},
}

@ARTICLE{Carr_2021,
       author = {{Carr}, Cody and {Scarlata}, Claudia and {Henry}, Alaina and {Panagia}, Nino},
        title = "{The Effects of Biconical Outflows on Ly{\ensuremath{\alpha}} Escape from Green Peas}",
      journal = {\apj},
         year = 2021,
        month = jan,
       volume = {906},
       number = {2},
          eid = {104},
        pages = {104},
          doi = {10.3847/1538-4357/abc7c3},
archivePrefix = {arXiv},
       eprint = {2011.02549},
 primaryClass = {astro-ph.GA},
       adsurl = {https://ui.adsabs.harvard.edu/abs/2021ApJ...906..104C}
}

@ARTICLE{Peng_2023,
       author = {{Peng}, Zixuan and {Martin}, Crystal L. and {Thibodeaux}, Pierre and {Zhang}, Jichen and {Hu}, Weida and {Li}, Yuan},
        title = "{Using KCWI to Explore the Chemical Inhomogeneities and Evolution of J1044+0353}",
      journal = {\apj},
         year = 2023,
        month = sep,
       volume = {954},
       number = {2},
          eid = {214},
        pages = {214},
          doi = {10.3847/1538-4357/ace9c0},
archivePrefix = {arXiv},
       eprint = {2308.00351},
 primaryClass = {astro-ph.GA},
       adsurl = {https://ui.adsabs.harvard.edu/abs/2023ApJ...954..214P}
}

@ARTICLE{McPherson_2023,
       author = {{McPherson}, Daniel K. and {Fisher}, Deanne B. and {Nielsen}, Nikole M. and {Kacprzak}, Glenn G. and {Reichardt Chu}, Bronwyn and {Cameron}, Alex J. and {Bolatto}, Alberto D. and {Chisholm}, John and {Fielding}, Drummond B. and {Berg}, Danielle and {Herrera-Camus}, Rodrigo and {Li}, Miao and {Vaught}, Ryan J. Rickards and {Sandstrom}, Karin},
        title = "{DUVET survey: mapping outflows in the metal-poor starburst Mrk 1486}",
      journal = {\mnras},
         year = 2023,
        month = nov,
       volume = {525},
       number = {4},
        pages = {6170-6181},
          doi = {10.1093/mnras/stad2685},
archivePrefix = {arXiv},
       eprint = {2308.06918},
 primaryClass = {astro-ph.GA},
       adsurl = {https://ui.adsabs.harvard.edu/abs/2023MNRAS.525.6170M}
}

@ARTICLE{Howatson_2025,
       author = {{Howatson}, Elliot L. and {Richings}, Alexander J. and {Roediger}, Elke and {Faucher-Gigu{\`e}re}, Claude-Andr{\'e} and {Theuns}, Tom and {Liu}, Yuankang and {Chan}, Tsang Keung and {Thompson}, Oliver and {Carr}, Cody and {Angl{\'e}s-Alc{\'a}zar}, Daniel},
        title = "{Emission line tracers of galactic outflows driven by stellar feedback in simulations of isolated disk galaxies}",
      journal = {\mnras},
         year = 2025,
        month = sep,
          doi = {10.1093/mnras/staf1641},
archivePrefix = {arXiv},
       eprint = {2509.21295},
 primaryClass = {astro-ph.GA},
       adsurl = {https://ui.adsabs.harvard.edu/abs/2025MNRAS.tmp.1570H}
}

@ARTICLE{Afruni_2026,
       author = {{Afruni}, Andrea and {Di Teodoro}, Enrico M. and {Armillotta}, Lucia and {Lynn}, Callum A. and {McClure-Griffiths}, Naomi M.},
        title = "{Modeling the Milky Way wind: Supernova-driven outflows accelerate HI clouds near the Galactic center}",
      journal = {\aap},
         year = 2026,
        month = feb,
       volume = {706},
          eid = {A297},
        pages = {A297},
          doi = {10.1051/0004-6361/202556813},
archivePrefix = {arXiv},
       eprint = {2601.05314},
 primaryClass = {astro-ph.GA},
       adsurl = {https://ui.adsabs.harvard.edu/abs/2026A&A...706A.297A}
}

@ARTICLE{NG_2024,
       author = {{Nikolis}, C. and {Gronke}, M.},
        title = "{Strength in numbers: A multiphase wind model with multiple cloud populations}",
      journal = {\mnras},
         year = 2024,
        month = jun,
       volume = {530},
       number = {4},
        pages = {4597-4613},
          doi = {10.1093/mnras/stae1169},
archivePrefix = {arXiv},
       eprint = {2404.19380},
 primaryClass = {astro-ph.GA},
       adsurl = {https://ui.adsabs.harvard.edu/abs/2024MNRAS.530.4597N}
}

@ARTICLE{Dutta_2025,
       author = {{Dutta}, Alankar and {Sharma}, Prateek and {Gronke}, Max},
        title = "{Fading in the flow: suppression of cold gas growth in expanding galactic outflows}",
      journal = {\mnras},
         year = 2025,
        month = dec,
       volume = {544},
       number = {4},
        pages = {4621-4650},
          doi = {10.1093/mnras/staf1845},
archivePrefix = {arXiv},
       eprint = {2506.08545},
 primaryClass = {astro-ph.GA},
       adsurl = {https://ui.adsabs.harvard.edu/abs/2025MNRAS.544.4621D}
}




\appendix

\section{O VI Emission in TRMLs as a Function of Pressure, Relative Mach Number, and Hot Phase Temperature}\label{appendix:ovi_flux_frac_trml}

\begin{figure*}
\centering
\includegraphics[width=\textwidth]{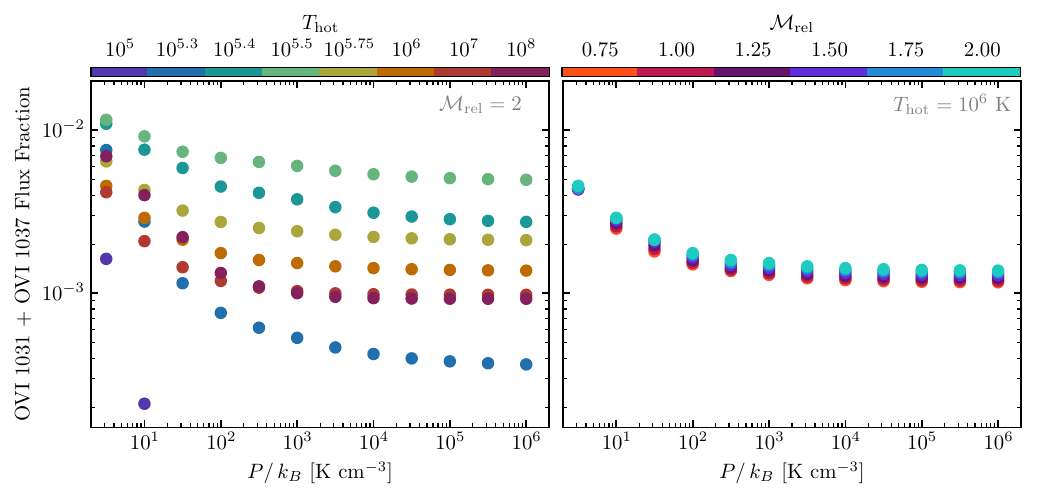}
\caption{O VI flux fraction as a function of $P$, $\mathcal{M}_{\rm rel}$, and $T_{\rm hot}$. The left panel shows that O VI flux fraction is highly sensitive to $T_{\rm hot}$. For $T_{\rm hot}$  below $10^5$ K, O VI flux fraction is practically 0 for most pressure values. As $T_{\rm hot}$ increases, O VI flux fraction increases as well until $T_{\rm hot}=10^{5.5}$ K, which is the temperature where O VI emissivity peaks. At this point, O VI flux fraction reaches a maximum value of $\sim 5\times 10^{-3} - 10^{-2}$. As we go to even higher $T_{\rm hot}$, O VI flux fraction starts to decrease and eventually stabilizes at $\sim 10^{-3}$. The right panel demonstrates that $\mathcal{M}_{\rm rel}$ has a much smaller impact on O VI flux fraction. Both panels show that O VI flux fraction is almost independent of pressure, especially for $\left.P\right/k_{\rm B} \gtrsim 10^3$ ${\rm K}$ ${\rm cm}^{-3}$. }
\label{fig:OVI_flux_fraction_vs_Thot_P_Mrel}
\end{figure*}

To calculate how TRMLs contribute to emission signatures in our multiphase galactic outflow framework, it is crucial to determine the flux fraction of different emission lines
(the fraction of emission flux originating from a specific emission line; see Appendix D.2 of \cite{Peng_2025} for calculation details). 
In general, a TRML can be characterized by three key parameters: $P$, $\mathcal{M}_{\rm rel}$, and $T_{\rm hot}$ (note that we keep the cold phase temperature at $10^4$ K, which is consistent with \citetalias{FB_2022}). All these parameters change non-trivially with radius in a galactic wind, which means it is useful to understand how each of them affects emission line flux fractions. Since this work primarily focuses on O\,{\footnotesize VI} emission, we will use O\,{\footnotesize VI} as an example for our investigation here. We note that the flux fraction of different emission lines can have different dependencies on $P$, $\mathcal{M}_{\rm rel}$, and $T_{\rm hot}$ (see Fig. 20 in \citealp{Peng_2025} for an example of [O\,{\footnotesize III}] $\lambda5007$).

In \autoref{fig:OVI_flux_fraction_vs_Thot_P_Mrel}, we demonstrate how O\,{\footnotesize VI} flux fraction depends on these parameters. O\,{\footnotesize VI} flux fraction appears to be relatively insensitive to $\mathcal{M}_{\rm rel}$ and pressure (especially for $\left.P\right/k_{\rm B} \gtrsim 10^3$ ${\rm K}$ ${\rm cm}^{-3}$). However, the dependency on $T_{\rm hot}$, as shown in the left panel of \autoref{fig:OVI_flux_fraction_vs_Thot_P_Mrel}, is non-trivial. 
For $T_{\rm hot}$  below $10^5$ K, O\,{\footnotesize VI} flux fraction is practically 0 for most pressure values. As $T_{\rm hot}$ increases, O\,{\footnotesize VI} flux fraction increases as well until $T_{\rm hot}=10^{5.5}$ K, which is the temperature where O\,{\footnotesize VI} emissivity peaks. At this point, O\,{\footnotesize VI} flux fraction reaches a maximum value of $\sim 5\times 10^{-3}-10^{-2}$. As we go to even higher $T_{\rm hot}$, O\,{\footnotesize VI} flux fraction starts to decrease and eventually stabilizes at $\sim 10^{-3}$. Since $T_{\rm hot}$ decreases with radius in a multiphase galactic outflow as the hot phase expands and cools, capturing how it affects the O\,{\footnotesize VI} flux fraction in TRMLs is crucial for our O\,{\footnotesize VI} SB profile calculation.

\section{O VI SB Detectability with Different Galactic wind Parameters and Sensitivity Limit}\label{appendix:ovi_sb_detect}

\begin{figure*}
\centering
\includegraphics[width=\textwidth]{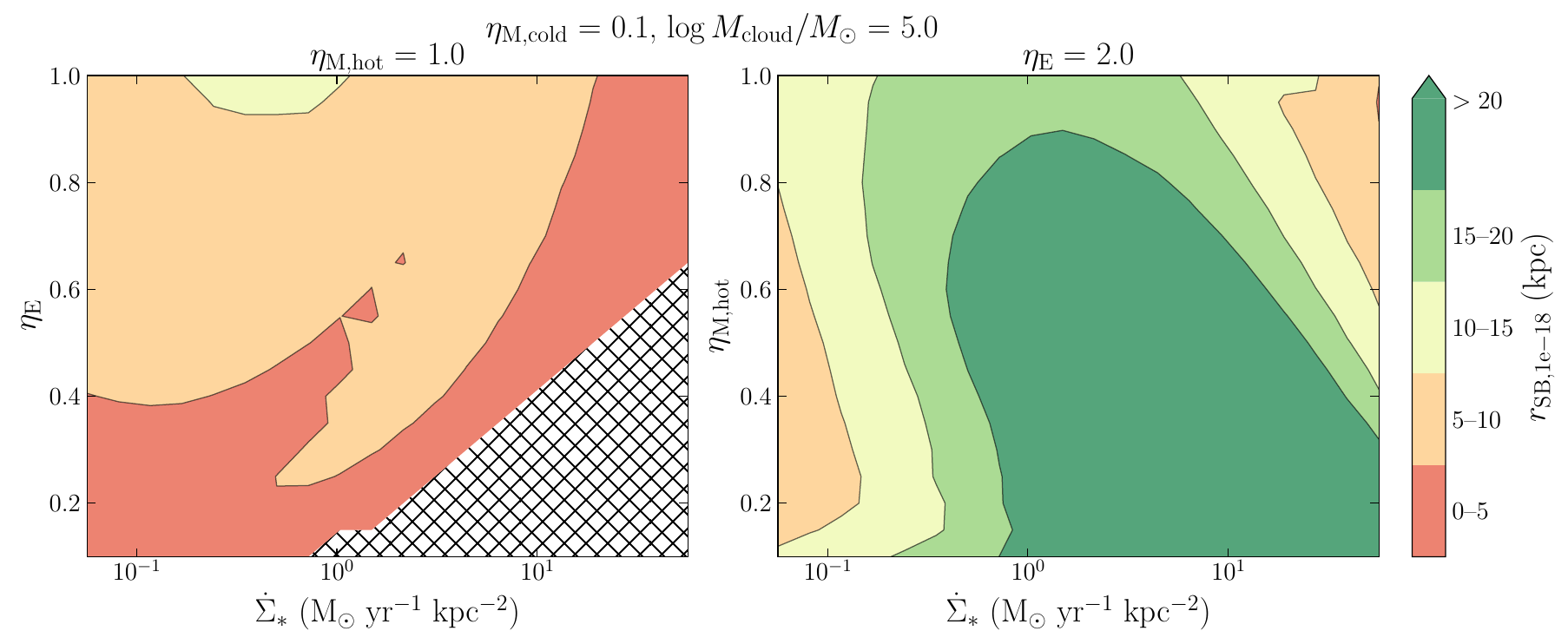}
\caption{Similar to \autoref{fig:ovi_detect_limit} but with $\eta_{\rm M,hot} = 1.0$ in the \textit{left} panel and $\eta_{\rm E} = 2.0$ in the \textit{right} panel. The left panel shows that when $\eta_{\rm M,hot}$ is removed from the sweet spot of $\sim 0.2-0.4$ we saw in \autoref{fig:ovi_detect_limit}, it is challenging to detect O VI emission regardless of the values of $\eta_{\rm E}$ and $\dot{\Sigma}_{\ast}$. 
The hatched region represents O VI profiles in high-$\dot{\Sigma}_{\ast}$ systems that terminate rapidly (i.e., $T_{\rm{mix}}$ and $T_{\rm{hot}}$ converge to $10^4 \ \rm{K}$) just beyond the sonic point.
The right panel shows that increasing $\eta_{\rm E}$ to 2.0 allows a larger portion of the parameter space to produce detectable emission.
}
\label{fig:ovi_detect_limit_a1}
\end{figure*}

\begin{figure*}
\centering
\includegraphics[width=\textwidth]{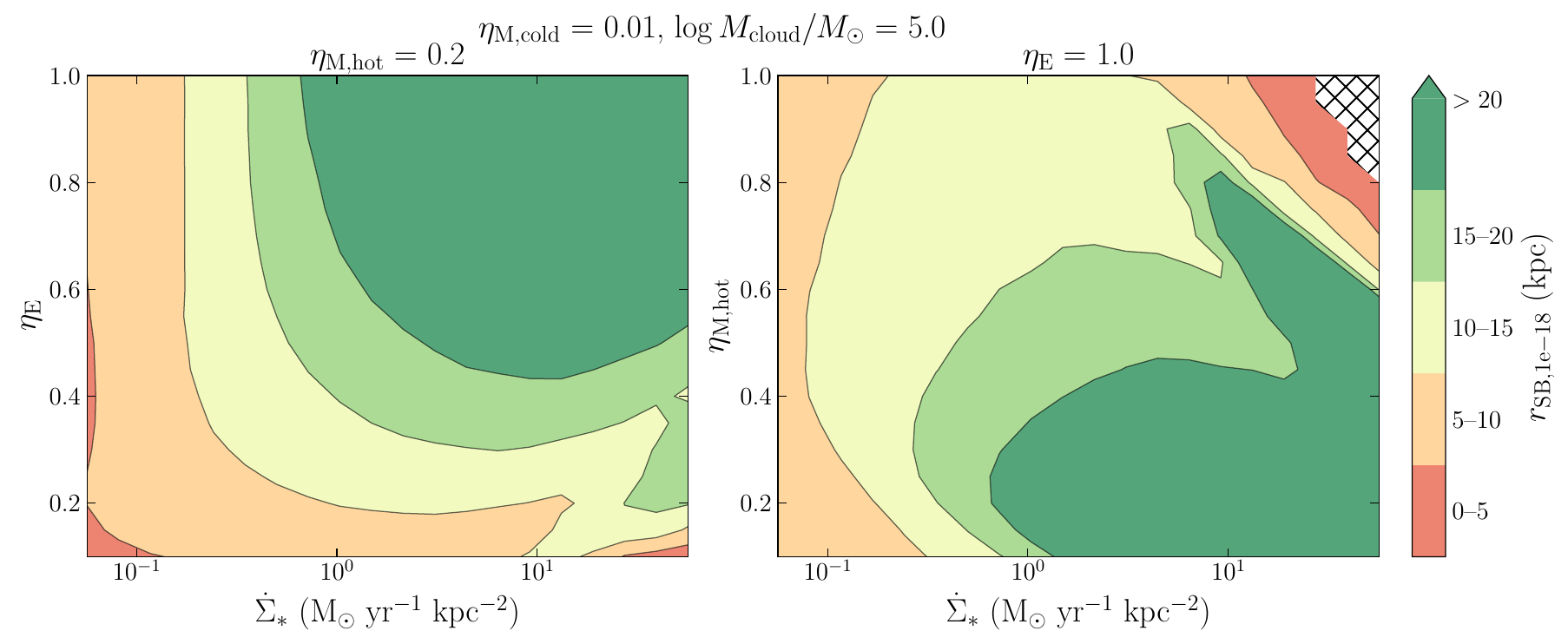}
\includegraphics[width=\textwidth]{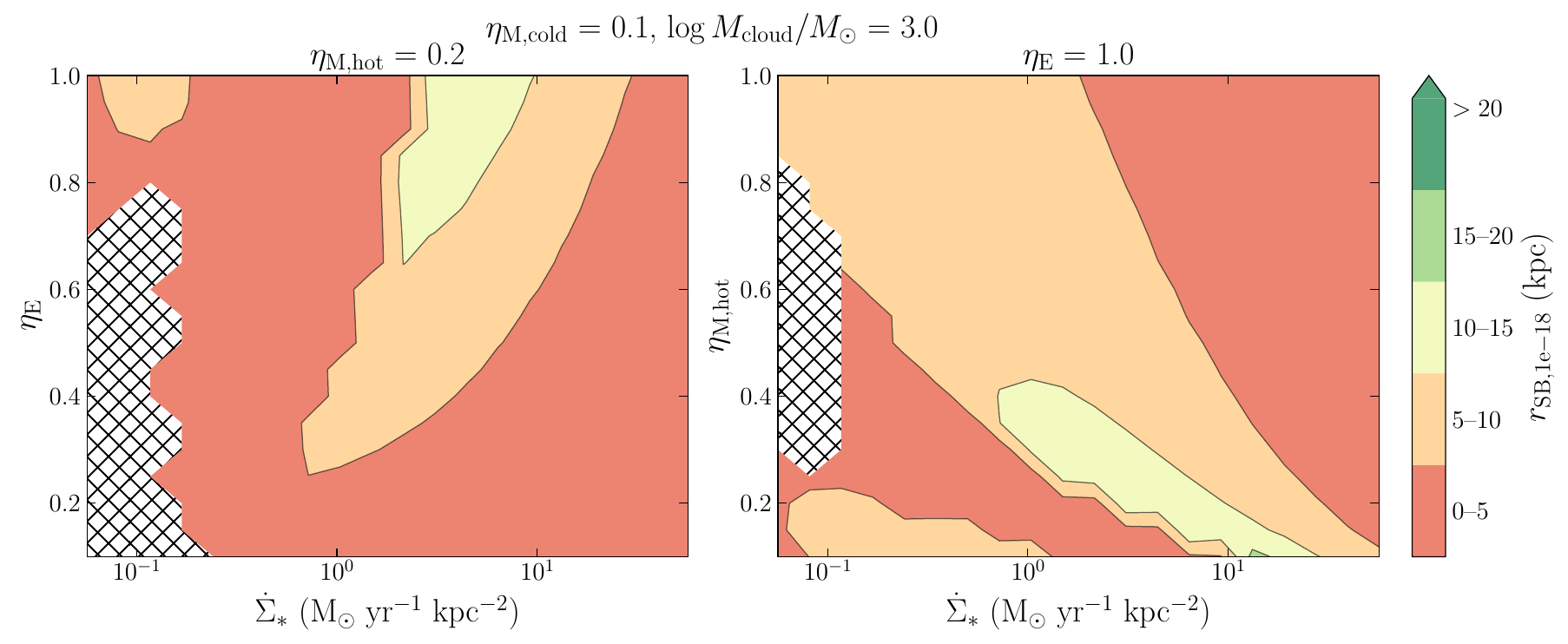}
\caption{Similar to \autoref{fig:ovi_detect_limit} but with $\eta_{\rm M,cold} = 0.01$ in the \textit{top} row and $\log M_{\rm{cloud}} / M_{\odot} = 3.0$ in the \textit{bottom} row. The \textit{top} row shows that lowering $\eta_{\rm M,cold}$ to 0.01 significantly boosts O VI detectability, as evident from the extended green regions. 
In \autoref{fig:SB_profile_vs_cold_phase_params}, we can see that lowering $\eta_{\rm M,cold}$ indeed makes the O VI SB profile shallower, but this effect is subtle and should not be capable of producing such a drastic improvement to detectability. Instead, the key difference maker here is the ram pressure. As shown in \autoref{fig:pram_over_pth_appendix}, the ram pressure is a factor of $\sim 20$ larger than the thermal pressure when $\eta_{\rm M,cold} = 0.01$. This is significantly higher than the factor of $\sim 6$ difference we saw in \autoref{fig:fb22_pressure_mrel_profiles} when $\eta_{\rm M,cold} = 0.1$. Thus, accounting for ram pressure provides a larger boost to O VI SB when $\eta_{\rm M,cold} = 0.01$, which explains the improved observability. 
As for the \textit{bottom} row, \autoref{fig:SB_profile_vs_cold_phase_params} shows that decreasing $M_{\rm{cloud}}$ steepens the SB profile at $r \lesssim 10$ kpc. This means the SB drops below the detection limit faster, hence the worsened detectability.}
\label{fig:ovi_detect_limit_a11}
\end{figure*}

\begin{figure*}
\centering
\includegraphics[width=\textwidth]{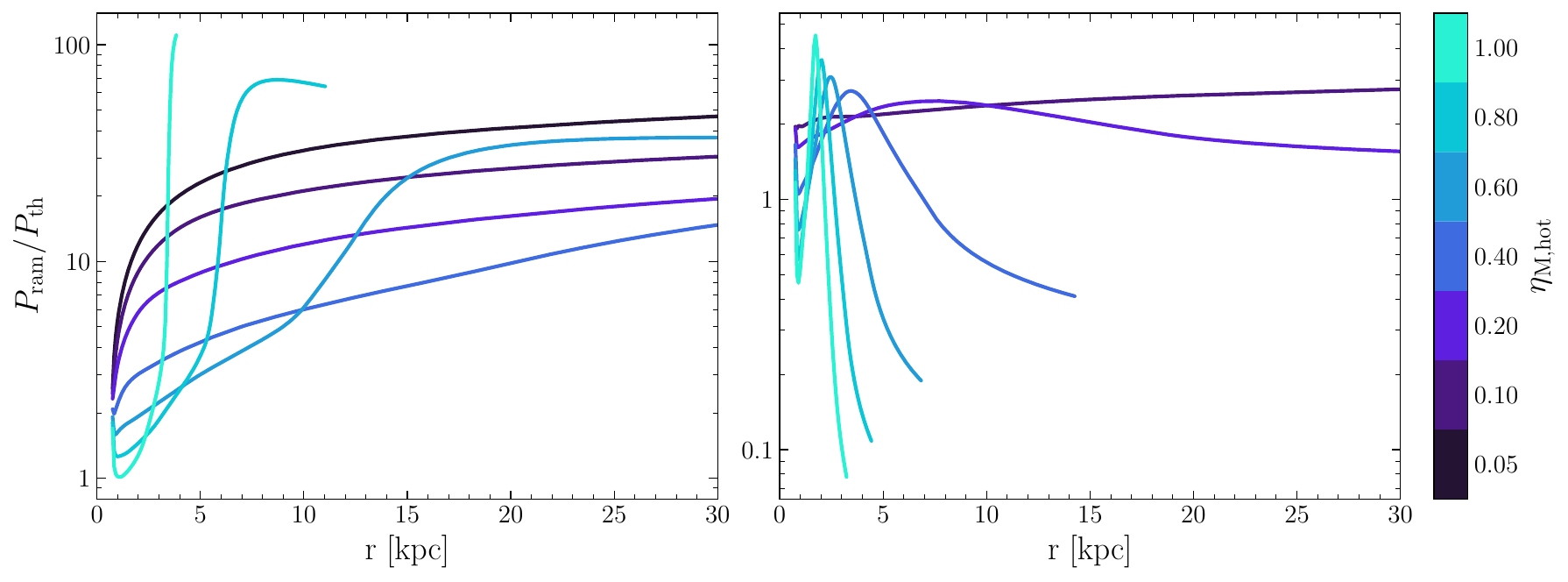}
\caption{Similar to \autoref{fig:fb22_pressure_mrel_profiles} but with $\eta_{\rm M,cold} = 0.01$ in the \textit{left} panel and $\log M_{\rm{cloud}} / M_{\odot} = 3.0$ in the \textit{right} panel. The \textit{left} and \textit{right} panels correspond to the \textit{top} and \textit{bottom} rows of \autoref{fig:ovi_detect_limit_a11}, respectively. In the \textit{left} panel, $\left. P_{\rm ram} \right/ P_{\rm th} \sim 20$, which is larger than the results in \autoref{fig:fb22_pressure_mrel_profiles} with fiducial parameter choices. This explains the significantly improved detectability in the \textit{top} row of \autoref{fig:ovi_detect_limit_a11}. The opposite is true for the \textit{right} panel and the \textit{bottom} row of \autoref{fig:ovi_detect_limit_a11}.}
\label{fig:pram_over_pth_appendix}
\end{figure*}

\autoref{fig:ovi_detect_limit_a1} and \autoref{fig:ovi_detect_limit_a11} show how O\,{\footnotesize VI} SB detectability depends on the hot-phase and cold-phase parameters, respectively, while \autoref{fig:ovi_detect_limit_a2} illustrates its dependence on the detection limit.

\autoref{fig:ovi_detect_limit_a1} shows that increasing $\eta_{\rm M,hot}$ from 0.2 to 1.0 (left panel) leads to a higher wind density and significantly enhanced radiative cooling within the hot phase. 
As a result, the O\,{\footnotesize VI} SB profile declines more rapidly and falls below the detection limit at smaller radii. 
This results in virtually no regimes with $r_{\rm SB,1e-18} \gtrsim 15 \ \rm{kpc}$ and a large parameter space (hatched region) in which the O\,{\footnotesize VI} profile terminates just outside the sonic point, where $T_{\rm{mix}}$ and $T_{\rm{hot}}$ converge to $10^4 \ \rm{K}$. 
In contrast, increasing $\eta_{\rm E}$ from 1.0 to 2.0 (right panel) enables most regions with $\eta_{\rm M,hot} \lesssim 0.5$ (except those with $\dot{\Sigma}_{\ast} \lesssim 1 \ M_{\odot} \ \rm{yr^{-1} \ kpc^{-2}}$) to remain detectable out to $\gtrsim 20 \ \rm{kpc}$.

The top (bottom) panels of \autoref{fig:ovi_detect_limit_a11} show that reducing $\eta_{\rm M,cold}$ ($M_{\rm cloud}$) from 0.1 to 0.01 ($10^5 \ M_{\odot}$ to $10^3 \ M_{\odot}$) significantly increases (decreases) the O\,{\footnotesize VI} detectability. Although both reductions produce shallower SB profiles with lower normalization (see \autoref{subsec:cold_phase_pars}), the dominant factor governing O\,{\footnotesize VI} detectability is ram pressure. The left panel of \autoref{fig:pram_over_pth_appendix} shows that lowering $\eta_{\rm M,cold}$ from 0.1 to 0.01 increases $P_{\rm ram}/P_{\rm th}$ from $\sim 5$ (fiducial setup; \autoref{fig:fb22_pressure_mrel_profiles}) to $\sim 20$ for $r \gtrsim 10 \ \rm{kpc}$ with $\eta_{\rm M,hot} = 0.2$. In contrast, the right panel shows that lowering $M_{\rm cloud}$ from $10^5 \ M_{\odot}$ to $10^3 \ M_{\odot}$ decreases $P_{\rm ram}/P_{\rm th}$ from $\sim 5$ to $\sim 2$.

Lowering $M_{\rm cloud}$ means the clouds are easier to accelerate, which reduces $v_{\rm rel}$ and $P_{\rm ram}/P_{\rm th} \propto \mathcal{M}_{\rm rel}^2$.
On the other hand, the impact of reducing $\eta_{\rm M,cold}$ on $P_{\rm ram}$ can be understood from Fig. 10 of \citetalias{FB_2022}, which shows the SFR dependence of a critical $\eta_{\rm M,cold}$ such that the hot phase (with $\eta_{\rm M,hot}=0.1$) retains $75\%$ of its energy flux at 1 kpc. They find this critical value scales as $\eta_{\rm M,cold} \propto \mathrm{SFR}^{-0.4}$, yielding $\eta_{\rm M,cold} \sim 0.1$ for $M_{\rm cloud} = 10^5 \ M_{\odot}$ and a $\dot{\Sigma}_{\ast}$ similar to \citetalias{Hayes_2016}. This scaling reflects the fact that high-SFR systems, with significantly enhanced cooling at smaller radii (\autoref{fig:SB_profile_vs_hot_phase_params}), require relatively small $\eta_{\rm M,cold}$ to exhibit a similar degree of cooling at 1 kpc compared to low-SFR systems. Therefore, since our adopted $\eta_{\rm M,cold} = 0.01$ is below this critical threshold, more energy is retained in the hot phase, enabling it to break out of the ISM with larger outflow velocity, leading to a larger $\mathcal{M}_{\rm rel}$ relative to the cold phase and a higher ram pressure.

As expected, \autoref{fig:ovi_detect_limit_a2} illustrates that the parameter space yielding detectable O\,{\footnotesize VI} emission at large radii (greenish region) is larger for the more sensitive detection limit and smaller for the less sensitive one. Despite these quantitative changes, the qualitative trends remain consistent with those shown in \autoref{fig:ovi_detect_limit}.

\begin{figure*}
\centering
\includegraphics[width=\textwidth]{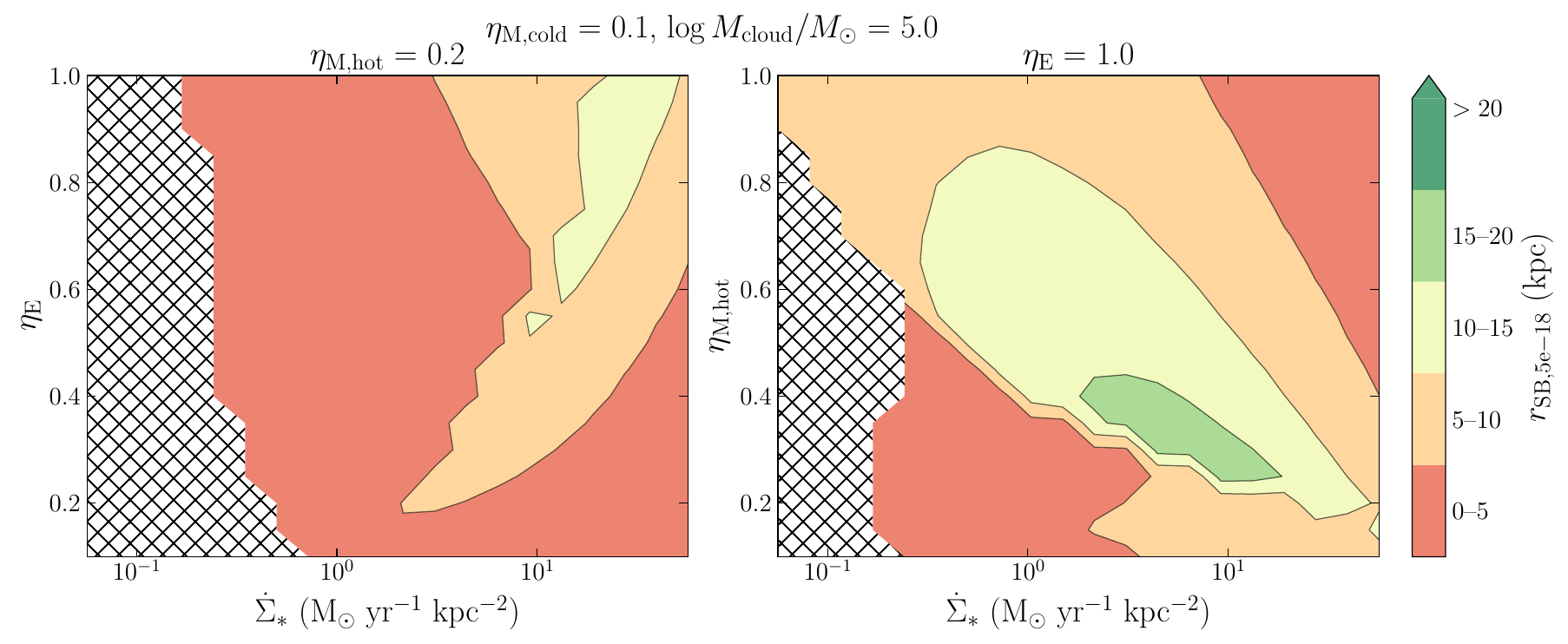}
\includegraphics[width=\textwidth]{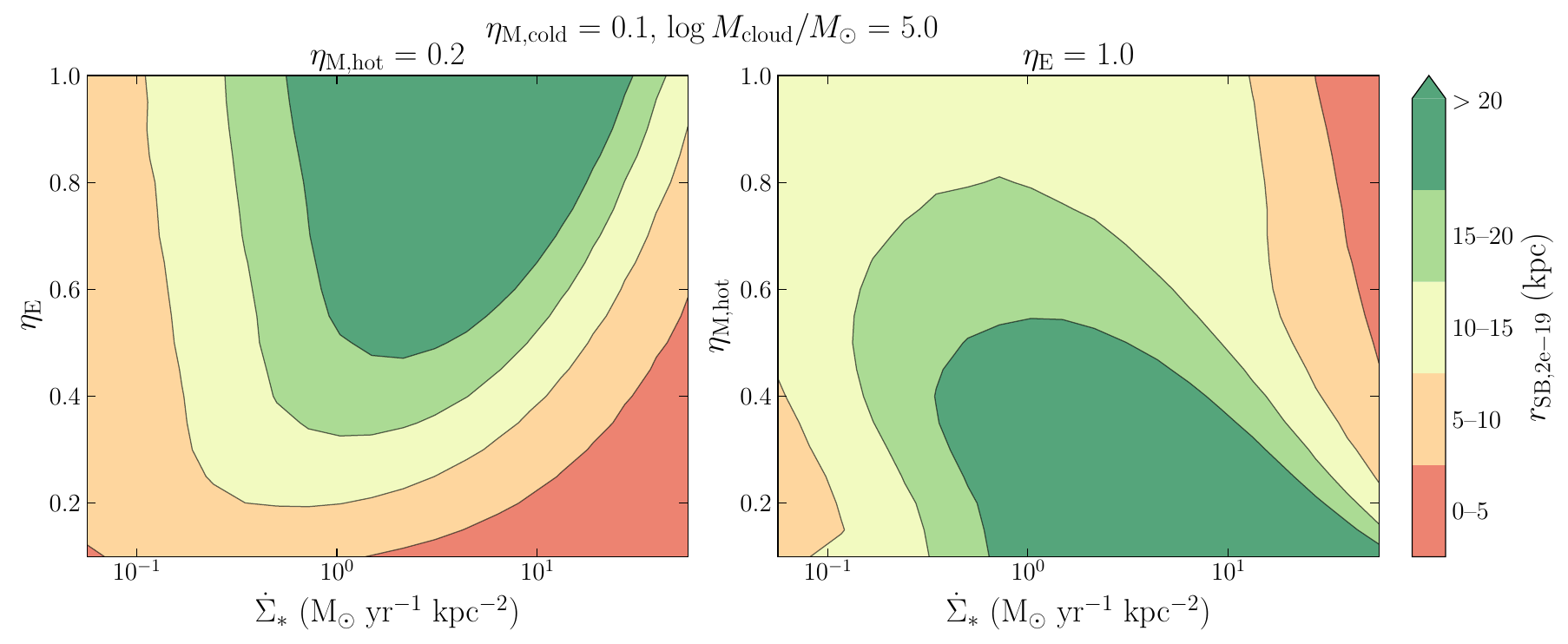}
\caption{Similar to \autoref{fig:ovi_detect_limit}, but with a detection limit that is a factor of 5 larger (\textit{top}) and smaller (\textit{bottom}) than that used in \autoref{fig:ovi_detect_limit}. As expected, a higher (lower) detection limit results in a smaller (larger) green region in the parameter space compared to \autoref{fig:ovi_detect_limit}. The qualitative trends seen in \autoref{fig:ovi_detect_limit} still hold.}
\label{fig:ovi_detect_limit_a2}
\end{figure*}


\bsp	
\label{lastpage}
\end{document}